 \documentclass[12pt]{article}
 \usepackage{amssymb}
 \usepackage{latexsym}
 \newcommand{\la}{\langle}
 \newcommand{\ra}{\rangle}
 \renewcommand{\r}{\right}
 \renewcommand{\l}{\left}
 \newcommand{\beq}{\begin{equation}}
 \newcommand{\eeq}{\end{equation}}
 \newcommand{\beqa}{\begin{eqnarray}}
 \newcommand{\eeqa}{\end{eqnarray}}
 \renewcommand{\d}[1]{\frac{\partial}{\partial #1}}
 \newcommand{\dd}[1]{\frac{\partial^2}{\partial #1^2}}
 \newcommand{\ab}[2]{\frac{\partial #1}{\partial #2}}
 \renewcommand{\lq}{{L^2(\Sigma,d^3\sigma)}}
 \newcommand{\co}[1]{{{\cal C}^\infty_0 (#1)}}
 
 \newcommand{\rf}[1]{(\ref{#1})}
 \renewcommand{\O}{{\cal O}}
 \newcommand{\U}{{\cal U}}
 
 \newcommand{\D}[1]{{\cal D}(#1)}
 \newcommand{\M}{{\cal M}}
 \newcommand{\Dp}[1]{{\cal D}'(#1)}
 \newcommand{\x}{(x_1,\xi_1;x_2,\xi_2)}
 \renewcommand{\H}{{{\cal H}}}
 \newcommand{\B}{{{\cal B}_1}}
 \newcommand{\A}{{\cal A}}

 \newcommand{\ph}{\phi_{\vec{k}}}
 \newcommand{\on}{{\Omega^{(n)}_k}}
 \newcommand{\dm}{\int\!d\mu(\vec{k})\,}
 \newcommand{\ol}{\overline}
 \newcommand{\vx}{\mathbf{x}}
 \newcommand{\vk}{{\vec{k}}}
 
 \newtheorem{thm}{Theorem}[section]
 \newtheorem{lemma}[thm]{Lemma}
 \newtheorem{dfn}[thm]{Definition}
 \newtheorem{cor}[thm]{Corollary}
 \newtheorem{prop}[thm]{Proposition}
 \newtheorem{remark}[thm]{Remark}
 \newenvironment{beweis}{{\em Proof:}}{\hfill $\rule{2mm}{2mm}$
 \vspace{3mm}     }

 \newcommand{\gk}{{\gtrless}}
\newcommand{\ci}{{\cal C}^\infty}
 \newcommand{\supp}{\mathrm{supp}\,}
  \def\C{\mathbb{C}}
  \def\N{\mathbb{N}}
  \def\R{\mathbb{R}}
  \def\Z{\mathbb{Z}}
\DeclareSymbolFont{ASMa}{U}{msa}{m}{n}
\DeclareMathSymbol{\hrist}{\mathord}{ASMa}{"16}

\begin{document}
\title{Adiabatic Vacuum States on General
Spacetime Manifolds: Definition, Construction, and Physical
Properties}

\author{ Wolfgang Junker\thanks{e-mail: junker@aei-potsdam.mpg.de} \quad
Elmar Schrohe\thanks{e-mail: schrohe@math.uni-potsdam.de}\\  \\
        ${}^*$Max-Planck-Institut f\"ur Gravitationsphysik\\
        Albert-Einstein-Institut\\
        Am M\"uhlenberg 1\\
        D-14476 Golm, Germany\\  \\
${}^\dag$Universit\"at Potsdam\\
Institut f\"ur Mathematik\\
Am Neuen Palais 10\\
D-14415 Potsdam, Germany
        }     
\date{{\small March 15, 2002\\AEI-2001-113\\{\it Ann.~Henri
Poincar{\'e}} {\bf 3} (2002) 1113--1181}}
\maketitle
\thispagestyle{empty}
%
\begin{abstract}  
Adiabatic vacuum states are a well-known class of physical states for
linear quantum fields on Robertson-Walker spacetimes. We extend the
definition of adiabatic vacua to general spacetime manifolds by using
the notion of the Sobolev wavefront set. This definition is also applicable
to interacting field theories. Hadamard states form a special subclass
of the adiabatic vacua. We analyze physical properties of adiabatic
vacuum representations of the Klein-Gordon field on globally
hyperbolic spacetime manifolds (factoriality, quasiequivalence, local
definiteness, Haag duality) and construct them explicitly, if
the manifold has a compact Cauchy surface.
\end{abstract}
\newpage
%
\tableofcontents
\section{Introduction}
It has always been one of the main problems of quantum field theory on
curved spacetimes to single out a class of physical states among the
huge set of positive linear functionals on the algebra of
observables. One prominent choice for linear field theories is the
class of Hadamard states. It has been much investigated in the past,
but only recently gained a deeper understanding due to the work of
Radzikowski \cite{Radzikowski96a}. He showed that the Hadamard states
are characterized by the wavefront set of their two-point functions
(see Definition \ref{Definition0}). This characterization immediately
allows for a generalization to interacting fields \cite{BFK96} and
puts all the techniques of microlocal analysis at our disposal
\cite{HormIII,HormIV,HormI}. They have made possible the construction of
the free field theory \cite{Junker96} and the perturbation theory
\cite{BF00} on general spacetime manifolds.\\
On the other hand, there is another well-known class of states for
linear field theories on Robertson-Walker spaces, the so-called
adiabatic vacuum states. They were introduced by Parker
\cite{Parker69} to describe the particle creation by the expansion of
cosmological spacetime models. Much work has also been devoted to the
investigation of the physical (for a review see \cite{Fulling89}) and
mathematical \cite{LR90} properties of these states, but it has never
been known how to extend their definition to field theories on general
spacetime manifolds. Hollands \cite{Hollands01} recently defined these
states for Dirac fields on Robertson-Walker spaces and observed that
they are in general not of the Hadamard form (correcting an erroneous
claim in \cite{Junker96}).\\
It has been the aim of the present work to find a microlocal definition of
adiabatic vacuum states which makes sense on arbitrary spacetime
manifolds and can be extended to interacting fields, in close analogy
to the Hadamard states. It turned out that the notion of the Sobolev
(or $H^s$-) wavefront set is the appropriate mathematical tool for
this purpose. In Appendix \ref{AppendixB} we review this notion and
the calculus related to it. After an introduction to the structure
of the algebra of observables of the Klein-Gordon quantum field on a
globally hyperbolic spacetime manifold $(\M,g)$ in Section
\ref{Section1} we present our definition of adiabatic states of order
$N$ (Definition \ref{Definition1}) in Section \ref{Section2}. It
contains the Hadamard states as a special case: they are adiabatic
states ``of infinite order''. To decide which order of adiabatic
vacuum is physically admissible we investigate the algebraic structure
of the corresponding GNS-representations. Haag, Narnhofer \& Stein
\cite{HNS84} suggested as a criterion for physical representations
that they should locally generate von Neumann factors that have all
the same set of normal states (in other words, the representations are
locally primary and quasiequivalent). We show in Section
\ref{Section3.1} (Theorem \ref{Theorem25} and Theorem \ref{Theorem29})
that this is generally the case if $N>5/2$. For the case of pure states
on a spacetime with compact Cauchy surface, which often occurs in
applications, we improve the admissible order to $N>3/2$. 
In addition, in Section 
\ref{Section3.2} we show that adiabatic vacua of order $N>5/2$ satisfy
the properties of local definiteness (Corollary \ref{Corollary32a})
and those of order $N>3/2$
Haag duality (Theorem \ref{Theorem34}). These results extend
corresponding statements for adiabatic vacuum states on Robertson-Walker
spacetimes due to L\"uders \& Roberts \cite{LR90}, and
for Hadamard states due to
Verch \cite{Verch97}; for their discussion in the framework of
algebraic quantum field theory we refer to \cite{Haag96}. In Section
\ref{Section5} we explicitly construct pure adiabatic vacuum states
on an arbitrary spacetime manifold with compact Cauchy surface
(Theorem \ref{Theorem5.7}). In Section \ref{Section6} we show that our
adiabatic states are indeed a generalization of the well-known
adiabatic vacua on Robertson-Walker spaces: Theorem \ref{Theorem6.2}
states that the adiabatic vacua of order $n$ (according to the
definition of \cite{LR90}) on a Robertson-Walker spacetime with
compact spatial section are adiabatic vacua of order $2n$ in the sense
of our microlocal Definition \ref{Definition1}. We conclude in Section
\ref{Section7} by summarizing the physical interpretation of our
mathematical analysis and calculating the response of an Unruh
detector to an adiabatic vacuum state. It allows in principle to
physically distinguish adiabatic states of different orders.
Appendix
\ref{AppendixA} provides a survey of the Sobolev spaces which are used
in this paper.
\section{The Klein-Gordon field in globally hyperbolic spacetimes}
\label{Section1}
We assume that spacetime is modeled by a 4-dimensional paracompact
$\ci$-manifold 
$\M$ without boundary endowed with a Lorentzian metric $g$ of
signature $(+---)$ such that $(\M,g)$ is globally hyperbolic. This
means that there is a 3-dimensional smooth spacelike hypersurface
$\Sigma$ (without boundary) which is intersected by each inextendible
causal (null or timelike) curve in $\M$ exactly once. As a consequence 
$\M$ is time-orientable, and we fix one orientation once and for all
defining ``future'' and ``past''. $\Sigma$ is also assumed to be
orientable. Our units are chosen such that $\hbar=c=G=1$.\\
In this work, we are concerned with the quantum theory of the linear 
Klein-Gordon field in globally hyperbolic spacetimes. We first 
present the properties of the classical scalar field in order to
introduce the phase space that underlies the quantization procedure.
Then we construct the Weyl algebra and define the set of quasifree
states on it. The material in this section is based on the papers
\cite{MV68,Dimock80,KW91}. Here, all function spaces are considered
to be spaces of {\it real}-valued functions.\\
Let us start with the Klein-Gordon equation
\beqa
(\Box_g+m^2)\Phi &=& (g^{\mu\nu}\nabla_\mu\nabla_\nu +m^2)\Phi
\label{1.1}\\
&=& \frac{1}{\sqrt {\mathfrak g}}\partial_\mu (g^{\mu\nu}\sqrt{\mathfrak g}\,
\partial_\nu \Phi)+m^2\Phi =0 \nonumber
\eeqa
for a scalar field $\Phi:\M\to {\R}$ on a globally hyperbolic 
spacetime $(\M,g)$ where $g^{\mu\nu}$ is the inverse matrix of
$g=(g_{\mu\nu})$, ${\mathfrak g}:=|\det (g_{\mu\nu})|$, $\nabla_\mu$ the
Levi-Civita connection associated to $g$ and $m >0$ the mass of
the field. Since \rf{1.1} is a 
hyperbolic differential equation, the Cauchy problem on a globally
hyperbolic space is well-posed. As a consequence (see
e.g.~\cite{Dimock80}) , there are two unique
continuous linear operators
\[ E^{R,A}:\D{\M}\to \ci(\M)
\]
with the properties
\beqa
&&(\Box_g+m^2)E^{R,A} f=E^{R,A} (\Box_g+m^2)f =f
\nonumber\\
&&\supp(E^A f)\subset J^-(\supp f)\nonumber\\
&&\supp(E^R f)\subset J^+(\supp f)\nonumber
\eeqa
for $f\in\D{\M}$ where $J^{+/-}(S)$ denotes the causal future/past of a
set $S\subset \M$, i.e.~the set of all points $x\in \M$ that can be
reached by future/past-directed causal 
(i.e.~null or timelike) curves emanating from $S$.
 They are called the advanced ($E^A$) and
retarded ($E^R$) fundamental solutions of the Klein-Gordon
equation \rf{1.1}. 
$E:=E^R-E^A$ is called the fundamental solution or classical
propagator of \rf{1.1}. It has the properties
\beqa
&&(\Box_g+m^2)Ef=E(\Box_g+m^2)f=0\label{1.3}\\
&&\supp(Ef)\subset J^+(\supp f)\cup J^-(\supp f)\nonumber
\eeqa
for $f\in\D{\M}$. 
$E^R,E^A$ and $E$ can be continuously extended to the
adjoint operators
\[ E^{R\prime}, E^{A\prime}, E':{\cal E}'(\M)\to\Dp{\M}
\]
by $E^{R\prime}=E^A,\; E^{A\prime}=E^R,\;E'=-E$.\\
Let $\Sigma$ be a given Cauchy surface of $\M$ with
future-directed unit normal field $n^\alpha$. Then we denote by
\beqa
\rho_0: &\ci(\M)&\to \ci(\Sigma) \nonumber\\
 &u& \mapsto u|_\Sigma \nonumber\\
\rho_1:&\ci(\M)&\to \ci(\Sigma) \label{1.4}\\
&u&\mapsto \partial_n u|_\Sigma:=(n^\alpha \nabla_\alpha u)|_\Sigma\nonumber
\eeqa
the usual restriction operators, while 
$\rho_0',\rho_1':{\cal E}'(\Sigma)\to
{\cal E}'(\M)$ denote their adjoints. 
Dimock \cite{Dimock80} proves the following
existence and uniqueness result for the Cauchy problem:
\begin{prop}\label{theorem1.1}
(a) $E\rho_0', E\rho_1'$ restrict to continuous operators from
$\D{\Sigma}$ ($\subset {\cal E}'(\Sigma)$) to ${\cal E}(\M)$ ($\subset
\Dp{\M}$), and the unique solution of the Cauchy problem \rf{1.1}
with initial data $u_0,u_1 \in \D{\Sigma}$ is given by
\beq
u=E\rho_0'u_1-E\rho_1'u_0. \label{1.5}
\eeq
(b) Furthermore, \rf{1.5} also holds in the sense of distributions, 
i.e.~given $u_0,u_1\in\Dp{\Sigma}$, there exists a unique distribution
$u\in\Dp{\M}$ which is a (weak) solution of \rf{1.1} and has initial
data $u_0=\rho_0u, u_1=\rho_1u$ (the restrictions in the sense of
Proposition~\ref{PropositionA6}). It is given by
\[
u(f)=-u_1(\rho_0Ef)+u_0(\rho_1Ef)
\]
for $f\in\D{\M}$.\\
(c) If $u$ is a smooth solution of \rf{1.1} with $\supp u_{0,1}$
contained in a bounded subset $\O\subset \Sigma$ then, for any open
neighborhood $\U$ of $\O$ in $\M$, there exists an $f\in \D{\U}$
with $u=Ef$.
\end{prop}
Inserting $u=Ef$ into both sides of Eq.~\rf{1.5} we get the identity
\beq
E= E\rho_0'\rho_1 E -E\rho_1'\rho_0 E \label{1.6}
\eeq
on $\D{\M}$. 
Proposition~\ref{theorem1.1} allows us to describe the phase
space of the classical field theory and the local observable algebras of the 
quantum field theory in two different 
(but equivalent) ways. One
uses test functions in $\D{\M}$, the other
the Cauchy data with compact support on $\Sigma$.
The relation between them is then established
with the help of the fundamental solution $E$ and Proposition 
\ref{theorem1.1}:\\

Let ($\tilde{\Gamma},\tilde{\sigma}$) be the real linear
symplectic space defined by $\tilde{\Gamma}:= \D{\M}/ker\,E$,
$\tilde{\sigma}([f_1],[f_2]) := \la f_1,E f_2\ra$. $\tilde{\sigma}$ is
independent of the choice of representatives $f_1,f_2\in
\D{\M}$ and defines a non-degenerate symplectic bilinear form
on $\tilde{\Gamma}$. For any open $\U\subset \M$ there is a local
symplectic subspace $(\tilde{\Gamma}(\U),\tilde{\sigma})$ of
$(\tilde{\Gamma},\tilde{\sigma})$ defined by
$\tilde{\Gamma}(\U):=\D{\U}/ker\,E$. To a symplectic space
$(\tilde{\Gamma},\tilde{\sigma})$ there is associated (uniquely up to
$*$-isomorphism) a Weyl algebra ${\cal A}[\tilde{\Gamma},
\tilde{\sigma}]$, which is a simple abstract 
$C^*$-algebra generated by the elements
$W([f]),\;[f]\in\tilde{\Gamma}$, that satisfy
\beqa
&&W([f])^*=W([f])^{-1}=W([-f])\quad\mbox{(unitarity)}\nonumber\\
&&W([f_1])W([f_2])=e^{-\frac{i}{2}\tilde{\sigma}([f_1],[f_2])}
W([f_1+f_2])\quad \mbox{(Weyl relations)} \label{1.9}
\eeqa
for all $[f],[f_1],[f_2]\in\tilde{\Gamma}$ (see e.g.~\cite{BW92}). The 
Weyl elements satisfy the ``field equation'' $W([(\Box_g +m^2) f])
=W(0) = {\bf 1}$.
(In a regular representation we can think of the elements $W([f])$ as 
the unitary operators $e^{i\hat{\Phi}([f])}$ where
$\hat{\Phi}([f])$ is the usual field operator smeared
with test functions $f\in\D{\M}$ and satisfying the field equation
$(\Box_g +m^2)\hat{\Phi}([f])=\hat{\Phi}([(\Box_g+m^2)f])=0$. 
\rf{1.9} then corresponds to the canonical commutation relations.)
A local subalgebra ${\cal A (\U)}$ (${\cal U}$ an open bounded subset of
$\M$) is then given by $\A[\tilde{\Gamma}(\U),\tilde{\sigma}]$. It is the
 $C^*$-algebra generated by the elements $W([f])$
with $\supp f\subset{\cal U}$ and contains the quantum
observables measurable in the spacetime region ${\cal U}$.
Then ${\cal A}[\tilde{\Gamma},\tilde{\sigma}]=C^*\l(\bigcup_{{\cal
U}}{\cal A(U)}\r)$.\\
Dimock \cite{Dimock80} has shown that ${\cal U}\mapsto {\cal A(U)}$
is a net of local observable algebras in the sense of Haag and Kastler
\cite{HK64}, i.e.~it satisfies\\
(i) ${\cal U}_1\subset{\cal U}_2\Rightarrow {\cal A(U}_1)\subset
{\cal A(U}_2)$ (isotony).\\
(ii) ${\cal U}_1$ spacelike separated from ${\cal U}_2 \Rightarrow
[{\cal A(U}_1),{\cal A(U}_2)]=\{0\}$ (locality).\\
(iii) There is a faithful irreducible representation of ${\cal A}$
(primitivity).\\
(iv) ${\cal U}_1\subset D({\cal U}_2)\Rightarrow {\cal A(U}_1)\subset
{\cal A(U}_2)$.\\
(v) For any isometry $\kappa:(\M,g)\to(\M,g)$ there is an isomorphism
$\alpha_\kappa:{\cal A}\to{\cal A}$ such that $\alpha_\kappa[{\cal
A(U)}]={\cal A}(\kappa({\cal U}))$ and $\alpha_{\kappa_1}\circ
\alpha_{\kappa_2}=\alpha_{\kappa_1\circ\kappa_2}$ (covariance).\\
In (iv), $D(\U)$ denotes the domain of dependence of $\U\subset \M$,
i.e.~the set of all points $x\in\M$ such that every inextendible causal 
curve through $x$ passes through ${\cal U}$.\\

Since we are dealing with a linear field equation we can equivalently
use the time zero algebras for the description of the quantum field
theory. To this end we pick a Cauchy surface $\Sigma$ with volume element
$d^3\sigma :=\sqrt{\mathfrak h}\,d^3x$, 
where ${\mathfrak h}:=\det (h_{ij})$ and 
$h_{ij}$ is the Riemannian metric induced on $\Sigma$ by $g$,  
and define a classical phase space $(\Gamma,\sigma)$ of the 
Klein-Gordon field by the space 
$\Gamma:=\D{\Sigma}\oplus\D{\Sigma}$ of real-valued initial
data with compact support and the real symplectic bilinear form
\beqa
\sigma:  \Gamma\times\Gamma &\to& {\R} \nonumber\\
(F_1,F_2) &\mapsto&- \int_\Sigma [q_1p_2-q_2p_1]\,d^3\sigma, \label{1.8}
\eeqa
$F_i:=(q_i, p_i)\in\Gamma, i=1,2$. In this case, the local subspaces
$\Gamma(\O):=\D{\O}\oplus \D{\O}$ are associated to bounded open
subsets $\O\subset\Sigma$. 
The next proposition establishes the equivalence between the two
formulations of the phase space:
\begin{prop}\label{Proposition1.1a}
The spaces $(\Gamma(\O),\sigma)$ and $(\tilde{\Gamma}(D(\O)),
\tilde{\sigma})$ are symplectically isomorphic. The isomorphism is given by
\begin{eqnarray*}
\rho_\Sigma: \tilde{\Gamma}(D(\O)) &\to& \Gamma(\O)\\
 {[f]} & \mapsto & (\rho_0 Ef,\rho_1 Ef).
\end{eqnarray*}
\end{prop}
The proof of the proposition is a simple application of Proposition
\ref{theorem1.1} and Eq.~\rf{1.6}. It shows in
particular that the symplectic form $\sigma$, Eq.~\rf{1.8}, is
independent of the choice of Cauchy surface $\Sigma$.\\
Now, to $(\Gamma,\sigma)$ we can associate the Weyl algebra
$\A[\Gamma,\sigma]$ with its local subalgebras
$\A(\O):=\A[\Gamma(\O),\sigma]$. By uniqueness, $\A(\O)$ is isomorphic 
(as a $C^*$-algebra) to $\A(D(\O))$
which should justify our misuse of the same letter $\A$. The
$*$-isomorphism is explicitly given by
\[\alpha: \A(D(\O))\to \A(\O),\quad
\alpha W([f]) := W(\rho_\Sigma([f])).\]
In the rest of the paper we will only have to deal with the net
$\O\mapsto \A(\O)$ of local time zero algebras, since they naturally
occur when one discusses properties of a linear quantum field
theory. Nevertheless, by the above isomorphism, one can translate all
properties of this net easily into statements about the net $\U\mapsto 
\A(\U)$ and vice versa. Let us only mention here that locality of the time
zero algebras means that $[\A(\O_1),\A(\O_2)]=\{0\}$ if $\O_1\cap
\O_2=\emptyset$. \\
The states on an observable algebra ${\cal A}$ are the linear 
functionals $\omega:{\cal A}\to {\bf C}$ satisfying $\omega({\bf 1})
=1$ (normalization) and $\omega(A^*A)\geq 0\;\forall A\in\A$
(positivity). The set of states on our Weyl algebra
$\A[\Gamma,\sigma]$ is by far too large to be tractable in a concrete
way. Therefore, for linear systems, one usually restricts oneself to
the quasifree states, all of whose truncated $n$-point functions vanish for
$n\not=2$:
\begin{dfn}\label{dfn1.2}
Let $\mu:\Gamma\times\Gamma\to{\R}$ be a real scalar product 
satisfying
\beq
\frac{1}{4}|\sigma(F_1,F_2)|^2\leq \mu(F_1,F_1)\mu(F_2,F_2)
\label{1.11}
\eeq
for all $F_1,F_2\in\Gamma$.
Then the {\rm\bf quasifree state} $\omega_\mu$ associated with $\mu$
is given by 
\[
\omega_\mu(W(F))=e^{-\frac{1}{2}\mu(F,F)}.
\]
If $\omega_\mu$ is pure it is called a {\rm\bf Fock state}.
\end{dfn}
The connection between this algebraic notion of a quasifree state
and the usual notion of ``vacuum state'' in a Hilbert space is
established by the following proposition which we cite from \cite{KW91}:
\begin{prop}\label{Proposition1.3}
Let $\omega_\mu$ be a quasifree state on $\A[\Gamma,\sigma]$.
\begin{enumerate}
\item[(a)] There exists a {\rm\bf one-particle Hilbert space structure},
i.e.~a Hilbert space ${\cal H}$ and a real-linear map $k:\Gamma\to
{\cal H}$ such that \\
(i) $k\Gamma +ik\Gamma$ is dense in ${\cal H}$,\\
(ii) $\mu(F_1,F_2)={\rm Re}\la kF_1,kF_2\ra_{\cal H}\;\forall
F_1,F_2\in \Gamma$,\\
(iii) $\sigma(F_1,F_2)=2{\rm Im}\la kF_1,kF_2\ra_{\cal H}\;
\forall F_1,F_2\in\Gamma$.\\
The pair $(k,{\cal H})$ is uniquely determined up to
unitary equivalence.\\ Moreover: $\omega_\mu$ is pure
$\Leftrightarrow\;k(\Gamma)$ is dense in ${\cal H}$.
\item[(b)] The GNS-triple $({\cal H}_{\omega_\mu},\pi_{\omega_\mu},\Omega_{
\omega_\mu})$ of the state $\omega_\mu$ can be represented as
$({\cal F}^s({\cal H}),\rho_\mu,\Omega^{{\cal F}})$, where\\
(i) ${\cal F}^s({\cal H})$ is the symmetric Fock space over the
one-particle Hilbert space ${\cal H}$,\\
(ii) $\rho_\mu [W(F)]=\exp\{-i\ol{[a^*(kF)+a(kF)]}\}$, where $a^*$
and $a$ are the standard creation and annihilation operators on
${\cal F}^s({\cal H})$ satisfying 
\[[a(u),a^*(v)]=\la u,v\ra_{\cal H}\;\mbox{and}\;a(u)\Omega^{\cal F}=0
\]
for $u,v\in{\cal H}$. (The bar over $a^*(kF)+a(kF)$ indicates that we
take the closure of this operator initially defined on the space of
vectors of finite particle number.)\\
(iii) $\Omega^{\cal F}:={\bf 1}\oplus{\bf 0}\oplus{\bf 0}\oplus\ldots$
is the (cyclic) Fock vacuum.\\
Moreover: $\omega_\mu$ is pure $\Leftrightarrow$ $\rho_\mu$ is irreducible.
\end{enumerate}
\end{prop}
Thus, $\omega_\mu$ can also be represented as $\omega_\mu(W(F))=
\exp\{-\frac{1}{2}||kF||^2_{\cal H}\}$ (by (a)) or 
$\omega_\mu(W(F))=\la \Omega^{\cal F},\rho_\mu (F)\Omega^{\cal F}\ra$
(by (b)). $\hat{\Phi}(F):=\ol{a^*(kF)+a(kF)}$ is the usual field
operator on ${\cal F}^s({\cal H})$ and we can determine the 
(``symplectically smeared'') two-point function as
\beqa
\lambda(F_1,F_2)&=& \la \Omega^{\cal F},
\hat{\Phi}(F_1)\hat{\Phi}(F_2)\Omega^{\cal F}\ra  \nonumber\\
&=& \la kF_1,kF_2\ra_{\cal H}  \nonumber\\
&=& \mu(F_1,F_2)+\frac{i}{2}\sigma (F_1,F_2) \label{1.13}
\eeqa
for $F_1,F_2\in \Gamma$,
resp.~the Wightman two-point function $\Lambda$ as
\beq
\Lambda(f_1,f_2)=\lambda\l({\rho_0Ef_1\choose \rho_1Ef_1}
,{\rho_0 Ef_2\choose \rho_1Ef_2}\r)\label{1.14}
\eeq
for $f_1,f_2\in\D{\M}$. 
The fact that the antisymmetric (= imaginary) part of $\lambda$
is the symplectic form $\sigma$ implies for $\Lambda$:
\beqa
{\rm Im}\,\Lambda(f_1,f_2)&=&-\frac{1}{2}\int_\Sigma
[f_1 E'\rho_0'\rho_1 E f_2-f_1E'\rho_1'\rho_0 E f_2]\,d^3\sigma \nonumber\\
&=& \frac{1}{2}\la f_1,Ef_2\ra \label{1.15}
\eeqa
by Eq.~\rf{1.6}.
All the other $n$-point functions can also be
calculated, one finds that they vanish if $n$ is odd and that the
$n$-point functions for $n$ even are sums of products of two-point
functions.\\

Once a (quasifree) state $\omega$ on the algebra $\A$ has been chosen the
GNS-representation $(\H_\omega,\pi_\omega,\Omega_\omega)$ of Proposition
\ref{Proposition1.3} allows us to represent all the algebras $\A(\O)$
as concrete algebras $\pi_\omega(\A(\O))$ of bounded operators on
$\H_\omega$. The weak closure of $\pi_\omega(\A(\O))$ in ${\cal
B}(\H_\omega)$, which, by von Neumann's double commutant theorem, is
equal to $\pi_\omega(\A(\O))''$ (the prime denoting the commutant of a 
subalgebra of ${\cal B}(\H_\omega)$), is denoted by
${\cal R}_\omega(\O)$. It is the net of von Neumann algebras $\O\mapsto {\cal
R}_\omega(\O)$ which contains all the physical information of the
theory and is therefore the main object of study in algebraic quantum
field theory (see e.g.~\cite{Haag96}). 
One of the most straightforward properties is the so-called
additivity. It states that if an open bounded subset $\O\subset\Sigma$ 
is the union of open subsets $\O=\bigcup_i \O_i$ then the von Neumann
algebra ${\cal R}_\omega(\O)$ is generated by the subalgebras ${\cal
R}_\omega(\O_i)$, i.e.
\beq
{\cal R}_\omega(\O) = \l(\bigcup_i {\cal R}_\omega(\O_i)\r)''.
\label{1.16}
\eeq
Additivity expresses the fact that the physical information contained
in ${\cal R}_\omega(\O)$ is entirely encoded in the observables that
are localized in arbitrarily small subsets of $\O$. The following
result is well-known:
\begin{lemma}\label{Lemma1.5}
Let $\omega$ be a quasifree state of the Weyl algebra, $\O$ an open
bounded subset of $\Sigma$. Then ${\cal R}_\omega(\O)$ is additive.
\end{lemma}
\begin{beweis}
Let $(k,\H)$ be the one-particle Hilbert space structure of $\omega$
(Proposition \ref{Proposition1.3}). According to results of Araki
\cite{Araki63,LRT78} Eq.~\rf{1.16} holds iff
\beq
\ol{k\Gamma(\O)}=\ol{\mbox{span}\,k\Gamma(\O_i)} \label{1.17}
\eeq
where the closure is taken w.r.t.~the norm in $\H$. With the help of a 
partition of unity $\{\chi_i;\;\supp\chi_i\subset \O_i\}$ it is clear
that any $u= k(F)\in k\Gamma(\O),\,F\in \Gamma(\O),$ can be written as 
$u= \sum_ik(\chi_iF)\in \mbox{span}\,k\Gamma(\O_i)$ (note that the sum is
finite since $F$ has compact support in $\O$), and therefore
$k\Gamma(\O)\subset \mbox{span}\,k\Gamma(\O_i)$. The converse inclusion is
obvious, and therefore also \rf{1.17} holds.
\end{beweis}

(More generally, additivity even holds for arbitrary states since already the 
Weyl algebra $\cal A(O)$ has an analogous property, cf.~\cite{BW92}.)
Other, more specific, properties of the net of von Neumann algebras
will not hold in such general circumstances, but will depend on a
judicious selection of (a class of) physically relevant states
$\omega$. For the choice of 
states we make in Section \ref{Section2} we will investigate the
properties of the local von Neumann algebras ${\cal R}_\omega(\O)$ in
Section \ref{Section3}.
 
\section{Definition of adiabatic states}\label{Section2}
As we have seen in the last section, the algebra of observables can 
easily be defined on any globally hyperbolic spacetime manifold. This is
essentially due to the fact that there is a well defined global causal 
structure on such a manifold, which allows to solve the classical
Cauchy problem and formulate the canonical commutation relations,
Eq.s \rf{1.9} and \rf{1.15}. Symmetries of the spacetime do not play
any role. This changes when one asks for the physical states of the
theory. For quantum field theory on Minkowski space the state space is 
built on the vacuum state which is defined to be the Poincar\'{e}
invariant state of lowest energy. A generic spacetime manifold however 
neither admits any symmetries nor the notion of energy, and it has
always been the main problem of quantum field theory on curved
spacetime to find a specification of the physical states of the theory 
in such a situation. \\
Using Hadamard's elementary solution of the wave
equation DeWitt \& Brehme \cite{DB60} wrote down an asymptotic
expansion of the singular kernel of a distribution which they called
the Feynman propagator of a quantum field on a generic spacetime
manifold. Since then quantum states whose two-point functions exhibit
these prescribed local short-distance singularities have been called Hadamard
states. Much work has been devoted to the investigation of the
mathematical and physical properties of these states (for the
literature see e.g.~\cite{KW91}), but only Kay \& Wald \cite{KW91}
succeeded in giving a rigorous mathematical definition of
them. Shortly later, in a
seminal paper Radzikowski \cite{Radzikowski96a} found a characterization of
the Hadamard states in terms of the wavefront set of their two-point
functions. This result proved to be fundamental to all ensuing work on 
quantum field theory in gravitational background fields.
Since we do not want to recall the old definition of
Hadamard states (it does not play any role in this paper) we
reformulate Radzikowski's main theorem as a {\it definition} of
Hadamard states:
\begin{dfn}\label{Definition0}
A quasifree state $\omega_H$ on the Weyl algebra $\A[\Gamma,\sigma]$
of the Klein-Gordon field on $(\M,g)$ is called an {\bf Hadamard state} if
its two-point function is a distribution $\Lambda_H\in \Dp{\M\times
\M}$ that satisfies the following wavefront set condition
\beq
WF'(\Lambda_H)=C^+. \label{2.0}
\eeq
\end{dfn}
Here, $C^+$ is the positive frequency component of the
bicharacteristic relation $C=C^+\dot{\cup}C^-$ that is associated to
the principal symbol of the Klein-Gordon operator $\Box_g+m^2$ (for
this notion see \cite{DH72}), more precisely
\beqa
C&:=& \{(\x\in T^*(\M\times\M)\setminus
0;\;g^{\mu\nu}(x_1)\xi_{1\mu}\xi_{1\nu}=0,\nonumber\\
 & &\quad g^{\mu\nu}(x_2)\xi_{2\mu}\xi_{2\nu} =0
,(x_1,\xi_1)\sim(x_2,\xi_2)\} \label{2.0a}\\
C^\pm &:=& \{\x\in C;\;\xi_1^0\gk 0,\xi_2^0\gk 0\}\label{2.0b}
\eeqa
where $(x_1,\xi_1)\sim (x_2,\xi_2)$ means that there is a null
geodesic $\gamma:\tau\mapsto x(\tau)$ such that
$x(\tau_1)=x_1,x(\tau_2)=x_2$ and $\xi_{1\nu}=\dot{x}^\mu
(\tau_1)g_{\mu\nu}(x_1),\xi_{2\nu}=\dot{x}^\mu(\tau_2)g_{\mu\nu}(x_2)$, 
i.e.~$\xi_1,\xi_2$ are cotangent to the null geodesic $\gamma$ at
$x_1$ resp.~$x_2$ and parallel transports of each other along
$\gamma$.\\
The fact that only positive frequencies occur in \rf{2.0} can be
viewed as a remnant of the spectrum condition in flat spacetime,
therefore \rf{2.0} (and its generalization to higher $n$-point functions 
in \cite{BFK96}) is also called microlocal spectrum
condition. However, condition \rf{2.0} does not fix a unique state,
but a class of states that generate locally quasiequivalent
GNS-representations \cite{Verch94}.\\
Now to which extent is condition
\rf{2.0} also {\it necessary} to characterize locally quasiequivalent states?
In \cite{Junker96} one of us gave a construction of Hadamard states by 
a microlocal separation of positive and negative frequency solutions
of the Klein-Gordon equation. From these solutions we 
observed that a truncation of the
corresponding asymptotic expansions destroys the microlocal spectrum
condition \rf{2.0} but preserves local quasiequivalence, at least if
the Sobolev order of the perturbation is sufficiently low (for Dirac
fields an analogous observation was made by Hollands
\cite{Hollands01}). In other words, the positive frequency condition in 
\rf{2.0} is not necessary to have local quasiequivalence, but can be
perturbed by non-positive frequency or even non-local 
singularities of sufficiently low
order. We formalize this observation by defining a new class of states 
with the help of the Sobolev (or $H^s$-) wavefront set. For a
definition and explanation of this notion see Appendix \ref{AppendixB}.

\begin{dfn}\label{Definition1}
A quasifree state $\omega_N$ on the Weyl algebra $\A[\Gamma,\sigma]$ 
of the Klein-Gordon
field on $(\M,g)$ is called an {\bf adiabatic state of order
$\mathbf{N} \in \R$} if its two-point function $\Lambda_N$ is a 
distribution that
satisfies the following $H^s$-wavefront set condition for all
$s<N+\frac{3}{2}$ 
\beq
WF'^s(\Lambda_N)\subset \begin{array}{ll} 
                        C^+.& \end{array}   \label{2.1}
\eeq
\end{dfn}
Note, that we did not specify $WF'^s$ for $s\geq N+\frac{3}{2}$ in the
definition. Hence every adiabatic state of order $N$ is also
one of order $N'\leq N$. In particular, every Hadamard state is also an
adiabatic state (of any order).
Now the task is to identify those adiabatic states that are physically 
admissible, i.e.~generate the same local quasiequivalence class as the 
Hadamard states. In \cite[Section 3.6]{Junker96} an example of an
adiabatic state of order $-1$ was given that does not satisfy this
condition. In Theorem \ref{Theorem29} we will prove that for $N>5/2$
(and in the special case of pure states on a spacetime with compact 
Cauchy surface already for $N>3/2$) 
the condition is satisfied (and the gap in between will remain
unexplored in this paper). For this purpose the following simple lemma 
will be fundamental:

\begin{lemma}\label{Lemma5}
Let $\Lambda_H$ and $\Lambda_N$ be the two-point functions of an arbitrary 
Hadamard state and an adiabatic state of order $N$, respectively, of the 
Klein-Gordon field on $(\M,g)$. Then 
\beq
WF^s(\Lambda_H-\Lambda_N)=\begin{array}{ll}\emptyset &\forall 
s<N+\frac{3}{2}.          \end{array} \label{2.2}
\eeq
\end{lemma}
\begin{beweis}
From Lemma \ref{Lemma5.1a} it follows that 
\[WF'^s(\Lambda_H)= \l\{\begin{array}{ll} \emptyset, & s< -\frac{1}{2} \\
                        C^+,& -\frac{1}{2} \leq s \end{array} \r. 
\]
and therefore
\beq
 WF'^s(\Lambda_H-\Lambda_N)\subset WF'^s(\Lambda_H)\cup
WF'^s(\Lambda_N) \subset \begin{array}{ll} 
                        C^+,& s<N+\frac{3}{2}. \end{array} \label{2.3} 
\eeq
On the other hand, since $\Lambda_H$ and $\Lambda_N$ have the same
antisymmetric part $\tilde{\sigma}$, $\Lambda_H-\Lambda_N$ must be a symmetric 
distribution, and thus also $WF^s(\Lambda_H-\Lambda_N)$ must be a
symmetric subset of $T^*(\M\times\M)$,
i.e.~$WF'^s(\Lambda_H-\Lambda_N)$ antisymmetric. However, the only 
antisymmetric
subset of the right hand side of \rf{2.3} is the empty set and hence
$WF^s(\Lambda_H-\Lambda_N)=\emptyset$ for $s<N+\frac{3}{2}$.

\end{beweis}

In the next section we will use this lemma to prove the result
mentioned above and some other algebraic properties of the Hilbert
space representations generated by our new states. In Section
\ref{Section5} we will explicitly construct these
states and in Section \ref{Section6} we will show that the old and
well-known class of adiabatic vacuum states on Robertson-Walker
spacetimes satisfies our Definition \ref{Definition1} (the comparison
with the order of these states led us to the normalization of $s$
chosen in Definition \ref{Definition1}). Contrary to an
erroneous claim in \cite{Junker96}, these states are in
general no Hadamard states, but in fact ``adiabatic states'' in our
sense. This justifies our naming of the new class of quantum
states on curved spacetimes in Definition
\ref{Definition1}. 

\section{The algebraic structure of adiabatic vacuum representations}
\label{Section3}
\subsection{Primarity and local quasiequivalence of adiabatic and
Hadamard states}\label{Section3.1}
Let $\A:=\A[\Gamma,\sigma]$ be the Weyl algebra associated to our
phase space $(\Gamma,\sigma)$ introduced in Section \ref{Section1} and 
$\A(\O):=\A[\Gamma(\O),\sigma]$ the subalgebra of observables
localized in an open, relatively compact subset $\O\subset \Sigma$.
Let $\omega_H$ denote some Hadamard state on
$\A$ and $\omega_N$ an adiabatic vacuum state of order $N$. It is the
main aim of this section to show that $\omega_H$ and $\omega_N$ are
locally quasiequivalent states for all sufficiently large $N$, i.e. the
GNS-representations $\pi_{\omega_H}$ and
$\pi_{\omega_N}$ are quasiequivalent when restricted to $\A(\O)$,
or, equivalently, there is an isomorphism $\tau$ between 
the von Neumann algebras $\pi_{\omega_H}(\A(\O))''$ and
$\pi_{\omega_N}(\A(\O))''$ such that $\tau\circ \pi_{\omega_H}
=\pi_{\omega_N}$ on $\A(\O)$ (see e.g.~\cite[Section 2.4]{BRI}).\\
 To prove this statement we will proceed as follows: 
We first notice that $\pi_{\omega_H}\hrist \A(\O)$ is quasiequivalent
to $\pi_{\omega_N}\hrist \A(\O)$ if
$\pi_{\omega_H}\hrist \A(\tilde{\O})$ is quasiequivalent to
$\pi_{\omega_N}\hrist \A(\tilde{\O})$ for some $\tilde{\O}\supset
\O$. Since to any open, relatively compact set $\O$ we can find an
open, relatively compact set $\tilde{\O}$ containing $\O$ and having a
smooth boundary we can assume without loss of generality that $\O$ has 
a smooth boundary. Under this assumption we first show
that $\pi_{\omega_N}(\A(\O))''$ is a factor (for $N>3/2$, Theorem
\ref{Theorem25}). Now we note that the GNS-representation
$(\pi_{\tilde{\omega}}, \H_{\tilde{\omega}},
\Omega_{\tilde{\omega}})$ of the partial state
$\tilde{\omega}:=\omega_N\hrist \A(\O)$ is a subrepresentation of 
$(\pi_{\omega_N}\hrist\A(\O), \H_{\omega_N},
\Omega_{\omega_N})$. This is easy to see: ${\cal
K}:=\ol{\{\pi_{\omega_N} (A)\Omega_{\omega_N};\;A\in\A(\O)\}}$ is a
closed subspace of $\H_{\omega_N}$ which is left invariant by
$\pi_{\omega_N}(\A(\O))$. Since for all $A\in \A(\O)$
\[ (\Omega_{\tilde{\omega}},
\pi_{\tilde{\omega}}(A)\Omega_{\tilde{\omega}})
=\tilde{\omega}(A)=\omega_N(A)=(\Omega_{\omega_N},\pi_{\omega_N}(A)
\Omega_{\omega_N}), \]
the uniqueness of the GNS-representation implies that
$\pi_{\tilde{\omega}}$ and $\pi_{\omega_N}\hrist\A(\O)$ coincide on
${\cal K}$ and
$(\pi_{\tilde{\omega}},\H_{\tilde{\omega}},\Omega_{\tilde{\omega}})$
can be identified with $(\pi_{\omega_N}\hrist \A(\O),{\cal
K},\Omega_{\omega_N})$ (up to unitary equivalence).\\
We recall that a primary representation (which means that the
corresponding von Neumann algebra is a factor) is quasiequivalent to
all its (non-trivial) subrepresentations (see
\cite[Prop.~5.3.5]{Dixmier69b}). Therefore,
$\pi_{\omega_N} \hrist \A(\O)$ is quasiequivalent to
$\pi_{\tilde{\omega}}=\pi_{(\omega_N\hrist \A(\O))}$, and analogously
$\pi_{\omega_H}\hrist\A(\O)$ is quasiequivalent to
$\pi_{(\omega_H\hrist\A(\O))}$. To prove that 
$\pi_{\omega_N}\hrist\A(\O)$ and $\pi_{\omega_H}\hrist\A(\O)$ are
quasiequivalent it is
therefore sufficient to prove the quasiequivalence of the
GNS-representations $\pi_{(\omega_N\hrist\A(\O))}$ and
$\pi_{(\omega_H\hrist\A(\O))}$ of the partial states. This will be done 
in Theorem \ref{Theorem29} for $N>5/2$.\\
To get started we have to prove in a first step that the real scalar
products $\mu_N$ and $\mu_H$ associated to the states $\omega_N$ and
$\omega_H$, respectively, induce the same topology on $\Gamma(\O)=
\co{\O}\oplus \co{\O}$. Let us denote by $\H_{\mu_N}(\O)$ and
$\H_{\mu_H}(\O)$ the completion of $\Gamma(\O)$ w.r.t.~$\mu_N$ and
$\mu_H$, respectively.
R.~Verch showed the following result \cite[Prop.~3.5]{Verch97}:
\begin{prop}\label{Proposition10}
For every open, relatively compact set ${\cal O} \subset \Sigma$ there
exist positive constants $C_1,C_2$ such that 
\[ C_1\l(\|q\|^2_{H^{1/2}({\cal O})} + \|p\|^2_{H^{-1/2}({\cal O})} \r)
\leq \mu_H\l({q\choose p},{q \choose p}\r) \leq
C_2\l(\|q\|^2_{H^{1/2}({\cal O})} +\|p\|^2_{H^{-1/2}({\cal O})} \r) \]
for all ${q\choose p}\in \Gamma(\O)$.
\end{prop}
\begin{thm}\label{Theorem20}
The topology of ${\cal H}_{\mu_N}({\cal O})$ coincides with that of
${\cal H}_{\mu_H}({\cal O})$ whenever $\Lambda_N$ satisfies \rf{2.1}
for $N>3/2$.
\end{thm}
\begin{beweis}
If $(\Sigma,h)$ is not a complete Riemannian manifold we can find a
function $f\in \ci(\Sigma),\;f>0$, with $f|_{\overline{\cal O}}=$ const.\
such that $(\Sigma, \tilde{h}:=fh)$ is complete \cite[Ch.~XX.18,
Problem 6]{DieudonneIV}.  
Then the Laplace-Beltrami operator $\Delta_{\tilde{h}}$ associated with
$\tilde{h}$ is essentially selfadjoint on $\co{\Sigma}$ \cite{Chernoff73}.
The topology on $\Gamma(\O)$ will not be
affected by switching from $h$ to $\tilde{h}$.
Without loss of generality we can therefore assume
that $\Delta$ is selfadjoint. Lemma \ref{Lemma5} shows that
\[ \Lambda_H-\Lambda_N\in H^s_{{loc}} (\M\times\M)
\quad\forall \;s < N+\frac{3}{2}. \]
In view of the fact that $\Sigma$ is a hyperplane, 
Proposition \ref{PropositionA6}
implies that, for $1< s < N+3/2$,
\beqa
(\Lambda_H-\Lambda_N)|_{\Sigma\times\Sigma} &\in&
H^{s-1}_{{loc}}(\Sigma\times \Sigma) \label{3.1}\\
\partial_{n_1}(\Lambda_H-\Lambda_N),
\partial_{n_2}(\Lambda_H-\Lambda_N)|_{\Sigma\times\Sigma} &\in&
H^{s-2}_{{loc}}(\Sigma\times \Sigma) \label{3.2}\\
\partial_{n_1}\partial_{n_2}(\Lambda_H-\Lambda_N)|_{\Sigma\times\Sigma} &\in&
H^{s-3}_{{loc}}(\Sigma\times \Sigma). \label{3.3}
\eeqa
Here, $\partial_{n_1}$ and $\partial_{n_2}$ denote the normal
derivatives with respect to the first and second variable,
respectively. We denote by $\lambda_H$ and $\lambda_N$ the scalar
products on $\Gamma$ induced via Eq.~\rf{1.14} by $\Lambda_H$ and $\Lambda_N$,
respectively. Since $\Lambda_H$ and $\Lambda_N$ have the same
antisymmetric parts we have
\beq
\l(\mu_H-\mu_N\r)\l({q_1\choose p_1},{q_2\choose p_2}\r) =
\l(\lambda_H-\lambda_N\r)\l({q_1\choose p_1},{q_2\choose p_2}\r)
=\l\la{q_1\choose p_1},{\mathbb M}{q_2\choose p_2}\r\ra_{L^2(\Sigma)\oplus
L^2(\Sigma)} \label{3.3a}
\eeq
for ${q_1\choose p_1}, {q_2\choose p_2}\in \Gamma$, where $\mathbb M$
is the integral operator with the kernel function 
\beq
M(x,y) = \l(\begin{array}{cc} \partial_{n_1}\partial_{n_2}
(\Lambda_H-\Lambda_N)|_{\Sigma\times \Sigma} &
-\partial_{n_1}(\Lambda_H-\Lambda_N)|_{\Sigma\times\Sigma} \\
-\partial_{n_2}(\Lambda_H-\Lambda_N)|_{\Sigma\times\Sigma} &
(\Lambda_H-\Lambda_N)|_{\Sigma\times\Sigma}
\end{array}\r). \label{3.3b}
\eeq
Note that $M(x,y)=M(y,x)^*$. We next fix a neighborhood $\tilde{\O}$
of $\O$ and a function $K=K(x,y) \in \co{\Sigma\times\Sigma}$ taking
values in $2\times 2$ real matrices such that $K(x,y)=K(y,x)^*,
x,y\in\Sigma,$ and the entries $K_{ij}$ of $K$ and $M_{ij}$ of $M$
satisfy the relations
\beq
\begin{array}{ll}\|K_{11}-M_{11}\|_{L^2(\tilde{\O}\times\tilde{\O})}<\epsilon
& \|K_{12}-M_{12}\|_{H^{1/2}(\tilde{\O}\times\tilde{\O})}<\epsilon\\
\|K_{21}-M_{21}\|_{H^{1/2}(\tilde{\O}\times\tilde{\O})}<\epsilon &
\|K_{22}-M_{22}\|_{H^1(\tilde{\O}\times\tilde{\O})}<\epsilon, \end{array}
\label{3.4} 
\eeq
where $\epsilon>0$ is to be specified lateron. By $\mathbb K$ we
denote the integral operator induced by $K$. We let
\begin{eqnarray*}
\mu_N'&:=&\mu_H+\la \cdot,({\mathbb K-\mathbb
M})\cdot\ra=\mu_N+\la\cdot, {\mathbb K}\cdot\ra\\ 
\lambda_N'&:=& \mu_N'+\frac{i}{2}\sigma.
\end{eqnarray*}
By $\Lambda_N'$ we denote the associated bilinear form on
$\co{\M}\times\co{\M}$ 
\beq
 \Lambda_N'(f,g):=\lambda_N'\l({\rho_0 \choose
\rho_1} Ef,  {\rho_0 \choose
\rho_1} Eg\r) \label{3.5a} 
\eeq
(note that, in spite of our notation, $\Lambda_N'$ is not the
two-point function of a quasifree state in general).
Recall from \rf{1.4} that $\rho_0,\rho_1$ 
are the usual restriction operators.
The definition of $\Lambda_N'$ makes sense, since both $\rho_0 Eg$ and 
$\rho_1 Eg$ have compact support in $\Sigma$ so that $\lambda_N'$ can
be applied. In view of the fact that ${\mathbb K}$ is 
an integral operator with a smooth kernel, also
$\lambda_N'-\lambda_N=\mu_N'-\mu_N$ is given by a smooth kernel. 
We claim that also $\Lambda_N'-\Lambda_N$ is smooth on $\M\times\M$:
In fact,
\[ (\Lambda_N'-\Lambda_N)(f,g) =\l\la {\rho_0 \choose \rho_1} Ef,
{\mathbb K}{\rho_0\choose \rho_1}Eg\r\ra \]
is given by the Schwartz kernel
\[ \l({\rho_0\choose \rho_1}E\r)^* K {\rho_0\choose\rho_1}E.\]
Since $E$ is a Lagrangian distribution of order $\mu=-3/2$ (for more
details see Section \ref{Section5} below), while $K$ is a compactly
supported smooth function, the calculus of Fourier integral operators
\cite[Thm.s 25.2.2, 25.2.3]{HormIV} show that the composition is also
smooth. \\
It follows from an argument of Verch \cite[Prop.~3.8]{Verch94} that
there are functions $\phi_j,\psi_j\in \co{\M},\;j=1,2,\ldots,$ such that
\[ \Lambda_N'(f,g)-\Lambda_N(f,g) =\sum_{j=1}^\infty
\sigma(f,\phi_j)\sigma(g,\psi_j) \]
for all $f,g\in \co{D({\cal O})}$, satisfying moreover
\[ \sum_{j=1}^\infty
\Lambda_N(\phi_j,\phi_j)^{1/2}\Lambda_N(\psi_j,\psi_j)^{1/2} < \infty.
\]
(An inspection of the proof of \cite[Prop.~3.8]{Verch94} shows that it
is sufficient for the validity of these  statements 
that $\Lambda_N$ is the two-point function of a
quasifree state, $\Lambda_N'$ need not be one.) It follows that
\begin{eqnarray*}
|\Lambda_N'(f,f)-\Lambda_N(f,f)|
&\leq&\sum_j|\sigma(f,\phi_j)\sigma(f,\psi_j)|\\
&\leq&\sum_j 4
\Lambda_N(f,f)^{1/2}\Lambda_N(\phi_j,\phi_j)^{1/2}\Lambda_N(f,f)^{1/2} 
\Lambda_N(\psi_j,\psi_j)^{1/2}\\
&=&4\Lambda_N(f,f)\sum_j\Lambda_N(\phi_j,\phi_j)^{1/2}
\Lambda_N(\psi_j,\psi_j)^{1/2}\\
&\leq& C\Lambda_N(f,f).
\end{eqnarray*}
Therefore
\[ |\Lambda_N'(f,f)|\leq (1+C)\Lambda_N(f,f).\]
Given $q,p\in \co{{\cal O}}$, we can find $f\in \co{D({\cal O})}$ such 
that $q=\rho_0 Ef,\;p=\rho_1 Ef$ (cf.~Proposition \ref{theorem1.1}). Hence
\begin{eqnarray}
\mu_N' \l({q\choose p},{q\choose p}\r) &=& \lambda_N'\l( {q\choose
p},{q\choose p}\r) = \Lambda_N'(f,f) \leq (1+C)
\Lambda_N(f,f)\nonumber\\
&=& (1+C) \lambda_N\l( {q\choose p},{q\choose p}\r)=(1+C) \mu_N\l(
{q\choose p},{q\choose p}\r). \label{3.5b}
\end{eqnarray}
We next claim that for all ${q\choose p}\in \Gamma(\O)$
\beq
\l|\l\la {q\choose p},{\mathbb M} {q\choose p}\r\ra\r| \leq C_3
\l(\|q\|^2_{H^{1/2}(\O)} + \|p\|^2_{H^{-1/2}(\O)}\r) \label{3.6a}
\eeq
and
\beq
\l|\l\la {q\choose p},(\mathbb{K-M}){q\choose p}\r\ra\r| \leq C_\epsilon 
\l(\|q\|^2_{H^{1/2}(\O)} + \|p\|^2_{H^{-1/2}(\O)}\r), \label{3.6b}
\eeq
where $C_3$ and $C_\epsilon$ are positive constants and $C_\epsilon$
can be made arbitrarily small by taking $\epsilon$ small in
\rf{3.4}. Indeed, in order to see this, we may first multiply the
kernel functions $M$ and $K-M$, respectively, by
$\varphi(x)\varphi(y)$ where $\varphi$ is a smooth function supported
in the neighborhood $\tilde{\O}$ of $\O$ and $\varphi \equiv 1$ on
$\O$. The above expressions \rf{3.6a} and \rf{3.6b} will not be
affected by this change. We may then localize the kernel functions to
$\R^3\times \R^3$ noting that the Sobolev regularity is
preserved. Now we can apply Lemma \ref{Lemma21} and Corollary
\ref{Corollary22} to derive \rf{3.6a} and \rf{3.6b}.\\
We finally obtain the statement of the theorem from the estimates
\begin{eqnarray*}
\frac{C_1}{2}\l(\|q\|_{H^{1/2}(\O)}+\|p\|_{H^{-1/2}(\O)}\r)
&\leq& (C_1-C_\epsilon)\l(\|q\|_{H^{1/2}(\O)}+\|p\|_{H^{-1/2}(\O)}\r)\\
& &\quad\mbox{if $\epsilon$ is sufficiently small}\\
&\leq& {\mu}_H\l( {q\choose p}, {q\choose p}\r)+\l\la {q\choose
p},({\mathbb K}-{\mathbb M}) {q\choose p}\r\ra\\
& &\quad\mbox{by Prop.\ \ref{Proposition10} and \rf{3.6b}}\\
&=&\mu_N'\l({q\choose p},{q\choose p}\r)\\
&\leq& (1+C){\mu}_N\l( {q\choose p},{q\choose p}\r)\quad\mbox{by
\rf{3.5b}}\\
&=&(1+C) \l({\mu}_H\l({q\choose p},{q \choose p}\r)- \l\la{q\choose
p},{\mathbb M} {q\choose p} \r\ra\r)\\
&\leq& (1+C)(C_2+C_3)\l(\|q\|_{H^{1/2}(\O)}+\|p\|_{H^{-1/2}(\O)}\r)\\
& &\quad\mbox{by Prop.\ \ref{Proposition10} and \rf{3.6a}}.
\end{eqnarray*}
\end{beweis}

\begin{lemma}\label{Lemma21}
Let $k\in H^{1/2}(\R^n\times\R^n)$. Then the integral operator
$\mathbb K$ with kernel $k$ induces an operator in ${\cal
B}(H^{1/2}(\R^n),H^{1/2}(\R^n))$ and ${\cal
B}(H^{-1/2}(\R^n),H^{-1/2}(\R^n))$. If we even have $k\in
H^1(\R^n\times \R^n)$, then $\mathbb K$ induces an operator in
${\cal B}(H^{-1/2}(\R^n),H^{1/2}(\R^n))$. In both cases, the operator norm of
$\mathbb K$ can be estimated by the Sobolev norm of $k$.
\end{lemma}
\begin{beweis}
The boundedness of ${\mathbb K}:H^{\pm 1/2}\to H^{\pm 1/2}$ 
is equivalent to the
boundedness of ${\mathbb L}:=\la D\ra^{\pm 1/2}{\mathbb K} \la
D\ra^{\mp 1/2}$ on $L^2(\R^n)$. (Here $\la
D\ra^{\pm 1/2}:=(1-\Delta)^{\pm 1/4}$, where $\Delta$ is the Euclidean
Laplacian.) This in turn will be true, if its integral kernel
$l(x,y):= \la D_x\ra^{\pm 1/2} \la D_y\ra^{\mp 1/2} k(x,y)$ is in
$L^2(\R^n\times \R^n)$. In this case
\[ \|{\mathbb L}\|_{{\cal B}(L^2(\R^n))} \leq \|l\|_{L^2(\R^n\times\R^n)}.\]
We know that
$\la D_x\ra^{\pm 1/2}\la D_y\ra^{\mp 1/2}$ are pseudodifferential
operators on $\R^n\times \R^n$ with symbols in
$S^{1/2}_{0,0}(\R^{2n}\times\R^{2n})$. By Calder\'{o}n and Vaillancourt's
Theorem (cf.\ \cite[Thm.~7.1.6]{Kumano-go81}), they yield bounded maps
$H^{1/2}(\R^n\times\R^n)\to L^2(\R^n\times\R^n)$. Hence $l\in
L^2(\R^n\times\R^n)$ and we obtain the first assertion. For the second
assertion we check that $\la D_x\ra^{1/2} \la D_y\ra^{1/2} k(x,y)\in
L^2(\R^n\times\R^n)$. Since the symbol of $\la D_x\ra^{1/2}\la
D_y\ra^{1/2}$ is in $S^1_{0,0}(\R^{2n}\times\R^{2n})$, this holds whenever
$k\in H^1$.
\end{beweis}

\begin{cor}\label{Corollary22}
If 
\[k=\l(\begin{array}{cc} k_{11} & k_{12}\\
k_{21}&k_{22}\end{array}\r)\] 
with 
\[k_{11}\in L^2(\R^n\times\R^n),\quad k_{12},k_{21}\in
H^{1/2}(\R^n\times\R^n), \quad k_{22}\in H^1(\R^n\times\R^n) ,\]
then the integral operator $\mathbb K$ with kernel $k$ induces a
bounded map
\[H^{1/2}(\R^n)\oplus H^{-1/2}(\R^n)\to H^{-1/2}(\R^n)\oplus H^{1/2}(\R^n).\]
Given $(q,p)\in \co{\R^n}\oplus\co{\R^n}$ we can estimate
\[\l|\l\la{q\choose p},{\mathbb K}{q\choose p}\r\ra_{L^2\oplus L^2}\r| \leq
\l\|{q\choose p}\r\|_{H^{1/2}\oplus H^{-1/2}} \l\|{\mathbb K}{q\choose
p}\r\|_{H^{-1/2}\oplus H^{1/2}} \leq \|{\mathbb K}\| \l\|{q\choose
p}\r\|^2_{H^{1/2}\oplus H^{-1/2}}.\]
(Here we used the fact that, for $u\in H^s(\R^n)$ and $v\in
H^{-s}(\R^n)$, $\la u,v\ra$ can be understood as the extension of the
$L^2$ bilinear form and $|\la u,v\ra|\leq \|u\|_{H^s} \|v\|_{H^{-s}}$.)
\end{cor}

We now apply Theorem \ref{Theorem20} to show that adiabatic vacua (of
order $N> 3/2$) generate primary representations. The proof is a
modification of the corresponding argument for Hadamard states due to
Verch \cite{Verch94}.
\begin{thm}\label{Theorem25}
Let $\omega_N$ be an adiabatic vacuum state of order $N>3/2$ on the
Weyl algebra ${\cal A}[\Gamma,\sigma]$ of the Klein-Gordon field on
$(\M,g)$ and $\pi_{\omega_N}$ its GNS-representation. Then, for any
open, relatively compact subset ${\cal O}\subset \Sigma$ with smooth
boundary, $\pi_{\omega_N}({\cal A}({\cal O}))''$ is a factor.
\end{thm}

In the proof of the theorem we will need the following lemma. Recall
that the metric $\tilde{h}$ introduced in the proof of Theorem
\ref{Theorem20} differs from $h$ only by a conformal factor which is
constant on $\O$.
\begin{lemma}\label{Lemma26}
$\co{\cal O} + \co{\Sigma \setminus \ol{\cal O}}$ is dense in
$\co{\Sigma}$ w.r.t.~the norm of $H^{1/2}(\Sigma,\tilde{h})$ (and hence 
also w.r.t.~the norm of $H^{-1/2}(\Sigma, \tilde{h})$).
\end{lemma}
\begin{beweis}  
Using a partition of unity we see that the problem is local. We can
therefore confine ourselves to a single relatively compact coordinate
neighborhood and work on Euclidean space. In view of the fact that
$\tilde{h}$ is positive definite, the topology of the Sobolev spaces
on $\Sigma$ locally yields the usual Sobolev topology. The problem
therefore reduces to showing that every function in $\co{\R^n}, n\in
\N$, can be approximated by functions in $\co{\R^n_+}+\co{\R^n_-}$ in
the topology of $H^{1/2}(\R^n)$. Following essentially a standard
argument \cite[2.9.3]{Triebel78} we proceed as follows. We choose a
function $\chi\in \ci(\R)$ with $\chi(t)=1$ for $|t|\geq 2$ and
$\chi(t)=0$ for $|t|\leq 1, 0\leq\chi\leq 1$. We define
$\chi_\epsilon:\R^n\to \R$ by
$\chi_\epsilon(x):=\chi(x_n/\epsilon)$. Given $f\in\co{\R^n}$ we have
\begin{eqnarray*}
\|f-\chi_\epsilon f\|_{L^2(\R^n)}&\leq& C_1 \sqrt{\epsilon}\\
\|f-\chi_\epsilon f\|_{H^1(\R^n)}&\leq& \frac{C_2}{\sqrt{\epsilon}}.
\end{eqnarray*}
Interpolation shows that $\{f-\chi_\epsilon f\}_{0<\epsilon<1}$ is
bounded in $H^{1/2}(\R^n)$ \cite[Thm.~1.9.3]{Triebel78}. Since
$H^{1/2}$ is a reflexive space, there is a sequence $\epsilon_j\to 0$
such that $f-\chi_{\epsilon_j}f$ converges weakly
\cite[Thm.~V.2.1]{Yosida80}. The limit necessarily is zero, since it
is zero in $L^2$. According to Mazur's Theorem
\cite[Thm.~V.1.2]{Yosida80} there is, for each $\delta>0$, a finite
convex combination $\sum_{j=1}^k \alpha_j(f-\chi_{\epsilon_j}f)$ (with 
$\alpha_j\geq 0, \sum_{j=1}^k\alpha_j=1$) such that
\[
\|\sum_{j=1}^k\alpha_j(f-\chi_{\epsilon_j}f)-0\|_{H^{1/2}}<\delta.\]
Since $\sum \alpha_j\chi_{\epsilon_j}f\in \co{\R^n_+}+\co{\R^n_-}$,
the proof is complete.
\end{beweis}

{\it Proof of Theorem \ref{Theorem25}:}
Let $(k_N,\H_N)$ be the one-particle Hilbert space structure of
$\omega_N$, let 
\beq
k_N(\Gamma(\O))^\vee := \{u\in
\H_N;\;\mbox{Im} \la u,v\ra_{\H_N}=0\;\forall v\in k_N(\Gamma(\O))\}
\label{7a} 
\eeq
denote the symplectic complement of $k_N(\Gamma(\O))$. It is a closed, 
real subspace of $\H_N$. According to results
of Araki \cite{Araki63, LRT78} $\pi_{\omega_N}(\A(\O))''$ is a factor iff 
\beq
\ol{k_N(\Gamma(\O))}\cap k_N(\Gamma(\O))^\vee = \{0\}, \label{8}
\eeq
where the closure is taken w.r.t.~the norm in $\H_N$.\\
In a first step we prove \rf{8} for the one-particle Hilbert space
structure $(\tilde{k},\tilde{\H})$ of an auxiliary quasifree state 
on $\A[\Gamma,
\tilde{\sigma}]$, where $\tilde{\sigma}$ is the symplectic form
w.r.t.~the metric $\tilde{h}$,
\begin{eqnarray}
\tilde{k}: \Gamma &\to& L^2(\Sigma,\tilde{h})=:\tilde{\H} \nonumber\\
 {q \choose p} &\mapsto& \frac{1}{\sqrt{2}}
\l(i\la{D}\ra^{1/2}q+\la {D}\ra^{-1/2}p\r)\label{8a}
\end{eqnarray}
(which, in general, induces neither an Hadamard state nor an
admissible adiabatic
vacuum state). As before, $\la D\ra :=(1-\Delta_{\tilde{h}})^{1/2}$. 
Since $\Delta_{\tilde{h}}$ is essentially selfadjoint on
$\co{\Sigma}$, $\sqrt{2}\,\tilde{k}(\Gamma) = i\la {D}\ra^{1/2}
\co{\Sigma} + \la {D}\ra^{-1/2}\co{\Sigma}$ 
is dense in
$L^2(\Sigma,\tilde{h})$ (since $\co{\Sigma}$ is dense in
$H^{1/2}(\Sigma)$ as well as in $H^{-1/2}(\Sigma)$), 
i.e.~$\tilde{k}$ describes a pure state.\\
Note that locally, i.e.~on $\Gamma(\O)$, the norm given by
\beq
\tilde{\mu}(F,F):=\la\tilde{k}F,\tilde{k}F\ra_{\tilde{\H}}
=\frac{1}{2}\l[ \|\la{D}\ra^{1/2} q \|^2_{\tilde{\H}}
+\|\la{D}\ra^{-1/2}p\|^2_{\tilde{\H}}\r],\quad 
F:=(q,p)\in\Gamma(\O), \label{8b}
\eeq
is independent of the choice of metric and
hence equivalent to the norm of $H^{1/2}(\O)\oplus
H^{-1/2}(\O)$. Also, since the conformal factor $f$ satisfies 
$f=C>0$ on $\ol{\O}$, we have
\begin{eqnarray*}
\tilde{\sigma}(F_1,F_2)&=& 2 \mbox{Im}\la
\tilde{k}F_1,\tilde{k}F_2\ra_{\tilde{\H}} =
\int_{\O}\!d^3\sigma_{\tilde{h}}\,(p_1q_2-q_1p_2)\\
&=& C^{3/2}\int_{\O}\!d^3\sigma_h\,(p_1q_2-q_1p_2)=C^{3/2}\sigma(F_1,F_2)
\end{eqnarray*}
for all $F_i=(q_i,p_i)\in \Gamma(\O),i=1,2$. \\
Define now $\tilde{k}(\Gamma(\O))^\vee :=
\{u\in\tilde{\H};\;\mbox{Im}\la u,v\ra_{\tilde{\H}}=0\;\forall v\in
\tilde{k}(\Gamma(\O))\}$ and let $u\in \ol{\tilde{k}(\Gamma(\O))}\cap
\tilde{k}(\Gamma(\O))^\vee$. 
Then
Im$\la u,\tilde{k}(F)\ra_{\tilde{\H}} =0$ for all $F\in \Gamma(\O)$ (by
the definition of $\tilde{k}(\Gamma(\O))^\vee$) and also Im$\la
u,\tilde{k}(F)\ra_{\tilde{\H}} =0$ for all $F\in
\Gamma(\Sigma\setminus\ol{\O})$ (since $\tilde{k}(F)\in
\tilde{k}(\Gamma(\O))^\vee$ for $F\in \Gamma(\Sigma\setminus \ol{\O})$). This,
together with the density statement of Lemma \ref{Lemma26}, implies
that Im$\la u,\tilde{k}(F)\ra_{\tilde{\H}} =0$ for all $F\in\Gamma$,
and, since $\tilde{k}(\Gamma)$ is dense in $\tilde{\H}$, it follows
that $u=0$, i.e.~\rf{8} is proven for the auxiliary state given by
$\tilde{k}$ on $\A[\Gamma,\tilde{\sigma}]$.\\
Let us now show \rf{8} for an adiabatic vacuum state $\omega_N,N>3/2$,
on $\A[\Gamma,\sigma]$. Let $u\in \ol{k_N(\Gamma(\O))}\cap 
k_N(\Gamma(\O))^\vee$, then there
is a sequence $\{F_n,n\in \N\}\subset \Gamma(\O)$ with $k_N(F_n)\to u$ 
in $\H_N$. Of course, $k_N(F_n)$ is in particular a Cauchy sequence in 
$\H_N$, i.e.
\[ \mu_N(F_n-F_m,F_n-F_m)= \|k_N(F_n)-k_N(F_m)\|^2_{\H_N}\to 0. \]
By Theorem \ref{Theorem20}, the norm given by
$\mu_N,N>3/2$, on $\Gamma(\O)$ is equivalent to the norm given by
$\tilde{\mu}$, namely that of $H^{1/2}(\O)\oplus H^{-1/2}(\O)$. Therefore we
also have
\[ \|\tilde{k}(F_n)-\tilde{k}(F_m)\|^2_{\tilde{\H}} =
\tilde{\mu}(F_n-F_m,F_n-F_m) \to 0 \]
and it 
follows that also $\tilde{k}(F_n)\to v$ in $\tilde{\H}$ for some $v\in 
\ol{\tilde{k}(\Gamma(\O))}$. For all $G\in \Gamma(\O)$ we have the equalities
\begin{eqnarray*}
0&=& \mbox{Im}\la u,k_N(G)\ra_{\H_N}=\lim_{n\to\infty}\mbox{Im}\la
k_N(F_n),k_N(G)\ra_{\H_N} \\
&=& \frac{1}{2}\lim_{n\to \infty} \sigma
(F_n,G)=\frac{1}{2}C^{-3/2}\lim_{n\to\infty}\tilde{\sigma}(F_n,G)\\
&=& C^{-3/2}\lim_{n\to\infty}\mbox{Im}\la
\tilde{k}(F_n),\tilde{k}(G)\ra_{\tilde{\H}} = C^{-3/2}\mbox{Im}\la v,
\tilde{k}(G)\ra_{\tilde{\H}}, 
\end{eqnarray*}
which imply that $v\in \ol{\tilde{k}(\Gamma(\O))}\cap 
\tilde{k}(\Gamma(\O))^\vee  =\{0\}$ and 
therefore $\tilde{k}(F_n)\to 0$ in $\tilde{\H}$. Since the norms given 
by $k_N$ and $\tilde{k}$ are equivalent on $\Gamma(\O)$ we also have
$u=\lim_{n\to\infty}k_N(F_n)= 0$ in $\H_N$,
which proves the theorem. \hfill $\blacksquare$\\

Our main theorem is the following:
\begin{thm}\label{Theorem29}
Let $\omega_N$ be an adiabatic vacuum state of order $N$ and
$\omega_H$ an Hadamard state on the Weyl algebra $\A[\Gamma,\sigma]$
of the Klein-Gordon field in the globally hyperbolic spacetime
$(\M,g)$, and let $\pi_{\omega_N}$ and $\pi_{\omega_H}$ be their
associated GNS-representations.\\
(i) If $N>5/2$, then 
$\pi_{\omega_N}\hrist\A(\O)$ and $\pi_{\omega_H}\hrist\A(\O)$ are
quasiequivalent for every open, relatively
compact subset $\O\subset \Sigma$.\\
(ii) If $\omega_N$ and $\omega_H$ are pure states on a spacetime with
compact Cauchy surface and $N>3/2$, then $\pi_{\omega_N}$ and
$\pi_{\omega_H}$ are unitarily equivalent.
\end{thm}
As explained at the beginning of this section it is sufficient to
prove the quasiequivalence of the GNS-representations of the partial
states $\omega_N\hrist \A(\O)$ and $\omega_H \hrist\A(\O)$ for part (i)
of the theorem, for part (ii) we can take $\O=\Sigma$. To this
end we shall use a result of Araki \& Yamagami \cite{AY82}.
To state it we first need some notation.\\
Given a bilinear form $\mu$ on a real vector space $K$ we shall
denote by $\mu^\C$ the extension of $\mu$ to the complexification
$K^\C$ of $K$ (such that it is antilinear in the first argument):
\[\mu^\C(F_1+iF_2,G_1+iG_2):= \mu (F_1,G_1)+\mu(F_2,G_2) +i \mu(F_1,G_2) 
-i\mu (F_2,G_1). \]
The theorem of Araki \& Yamagami gives necessary and sufficient
conditions for the quasiequivalence of two quasifree states
$\omega_{\mu_1}$ and $\omega_{\mu_2}$ on the Weyl algebra
$\A[K,\sigma]$ of a phase space
$(K,\sigma)$ in terms of the complexified data $K^\C,\sigma^\C$, and
$\mu_i^\C,\, i=1,2$. Assuming that $\mu_1^\C$ and $\mu_2^\C$ induce the 
same topology on $K^\C$, denote by $\bar{K}^\C$ the completion. Then
$\mu_1^\C, \mu_2^\C$, and $\lambda_1^\C:=\mu_1^\C+\frac{i}{2}
\sigma^\C, \lambda_2^\C:=\mu_2^\C+\frac{i}{2}\sigma^\C$ extend to
$\bar{K}^\C$ by continuity ($\sigma^\C$ extends due to \rf{1.11}). 
We define bounded positive selfadjoint 
operators $S_1,S_2$, and $S'_2$ on $\bar{K}^\C$ by
\begin{eqnarray}
\lambda_j^\C(F,G) &=& 2\mu_j^\C(F,S_j G),\quad j=1,2,\nonumber\\
\lambda_2^\C(F,G) &=& 2\mu_1^\C(F,S'_2 G),\quad F,G\in \bar{K}^\C.\label{8c}
\end{eqnarray}
Note that $S_j$ is a projection operator if and only if
$\omega_{\mu_j}$ is a Fock state. 
The theorem of Araki \& Yamagami \cite{AY82} then 
states that the corresponding GNS-representations $\pi_{\omega_1}$ and 
$\pi_{\omega_2}$ are quasiequivalent if and only if both of the
following two conditions are satisfied:
\begin{itemize}
\item[(AY1)] $\mu_1^\C$ and $\mu_2^\C$ induce the same topology on
$K^\C$,
\item[(AY2)] $S_1^{1/2}-S_2^{\prime 1/2}$ is a Hilbert-Schmidt operator on
$(\bar{K}^\C,\mu_1^\C)$. 
\end{itemize}
{\it Proof of Theorem \ref{Theorem29}:} (i) We choose $K=
\Gamma(\O) =\co{\O}\oplus \co{\O}$, 
$\sigma$ our real symplectic form \rf{1.8}, $\mu_H$ and $\mu_N$
the real scalar products on $K$ defining an Hadamard state and an 
adiabatic vacuum state of order $N>5/2$, respectively, and check (AY1) 
and (AY2) for the data $K^\C,\sigma^\C,\mu_H^\C$ and $\mu_N^\C$. From Theorem
\ref{Theorem20} we know that the topologies induced by $\mu_H$ and
$\mu_N$ on $\Gamma(\O)$ coincide. In view of the fact that
\[ \mu_H^\C(F_1+iF_2, F_1+iF_2)=\mu_H(F_1,F_1)+\mu_H(F_2,F_2),\quad
F_1,F_2\in \Gamma(\O),\] 
(and the corresponding relation for $\mu_N$), 
we see that the topologies coincide also on the
complexification. Hence (AY1) holds.\\
In order to prove (AY2), we first note that the difference
$S_H^{1/2}-S_N^{\prime 1/2}$ for the operators $S_H$ and $S'_N$ induced by
$\mu_H$ and $\mu_N$ via \rf{8c} 
will be a Hilbert-Schmidt operator provided that 
$S_H-S'_N$ is of trace class, cf.~\cite[Lemma 4.1]{PS70}. 
By definition, 
\beq
\mu_H^\C (F,(S_H-S_N')G)=\frac{1}{2}\l(\lambda_H^\C-\lambda_N^\C\r)(F,G)
=\frac{1}{2}\l(\mu_H^\C-\mu_N^\C\r)(F,G).\label{9}
\eeq
As in \rf{3.3a}, \rf{3.3b}, our assumption $N>5/2$ and Lemma
\ref{Lemma5} imply that there is an integral kernel $M=M(x,y)$ on
$\O\times\O$, given by \rf{3.3b} with entries satisfying
\rf{3.1}--\rf{3.3}, such that
\beq
\frac{1}{2}\l(\mu_H^\C-\mu_N^\C\r)(F,G) = \la F, {\mathbb
M}G\ra_{L^2(\O)\oplus L^2(\O)},\quad F,G\in \Gamma(\O), \label{9a}
\eeq
where $\mathbb M$ is the integral operator with kernel $M$. We may
multiply $M$ by $\varphi(x)\varphi(y)$ where $\varphi$ is a smooth
function supported in a relatively compact neighborhood $\tilde{\O}$ of
$\O$ with $\varphi \equiv 1$ on $\O$. Equality \rf{9a} is not affected
by this change. Moreover, as we saw in the beginning of this section we
may suppose that $\O$ and $\tilde{\O}$ have smooth boundary. 
Using a partition of unity it is no loss of generality to 
assume that $\tilde{\O}$ is contained in a
single coordinate neighborhood. We then denote by $\O_*\subset \R^3$
the image of $\O$ under the coordinate map. We shall use the notation
$\mu_H^\C,\mu_N^\C$, and $M,{\mathbb M}$ also for the push-forwards of
these objects. We note that the closure of $\Gamma(\O_*)$ with respect
to the topology of $H^{1/2}(\R^3)\oplus H^{-1/2}(\R^3)$ is
$H^{1/2}_0(\ol{\O_*}) \oplus H_0^{-1/2}(\ol{\O_*}) =:\H$, cf. Appendix
\ref{AppendixA} for the notation. The dual
space $\H'$ w.r.t.\ the extension of $\la\cdot,\cdot\ra_{L^2(\O_*)\oplus
L^2(\O_*)}$, denoted by $\la\cdot,\cdot\ra$, 
is $H^{-1/2}(\O_*)\oplus H^{1/2}(\O_*)$. The inner product
$\mu_H^\C$ extends to $\H$. By Riesz' theorem, $\mu_H^\C$ induces an
antilinear isometry $\tilde{\theta}:\H\to\H'$ by $\la \tilde{\theta}
F,G\ra =\mu_H^\C(F,G)$. Defining instead
\beq
\la F,\theta G\ra =\mu_H^\C(F,G) \label{9b}
\eeq
we obtain a linear isometry $\theta$ from $\H$ to the space $\tilde{\H}'$ of
antilinear functionals on $\H$. Complex conjugation provides a
(real-linear) isometry between $\H'$ and $\tilde{\H}'$, hence
$\tilde{\H}'=H^{-1/2}(\O_*)\oplus H^{1/2}(\O_*)$ as a normed space
(and hence as a Hilbert space). We deduce from Lemma \ref{Lemma30}
and Corollary \ref{Corollary30a} below, in connection with the
continuity of the extension operator $\H\to H^{1/2}(\R^3)\oplus
H^{-1/2}(\R^3)$ and the restriction operator $H^{-1/2}(\R^3)\oplus
H^{1/2}(\R^3)\to H^{-1/2}(\O_*)\oplus H^{1/2}(\O_*)$, 
that $M$ induces a mapping
\[{\mathbb M}:\H\to H^{-1/2}(\O_*)\oplus H^{1/2}(\O_*) \]
which is trace class. In particular, for $G\in \H$, ${\mathbb M}G$
defines an element of $\tilde{\H}'$ by $F\mapsto \la F, {\mathbb
M}G\ra$. Combining \rf{9}--\rf{9b}, we see that, for $F,G\in \H$,
\[ \la F,{\mathbb M}G\ra = \la F, \theta (S_H-S_N')G\ra.\]
Hence 
\[\theta (S_H-S_N') ={\mathbb M} \quad\mbox{in}\;{\cal
B}(\H,\tilde{\H}'),\]
so that 
\[S_H-S_N'=\theta^{-1}{\mathbb M}\quad\mbox{in}\;{\cal 
B}(\H).\]
As a consequence of the fact that $\theta^{-1}: \tilde{\H}'\to \H$ is
an isometry while ${\mathbb M}:\H\to\tilde{\H}'$ is trace class, this
implies that $S_H-S_N'$ is trace class.\\
(ii) To prove (ii) we apply the technique of Bogoljubov
transformations (we follow \cite{LR90} and
\cite[p.~68f.]{Wald94}). Assume that $\Sigma$ is compact and let $S_H,
S_N$, and $S'_N$ be the operators induced by a pure Hadamard state
$\omega_H$ resp.\ a pure adiabatic state $\omega_N$ of order $N>3/2$
via \rf{8c}. As remarked above, $S_H$ and $S_N$ are projection operators
on $\bar{K}^\C$, the closure of the complexification of
$K:=\Gamma(\Sigma)$ w.r.t.~$\mu_H^\C$ or $\mu_N^\C$ (since $\Sigma$ is
compact, $\mu_H^\C$ and $\mu_N^\C$ are equivalent on all of
$\Gamma(\Sigma)$, Theorem \ref{Theorem20}). We make a direct sum
decomposition of $\bar{K}^\C$ into
\beq
\bar{K}^\C= \begin{array}{c}\H_N^+\\ \oplus\\ \H_N^- \end{array}
=\begin{array}{c}\H_H^+\\ \oplus\\ \H_H^-\end{array}
\label{35a}
\eeq
such that $S_{H/N}$ has the eigenvalue $1$ on $\H^+_{H/N}$ and $0$ on
$\H^-_{H/N}$, and the first decomposition is orthogonal w.r.t.\
$\mu_N^\C$, the second w.r.t.\ $\mu_H^\C$. We also denote the
corresponding orthogonal projections of $\bar{K}^\C$ onto $\H^+_{H/N}$
resp.\ $\H^-_{H/N}$ by $P^+_{H/N}:= S_{H/N}$ resp.\
$P^-_{H/N}:=1-S_{H/N}$. From Eq.s \rf{1.13} and \rf{8c} we obtain for
$j\in\{H,N\}$
\beqa
2\mu_j^\C(F,S_j G)
&=&\lambda_j^\C(F,G)=\mu_j^\C(F,G)-\frac{i}{2}\sigma^\C(F,G)\nonumber\\
\Rightarrow \sigma^\C(F,G) &=& 2\mu_j^\C(F,i(2S_j-1)G) =2\mu_j^\C(F,J_jG)
\label{35b}
\eeqa
where $J_j:=i(2S_j-1)$ is a bounded operator on $\bar{K}^\C$ with the
properties $J^2_j=-1,\,J_j^*=-J_j$ (w.r.t.\ $\mu_j^\C$). It has
eigenvalue $+i$ on $\H^+_j$ and $-i$ on $\H_j^-$ and is called the
complex structure associated to $\mu_j$. Because of \rf{35b} both
decompositions in \rf{35a} are orthogonal w.r.t.\ $\sigma^\C$. We now
define the Bogoljubov transformation
\beq
\l(\begin{array}{cc} A&C\\B&D\end{array}\r): 
\begin{array}{c}\H_N^+\\ \oplus\\ \H_N^- \end{array} \to
\begin{array}{c}\H_H^+\\ \oplus \\ \H_H^-\end{array} \label{35c}
\eeq
by the bounded operators
\[ A:=P^+_H|_{\H_N^+},\;B:=P^-_H|_{\H_N^+},\; C:=P^+_H|_{\H^-_N},\;D:=
P^-_H|_{\H_N^-}.\]
Taking into account Eq.~\rf{35b} and the fact that the decomposition
\rf{35a} is orthogonal w.r.t.\ $\sigma^\C$ we obtain for $F,G\in
\H^+_N$
\begin{eqnarray*}
\mu_N^\C(F,G) &=& \mu_N^\C(F,-iJ_NG) = -\frac{i}{2}\sigma^\C(F,G) \\
&=& -\frac{i}{2}\sigma^\C(P_H^+F,P_H^+G)
-\frac{i}{2}\sigma^\C(P^-_HF,P_H^-G) \\
&=& -\frac{i}{2}\sigma^\C(AF,AG) -\frac{i}{2}\sigma^\C(BF,BG) \\
&=& -i\mu^\C_H(AF,J_HAG) -i\mu_H^\C(BF,J_HBG) \\
&=& \mu_H^\C(AF,AG)-\mu_H^\C(BF,BG), 
\end{eqnarray*}
similarly for $F,G\in \H_N^-$
\[\mu_N^\C(F,G)=\mu_H^\C(DF,DG)-\mu_H^\C(CF,CG),\]
and for $F\in \H_N^+, G\in \H^-_N$
\begin{eqnarray*}
0=\mu_N^\C(F,G) &=& \mu_N^\C(F,iJ_N G)=\frac{i}{2}\sigma^\C(F,G) \\
&=& \frac{i}{2}\sigma^\C(P_H^+F,P_H^+G)
+\frac{i}{2}\sigma^\C(P_H^-F,P^-_HG) \\
&=&\frac{i}{2}\sigma^\C(AF,CG)+\frac{i}{2}\sigma^\C(BF,DG) \\
&=&i \mu^\C_H(AF,J_HCG)+i \mu_H^\C(BF,J_HDG)\\
&=&-\mu_H^\C(AF,CG)+\mu_H^\C(BF,DG),
\end{eqnarray*}
hence
\beqa
A^*A-B^*B =1 &\mbox{in}&{\cal B}(\H_N^+,\H_N^+)\nonumber\\
D^*D-C^*C=1 &\mbox{in}&{\cal B}(\H_N^-,\H_N^-)\label{35d}\\
B^*D-A^*C =0 &\mbox{in}&{\cal B}( \H^-_N,\H_N^+). \nonumber
\eeqa
In a completely analogous way we can define the inverse Bogoljubov
transformation 
\beq
\l(\begin{array}{cc} \tilde{A}&\tilde{C}\\\tilde{B}&\tilde{D}\end{array}\r): 
\begin{array}{c}\H_H^+\\ \oplus\\ \H_H^- \end{array} \to
\begin{array}{c}\H_N^+\\ \oplus \\ \H_N^-\end{array} \label{35e}
\eeq
by
\[ \tilde{A}:=P^+_N|_{\H_H^+},\;\tilde{B}:=P^-_N|_{\H_H^+},\; 
\tilde{C}:=P^+_N|_{\H^-_H},\;\tilde{D}:= P^-_N|_{\H_H^-}.\]
These operators satisfy relations analogous to \rf{35d}. Moreover, for
$F\in\H_N^+, G\in\H_H^+$
\begin{eqnarray*}
\mu_N^\C(F,\tilde{A}G)&=& \mu_N^\C(F,-iJ_N\tilde{A}G)
=-\frac{i}{2}\sigma^\C(F,\tilde{A}G)
=-\frac{i}{2}\sigma^\C(F,P_N^+ G)\\
& =&-\frac{i}{2}\sigma^\C(F,G)
=-\frac{i}{2}\sigma^\C(P^+_H F,G) =-i\mu_H^\C(AF,J_HG)\\
&=& \mu_H^\C(AF,G),
\end{eqnarray*}
\beqa
\mbox{i.e.}\quad \tilde{A}&=&A^*:\H_H^+\to \H_N^+ \nonumber\\
\mbox{and similarly}\quad\tilde{B}&=&-C^*:\H_H^+\to \H_N^-\nonumber\\
\tilde{C} &=&-B^*:\H_H^-\to \H_N^+\nonumber\\
\tilde{D} &=&D^* : \H^-_H \to \H_N^-.\label{35f}
\eeqa
From \rf{35d} and \rf{35f} one easily finds that
\beqa
AA^*-CC^*=1&\mbox{in}&{\cal B}(\H_H^+,\H_H^+)\nonumber\\
DD^*-BB^*=1 &\mbox{in}& {\cal B}(\H_H^-,\H_H^-)\label{35g}\\
AB^*-CD^*=0&\mbox{in}&{\cal B}(\H_H^-,\H_H^+)\nonumber
\eeqa
and that \rf{35e} is the inverse of \rf{35c}. Moreover, $A$ is
invertible with bounded inverse: It follows from the
first Eq.s in \rf{35d} and \rf{35g} that $A^*A\geq 1$ on $\H_N^+$ and
$AA^*\geq 1$ on $\H_H^+$, hence $A$ and $A^*$ are injective. Since
$\{0\}=$ Ker$(A^*)= \mbox{Ran} (A)^\perp$, $A$ has dense range in
$\H_H^+$. For $F=AG\in \mbox{Ran}(A)$ we have $\|A^{-1}F\|^2_{\H_N^+} =
\|G\|^2_{\H_N^+}
\leq \la G,A^*AG\ra_{\H_N^+}=\|AG\|^2_{\H_H^+} =\|F\|^2_{\H_H^+}$, i.e.\
$A^{-1}$ is bounded and can be defined on all of $\H_H^+$.\\
We are now prepared
to show that $S_H^{1/2}-S_N^{\prime 1/2}$ is a Hilbert-Schmidt operator on
$(\bar{K}^\C,\mu_H^\C)$: We write $F\in\bar{K}^\C$ as a column vector
w.r.t.\ the decomposition of $\bar{K}^\C$ w.r.t.\ $\mu_H^\C$:
\[ F=\l(\begin{array}{c}P_H^+ F\\P_H^-F\end{array}\r)
=:\l(\begin{array}{c}F^+\\F^-\end{array}\r) \in
\begin{array}{c}\H_H^+\\ \oplus \\ \H_H^-\end{array}. \]
Then 
\beq
S_HF=P_H^+F=\l(\begin{array}{cc}1&0\\0&0\end{array}\r)
\l(\begin{array}{c} F^+\\F^-\end{array}\r). \label{35h} 
\eeq
For $S'_N$ we get by the basis transformation \rf{35e} for all
$F,G\in\bar{K}^\C$
\begin{eqnarray*}
\mu_H^\C(G,S'_N F) &=& \frac{1}{2}\lambda_N^\C(G,F) =\mu_N^\C(G,S_NF)
\\
&=&\mu_N^\C\l(\l(\begin{array}{c}P_N^+G\\P_N^-G\end{array}\r),
\l(\begin{array}{cc}1&0\\0&0\end{array}\r)
\l(\begin{array}{c}P_N^+F\\P_N^-F\end{array}\r)\r) \\
&=&\mu_N^\C\l(\l(\begin{array}{cc}\tilde{A}&\tilde{C}\\
\tilde{B}&\tilde{D}\end{array}\r) 
\l(\begin{array}{c}P_H^+G\\P_H^-G\end{array}\r),
\l(\begin{array}{cc}1&0\\0&0\end{array}\r)
\l(\begin{array}{cc}\tilde{A}&\tilde{C}\\
\tilde{B}&\tilde{D}\end{array}\r)
\l(\begin{array}{c}P_H^+F\\P_H^-F\end{array}\r)\r) \\
&=& \mu_H^\C\l(\l(\begin{array}{c}G^+\\G^-\end{array}\r),
\l(\begin{array}{cc}\tilde{A}^*&\tilde{B}^*\\
\tilde{C}^*&\tilde{D}^*\end{array}\r) 
\l(\begin{array}{cc}1&0\\0&0\end{array}\r)
\l(\begin{array}{cc}\tilde{A}&\tilde{C}\\
\tilde{B}&\tilde{D}\end{array}\r)
\l(\begin{array}{c}F^+\\F^-\end{array}\r)\r),
\end{eqnarray*}
and hence, utilizing \rf{35f},
\beqa
S'_N\l(\begin{array}{c}F^+\\F^-\end{array}\r) &=& 
\l(\begin{array}{cc}{A}&-C\\
-B&D\end{array}\r) 
\l(\begin{array}{cc}1&0\\0&0\end{array}\r)
\l(\begin{array}{cc}{A}^*&-B^*\\
-C^*&D^*\end{array}\r)
\l(\begin{array}{c}F^+\\F^-\end{array}\r)\nonumber\\
&=& \l(\begin{array}{cc}AA^*&-AB^*\\-BA^*&BB^*\end{array}\r)
\l(\begin{array}{c}F^+\\F^-\end{array}\r).\label{35j}
\eeqa
From \rf{35h} and \rf{35j} we have now on $\H_H^+ \oplus \H_H^-$ 
\beqa
S^{1/2}_H-S_N^{\prime 1/2}&=&
\l(\begin{array}{cc}1&0\\0&0\end{array}\r)^{1/2} -
\l(\begin{array}{cc}AA^*&-AB^*\\-BA^*&BB^*\end{array}\r)^{1/2}\nonumber\\
&=& \l(\begin{array}{cc}1&0\\0&0\end{array}\r) -
\l(\begin{array}{cc}AZ^{-1/2}A^*&-AZ^{-1/2}B^*\\-BZ^{-1/2}A^*&
BZ^{-1/2}B^*\end{array}\r),\label{35k}
\eeqa
where $Z:=A^*A+B^*B=1+2B^*B$ is a bounded selfadjoint positive
operator on $\H_N^+$, which has a bounded inverse due to the fact that
$Z\geq 1$. In Lemma \ref{Lemma31} below we will show that \rf{35k} is
a Hilbert-Schmidt operator on $\H_H^+
\oplus \H_H^-$ if and only if the operator
\[Y:=\l(\begin{array}{cc}1&0\\0&0\end{array}\r) -
\l(\begin{array}{cc}AA^*&-AB^*\\-BA^*&BB^*\end{array}\r) \]
is Hilbert-Schmidt. From \rf{35j} and \rf{8c} we see that
\begin{eqnarray*}
\mu_H^\C(G,YF) &=& \mu_H^\C(G,(S_H-S'_N)F)
=\frac{1}{2} \l(\lambda_H^\C(G,F) -\lambda_N^\C(G,F)\r)\\
&=&\frac{1}{2}\l(\mu_H^\C(G,F)-\mu_N^\C(G,F)\r).
\end{eqnarray*}
Now we argue as in the proof of part (i): $\mu_H^\C-\mu_N^\C$ is given
by an integral operator $\mathbb{M}$ with kernel $M$, where $M$ has
the form \rf{3.3b} with entries satisfying \rf{3.1}--\rf{3.3}. Using a
partition of unity we can transfer the problem to $\R^n$ with the
Sobolev regularity of the entries preserved. For $N>3/2$ the
conditions in Remark \ref{Remark30b} are satisfied. Hence $\mathbb{M}$
and also $Y$ are Hilbert-Schmidt operators, i.e.\
$\omega_H$ and $\omega_N$ are quasiequivalent on ${\cal
A}[\Gamma,\sigma]$, which, in turn, is equivalent to the unitary
equivalence of the representations $\pi_{\omega_H}$ and
$\pi_{\omega_N}$, if $\omega_H$ and $\omega_N$ are pure states.
\hfill $\blacksquare$\\

\begin{lemma}\label{Lemma30}
Let $M\in H^s_{comp}(\R^n\times \R^n),\; s\geq 0$, and consider 
the integral operator $\mathbb{M}$ with kernel $M$, 
defined by 
\[(\mathbb{M}u)(x) = \int
M(x,y)u(y)\,dy,\quad u\in \co{\R^n}.\]
If $s>\frac{n-1}{2}, \frac{n}{2}, \frac{n+1}{2},$ and $\frac{n}{2}+1$,
respectively, then $\mathbb{M}$ yields trace class operators in 
\[{\cal
B}(H^{1/2}(\R^n),H^{-1/2}(\R^n)),\;{\cal
B}(H^{-1/2}(\R^n)),\;{\cal
B}(H^{1/2}(\R^n)),\;{\cal
B}(H^{-1/2}(\R^n),H^{1/2}(\R^n)),\] 
respectively.
\end{lemma}
\begin {beweis}
For the first case, $s>\frac{n-1}{2}$, write 
\[\mathbb{M}=\l(\la D\ra^{-s-\frac{1}{2}} \la x\ra^{-s-\frac{1}{2}}\r) 
 \l(\la x\ra^{s+\frac{1}{2}}\la D\ra^{s+\frac{1}{2}} \mathbb{M}\r)\]
where 
$\la x\ra^s$ is the operator of multiplication by $\la x\ra^s$ and $\la
D\ra^s=op(\la \xi\ra^s)$ w.r.t.~the flat Euclidean metric of $\R^n$.
The first factor is known to be a Hilbert-Schmidt operator on
$H^{-1/2}(\R^n)$. 
Since $M$ has compact support, it is sufficient to check the
Hilbert-Schmidt property of $\la D\ra^{s+\frac{1}{2}}\mathbb{M}$ in 
${\cal B}(H^{1/2}(\R^n),H^{-1/2}(\R^n))$ or, equivalently, of $\la
D\ra^{s}\mathbb{M}\la D\ra^{-1/2}$ on $L^2(\R^n)$. This operator,
however, has the integral kernel $\la D_x\ra^s\la D_y\ra^{-1/2}
M(x,y)$. We may consider $\la D_x\ra^s$  
as the pseudodifferential operator $\la D_x\ra^s \otimes I$ on
$\R^n\times \R^n$ with a symbol in the class $S^s_{0,0}(\R^{2n}\times
\R^{2n})$ and $\la D_y\ra^{-1/2}$ as the pseudodifferential operator
$I\otimes \la D_y\ra^{-1/2}$ on $\R^n\times \R^n$ with symbol in
$S_{0,0}^0(\R^{2n}\times \R^{2n})$. 
By Calder{\'o}n and Vaillancourt's Theorem, $\la D_x\ra^s\la
D_y\ra^{-1/2}$ maps
$H^s(\R^n\times \R^n)$ to $L^2(\R^n\times\R^n)$,
hence $\la D\ra^s\mathbb{M}\la D\ra^{-1/2}$ is an
integral operator with a square integrable kernel, hence
Hilbert-Schmidt,  and $\mathbb{M}$ is
the composition of two Hilbert-Schmidt operators, hence trace class.\\
The proofs of the other cases are similar.
\end{beweis}

\begin{cor}\label{Corollary30a}
It is well-known that the operator
\[ \mathbb{M}=\l(\begin{array}{cc}\mathbb{M}_{11} & \mathbb{M}_{12} \\
\mathbb{M}_{21} & \mathbb{M}_{22} \end{array}\r): 
\begin{array}{c}H^{1/2}(\R^n)\\ \oplus \\
H^{-1/2}(\R^n)\end{array} \to \begin{array}{c}H^{-1/2}(\R^n)\\ \oplus \\
H^{1/2}(\R^n)\end{array}\]
is trace class if and only if each of the entries $\mathbb{M}_{ij}$ of
the matrix is a trace class operator between the respective spaces,
cf.\ e.g.\ \cite[Sect.~4.1.1.2, Lemma 2]{RS82}. Denoting by $M_{ij}$
the integral kernel of $\mathbb{M}_{ij}$, $\mathbb{M}$ will be trace
class if
\begin{eqnarray*}
M_{11} &\in& H^s(\R^n\times\R^n),\quad s>\frac{n-1}{2}\\
M_{12} &\in& H^s(\R^n\times\R^n),\quad s>\frac{n}{2}\\
M_{21}&\in& H^s(\R^n\times\R^n),\quad s>\frac{n+1}{2}\\
M_{22}&\in& H^s(\R^n\times\R^n),\quad s>\frac{n}{2}+1.
\end{eqnarray*}
\end{cor}

\begin{remark}\label{Remark30b}
In the situation of Corollary \ref{Corollary30a}, $\mathbb{M}$ will be
a Hilbert-Schmidt operator if each of its entries has this
property. Using the fact that an integral operator on $L^2(\R^n)$ is
Hilbert-Schmidt if its kernel is in $L^2(\R^n\times\R^n)$, we easily
see that it is sufficient for the Hilbert-Schmidt property of
$\mathbb{M}$ that 
\begin{eqnarray*}
M_{11}&\in& L^2(\R^n\times\R^n)\\
M_{12},M_{21}&\in& H^{1/2}(\R^n\times\R^n)\\
M_{22} &\in& H^1(\R^n\times\R^n).
\end{eqnarray*}
\end{remark}

\begin{lemma}\label{Lemma31}
In the notation of above, the following statements are equivalent:\\
(i) The operator
\[ X:=\l(\begin{array}{cc}1-AZ^{-1/2}A^* &AZ^{-1/2}B^*\\BZ^{-1/2}A^*&
-BZ^{-1/2}B^*\end{array}\r):\begin{array}{c}\H_H^+\\ \oplus \\
\H_H^-\end{array} \to \begin{array}{c}\H_H^+\\ \oplus \\
\H_H^-\end{array}\]
is Hilbert-Schmidt.\\
(ii) The operator
\[ Y:=\l(\begin{array}{cc}1-AA^* &AB^*\\BA^*&
-BB^*\end{array}\r):\begin{array}{c}\H_H^+\\ \oplus \\
\H_H^-\end{array} \to \begin{array}{c}\H_H^+\\ \oplus \\
\H_H^-\end{array}\]
is Hilbert-Schmidt.\\
(iii) The operator $BB^*:\H_H^-\to\H_H^-$
is of trace class. 
\end{lemma}
\begin{beweis}
Using again the fact that a $2\times 2$-matrix of operators is trace
class if and only if each of its entries is a trace class operator
\cite[Sec.~4.1.1.2, Lemma 2]{RS82} it is sufficient to show the
equivalence of the following statements:\\
(i) Each of the entries of the operator 
\[X^*X= \l(\begin{array}{cc}1-A(2Z^{-1/2}-1)A^*& A(Z^{-1/2}-1)B^*\\
B(Z^{-1/2}-1)A^* & BB^*\end{array}\r): \begin{array}{c}\H_H^+\\ \oplus \\
\H_H^-\end{array} \to \begin{array}{c}\H_H^+\\ \oplus \\
\H_H^-\end{array}\]
is trace class.\\
(ii) Each of the entries of the operator
\[Y^*Y= \l(\begin{array}{cc}1-A(2-Z)A^*& A(1-Z)B^*\\
B(1-Z)A^* & BZB^*\end{array}\r): \begin{array}{c}\H_H^+\\ \oplus \\
\H_H^-\end{array} \to \begin{array}{c}\H_H^+\\ \oplus \\
\H_H^-\end{array}\]
is trace class.\\
(iii) $BB^*:\H_H^-\to \H_H^-$ is trace class.\\

Remember that a compact operator $T:\H_1\to\H_2$ acting between two
(possibly different) Hilbert spaces $\H_1$ and $\H_2$ is said to be
trace class, $T\in \B(\H_1,\H_2)$, if it has finite trace norm
$\|T\|_1:=\sum_{i=1}^\infty s_i <\infty$, where $s_i$ are the
eigenvalues of $|T|:=(T^*T)^{1/2}$ on $\H_1$.\\
Note first, that $BB^*\in \B(\H_H^-,\H_H^-)\Leftrightarrow
B:\H_N^+\to\H_H^-$ is Hilbert-Schmidt $\Leftrightarrow
B^*:\H_H^-\to\H_N^+$ is Hilbert-Schmidt $\Leftrightarrow
B^*B\in\B(\H_N^+,\H_N^+)$.\\
Since $Z:=1+2B^*B:\H_N^+\to\H_N^+$ is a bounded operator with bounded
inverse we have
\[B^*B\in \B(\H_N^+,\H_N^+)\Leftrightarrow ZB^*B\in
\B(\H_N^+,\H_N^+)\Leftrightarrow BZB^*\in \B(\H_N^+,\H_N^+)\]
which proves the assertion for the $22$-components of $X^*X$ and
$Y^*Y$.\\
For the $12$-components we note that
\beq
-2B^*B=1-Z=(Z^{-1/2}-1)(Z^{1/2}+Z) \label{35l}
\eeq
where $Z^{1/2}+Z$ is a bounded operator on $\H_N^+$ with bounded
inverse. As shown after Eq.~\rf{35g}, also $A:\H_N^+\to\H_H^+$ is a bounded
operator with bounded inverse, therefore
\begin{eqnarray*}
B^*B\in\B(\H_N^+,\H_N^+)&\Leftrightarrow& A(1-Z)
=-2AB^*B\in\B(\H_N^+,\H_H^+)\\
& &\l\{\begin{array}{l}\Rightarrow A(1-Z)B^*\in \B(\H_H^-,\H_H^+)\\
\Rightarrow A(Z^{-1/2}-1)B^* = A(1-Z)(Z^{1/2}+Z)^{-1}B^* \in
\B(\H_H^-,\H_H^+). \end{array}\r.
\end{eqnarray*}
The argument for the $21$-component is analogous.\\
As for the $11$-component of $Y^*Y$ we note, using the invertibility
of $A$, the identity $Z=1+2B^*B$, and \rf{35d}, that
\begin{eqnarray*}
& &1-A(2-Z)A^* \in \B(\H_H^+,\H_H^+)\Leftrightarrow A^*A-A^*A(2-Z)A^*A
\in \B(\H_N^+,\H_N^+)\\
& &\Leftrightarrow (1+B^*B)B^*B(1+2B^*B)\in \B(\H_N^+,\H_N^+)
\Leftrightarrow B^*B\in\B(\H_N^+,\H_N^+).
\end{eqnarray*}
Similarly, using $A^*A=1+B^*B=\frac{1}{2}(1+Z)$, we rewrite the
$11$-component of $X^*X$ in terms of $Z$ and obtain
\begin{eqnarray}
& &1-A(2Z^{-1/2}-1)A^*\in \B(\H_H^+,\H_H^+)\Leftrightarrow
A^*A-A^*A(2Z^{-1/2}-1)A^*A \in \B(\H_N^+,\H_N^+)\nonumber\\
& &\Leftrightarrow (1+Z)(1-Z^{-1/2})(2-Z^{1/2}+Z)\in
\B(\H_N^+,\H_N^+). \label{35m}
\end{eqnarray}
Taking into account \rf{35l} and the identity
\[ (2-Z^{1/2}+Z)(Z^{1/2}+Z+2)=4+3Z+Z^2, \]
where both $Z^{1/2}+Z+2$ and $4+3Z+Z^2$ are bounded operators with
bounded inverse, we note that \rf{35m} is equivalent to
\begin{eqnarray*}
&&(1+Z)(Z-1)(4+3Z+Z^2)\in\B(\H_N^+,\H_N^+)\\
&&\Leftrightarrow Z-1\in\B(\H_N^+,\H_N^+)\Leftrightarrow
B^*B\in\B(\H_N^+,\H_N^+).
\end{eqnarray*}
This finishes the proof.
\end{beweis}

Theorem \ref{Theorem25} and Theorem \ref{Theorem29} imply that, for
$N>5/2$, $\pi_{\omega_N}(\A(\O))''$ and $\pi_{\omega_H}(\A(\O))''$, and,
if $\Sigma$ is compact, for $N>3/2$, $\pi_{\omega_N}(\A[\Gamma,\sigma])''$ and
$\pi_{\omega_H}(\A[\Gamma,\sigma])''$ 
are isomorphic von Neumann factors. Therefore it follows from the
corresponding results for Hadamard representations due to Verch
\cite[Thm.~3.6]{Verch97} that $\pi_{\omega_N}(\A(\O))''$ is isomorphic 
to the unique hyperfinite type $III_1$ factor if $\O^c$ is non-empty,
and is a type $I_\infty$ factor if $\O^c=\emptyset$ (i.e.~$\Sigma=\O$
is a compact Cauchy surface).\\
Our Theorem \ref{Theorem29} is the analogue of Theorem 3.3 of
L{\"u}ders \& Roberts \cite{LR90} extended to our definition of
adiabatic states on arbitrary curved spacetime manifolds. 
The loss of order $3/2+\epsilon$ in the compact case and
$1/2+\epsilon$ in the non-compact case ($\epsilon>0$ arbitrary)
compared to their result is probably due to the fact that
we use the regularity of $\Lambda_H-\Lambda_N$ rather generously in
the part of the proof of Theorem \ref{Theorem20} between Eq.s \rf{3.1}
and \rf{3.4}.

\subsection{Local definiteness and Haag duality}\label{Section3.2}
The next property of adiabatic vacua we check is that of local
definiteness. It says that any two adiabatic vacua (of order $>5/2$)
get indistinguishable upon measurements in smaller and smaller
spacetime regions. In a first step let us show that
in the representation $\pi_{\omega_N}$
generated by an adiabatic vacuum state $\omega_N$ (of order $N>3/2$)
there are no nontrivial observables which are localized at a single
point, more precisely:
\begin{thm}\label{Theorem32}
Let $x\in\Sigma$. Then, for $N>3/2$,
\[ \bigcap_{\O\ni x} \pi_{\omega_N}(\A(\O))''=\C \mathbf{1}, \]
where the intersection is taken over all open bounded subsets
$\O\subset \Sigma$.
\end{thm}
Before we prove the theorem let us recall how this, combined with
Theorem \ref{Theorem29}, implies the property of local definiteness:
\begin{cor}\label{Corollary32a}
Let $\omega_N$ be an adiabatic vacuum state of order $N>5/2$ and
$\omega_H$ an Hadamard state. Let $\O_n,\,n\in\N_0$, be a sequence of
open bounded subsets of $\Sigma$ shrinking to a point $x\in\Sigma$,
i.e.~$\O_{n+1}\subset\O_n$ and $\bigcap_{n\in\N_0}\O_n =\{x\}$. Then
\[\|(\omega_N-\omega_H)|_{\A(\O_n)}\|\to
0\quad\mbox{as}\;n\to\infty.\]
\end{cor}
\begin{beweis}
Let $(\pi_{\omega_N},\H_{\omega_N},\Omega_{\omega_N})$ be the
GNS-triple generated by $\omega_N$, and let ${\cal
R}_N(\O_n):=\pi_{\omega_N}(\A(\O_n))''$ be the corresponding von Neumann
algebras associated to the regions $\O_n\subset\Sigma$. Due to Theorem
\ref{Theorem29} and the remarks at the beginning of Section
\ref{Section3.1} $\pi_{(\omega_H\hrist\A(\O_0))}$ is quasiequivalent to
$\pi_{\omega_N}\hrist \A(\O_0)$. This implies \cite[Thm.~2.4.21]{BRI}
that $\omega_H\hrist\A(\O_0)$ can be represented in $\H_{\omega_N}$ as
a density matrix, i.e.~there is a sequence $\psi_m\in\H_{\omega_N}$
with $\sum_m\|\psi_m\|^2=1$ such that $\omega_H(A)=\sum_m \la\psi_m,
A\psi_m\ra$ for all $A\in \A(\O_0)$.\\
Let now $A_n\in {\cal R}_N(\O_n)\subset {\cal R}_N(\O_0)$ be a
sequence of observables with $\|A_n\|=1$. From Theorem \ref{Theorem32}
it follows that $A_n\to c \mathbf{1}$ in the
topology of ${\cal R}_N(\O_0)$ for some $c\in\C$. 
In particular, $A_n\to c\mathbf{1}$ in
the weak topology, thus
\[|\la \Omega_{\omega_N},(A_n-c\mathbf{1})\Omega_{\omega_N}\ra|\to
0\quad\mbox{as}\,n\to\infty,\]
and $A_n\to c\mathbf{1}$ in the $\sigma$-weak topology, thus
\[\sum_m|\la\psi_m,(A_n-c\mathbf{1})\psi_m\ra|\to
0\quad\mbox{as}\,n\to\infty.\]
From this we can now conclude
\beqa
|(\omega_N-\omega_H)(A_n)|&=& |\la \Omega_{\omega_N},A_n\Omega_{\omega_N}\ra
-\sum_m\la\psi_m, A_n\psi_m\ra|\nonumber\\
&=&|\la\Omega_{\omega_N},(A_n-c\mathbf{1})\Omega_{\omega_N}\ra
-\sum_m\la\psi_m,(A_n-c\mathbf{1})\psi_m\ra|\nonumber\\
&\leq&
|\Omega_{\omega_N},(A_n-c\mathbf{1})\Omega_{\omega_N}
\ra|+\sum_m|\la\psi_m,(A_n-c\mathbf{1})
\psi_m\ra|\nonumber\\
&\to& 0\quad \mbox{as}\,n\to\infty,\label{10}
\eeqa
i.e. $(\omega_N-\omega_H)(A_n)$ converges to $0$ pointwise for each
sequence $A_n$. To show the uniform convergence we note that due to
$\A(\O_{n+1})\subset \A(\O_n)$
\[r_n:=\sup\{|(\omega_N-\omega_H)(A)|;\;A\in \A(\O_n),\|A\|=1\}\]
is a bounded monotonically decreasing sequence in $n\in\N_0$ with
values in $\R_0^+$. Hence $r_n\to r$ for some $r\in\R_0^+$.
To show that $r=0$ 
let $\epsilon >0$. For all $n\in\N_0$ there is an $A_n\in\A(\O_n)$
with $\|A_n\|=1$ such that
\[ 0\leq r_n-|(\omega_N-\omega_H)(A_n)|\leq \epsilon.\]
Furthermore, due to \rf{10} there is an $n_o\in\N_0$ such that for all
$n\geq n_o$
\[|(\omega_N-\omega_H)(A_n)|\leq \epsilon.\]
From these inequalities we obtain for $n\geq n_o$
\[0\leq r\leq r_n\leq \epsilon +|(\omega_N-\omega_H)(A_n)|\leq
2\epsilon\] 
and hence $r=0$. This proves the assertion.
\end{beweis}

To prove Theorem \ref{Theorem32} we show an even stronger statement, namely
\beq
\bigcap_{\O\supset S} \pi_{\omega_N}(\A(\O))''=\C\mathbf{1} \label{11}
\eeq
for any smooth $2$-dim.~closed submanifold $S$ of $\Sigma$. The
statement of Theorem \ref{Theorem32} then follows if we choose $x\in
S$.
In the proof we will need the following lemma:
\begin{lemma}\label{Lemma33}
\[ \bigcap_{\O\supset S} \ol{\co{\O}}=\{0\}, \]
where the closure is taken w.r.t.~the norm of $H^{-1/2}(\Sigma)$ (and
hence it also holds w.r.t.~the norm of $H^{1/2}(\Sigma)$).
\end{lemma}
Note that we can confine the intersection to all sets $\O$ contained
in a suitable compact subset of $\Sigma$. Hence we can assume that
$(\Sigma,h)$ is a complete Riemannian manifold (otherwise we modify
$h$ as in the proof of Theorem \ref{Theorem20}), so that
$H^{\pm 1/2}(\Sigma)$ is well-defined.\\   \\
{\em Proof of Lemma \ref{Lemma33}:}
The problem is local, so it suffices to consider the case
$\Sigma=\R^n, S=\R^{n-1}\times\{0\}$. Suppose the above intersection
contains some $f\in H^{-1/2}(\R^n)$, say $\|f\|_{H^{-1/2}}=1$. Fix
$0<\epsilon<1/2$. Since
\[ \|f\|_{H^{-1/2}}=\sup \{|f(F)|;\;F\in H^{1/2},\|F\|_{H^{1/2}}=1\}
\]
we find some $F\in H^{1/2}(\R^n)$ such that $\|F\|_{H^{1/2}}=1$ and
$f(F)>1-\epsilon$. According to Lemma \ref{Lemma26} there exists an
$F_0\in \co{\R^n\setminus(\R^{n-1}\times \{0\})}$ such that
$\|F-F_0\|_{H^{1/2}} <\epsilon$ and therefore $f(F_0)= f(F) + f(F_0-F) 
>1-2\epsilon$. Clearly there is a $\delta>0$ such that
$|x_n|>2\delta$ for each $x=(x',x_n)\in \supp F_0$.\\
On the other hand, $f\in \ol{\co{\R^{n-1}\times
(-\delta,\delta)}}^{H^{-1/2}}$, hence $\supp f\subset \R^{n-1}\times
[-\delta,\delta]$ (in order to see this, use the fact that the closure 
of $\co{\R^n_+}$ in the topology of $H^s(\R^n)$ is equal to $\{u\in
H^s(\R^n);\; \supp u\subset \ol{\R^n_+}\}$ for $s\in \R$,
cf.~\cite[2.10.3]{Triebel78}). Denoting by $\chi_\delta$ a smooth
function, equal to $1$ on $\R^{n-1}\times [-\delta,\delta]$ and
vanishing outside $\R^{n-1}\times (-2\delta,2\delta)$, we have
$f=\chi_\delta f$ and therefore
\[ 1-2\epsilon < f(F_0)=(\chi_\delta f)(F_0) =f(\chi_\delta F_0) =
f(0) =0,\]
a contradiction. \hfill$\blacksquare$\\

{\it Proof of Theorem \ref{Theorem32}:}
Let $(k_N,\H_N)$ be the one-particle Hilbert space structure of
$\omega_N$. According to results of Araki \cite{Araki63,LRT78} \rf{11} 
holds iff
\beq
\bigcap_{\O\supset S} \ol{k_N(\Gamma(\O))} =\{0\},\label{12}
\eeq
where the closure is taken w.r.t.~the norm in $\H_N$.\\
As in the proof of Theorem \ref{Theorem25}, let us define a
one-particle Hilbert space structure $(\tilde{k}, \tilde{\H})$ of an
auxiliary pure quasifree state on $\A[\Gamma,\sigma]$ by
\begin{eqnarray*}
\tilde{k}:\Gamma &\to& L^2(\Sigma,h)=:\tilde{\H}\\
{q\choose p}&\mapsto& \frac{1}{\sqrt{2}}\l(i\la D\ra^{1/2}q +\la
D\ra^{-1/2} p\r);
\end{eqnarray*}
as before, we may change $h$ near infinity to obtain completeness. 
Note that the norm given by
\[ \tilde{\mu}(F,F) := \la \tilde{k} F,\tilde{k}
F\ra_{\tilde{\H}}=\frac{1}{2} \l[\|\la D\ra^{1/2} q\|^2_{L^2} +\|
\la D\ra^{-1/2} p\|^2_{L^2}\r], \quad
F:=(q,p)\in\Gamma, \]
is equivalent to the norm of
$H^{1/2}(\Sigma)\oplus H^{-1/2}(\Sigma)$.\\
Let $u\in \ol{k_N(\Gamma(\O))}$ for all $\O\supset S$. Thus for every
$\O$ there is a sequence $\{F_n^\O,\;n\in \N\}\subset \Gamma(\O)$ with 
$k_N(F_n^\O)\to u$ in $\H_N$. By Theorem \ref{Theorem20} the norm
given by $\mu_N,\,N>3/2$, on $\Gamma(\O)$ is equivalent to the norm
given by $\tilde{\mu}$, namely that of $H^{1/2}(\O)\oplus
H^{-1/2}(\O)$. Therefore it follows that also $\tilde{k}(F_n^\O)\to
v^\O$ in $\tilde{\H}$ for some $v^\O\in
\ol{\tilde{k}(\Gamma(\O))}$. Moreover, $v^\O$ must be independent of
$\O$: To see this, suppose that $\O_1$ and $\O_2$ are contained in a
common open, bounded set $\tilde{\O}\subset\Sigma$, and let $\epsilon
>0$. Then there is an $n\in \N$ such that
\begin{eqnarray*}
\|v^{\O_1} - v^{\O_2}\|_{\tilde{\H}}&\leq&
\|v^{\O_1}-\tilde{k}(F_n^{\O_1})\|_{\tilde{\H}}
+\|\tilde{k}(F_n^{\O_1})-\tilde{k}(F_n^{\O_2})\|_{\tilde{\H}} +
\|\tilde{k}(F_n^{\O_2})-v^{\O_2}\|_{\tilde{\H}}\\
&\leq& 2\epsilon +\|\tilde{k}(F_n^{\O_1}-F_n^{\O_2})\|_{\tilde{\H}} 
=2\epsilon
+\tilde{\mu}(F_n^{\O_1}-F_n^{\O_2},F_n^{\O_1}-F_n^{\O_2})^{1/2}\\ 
&\leq& 2\epsilon
+C(\tilde{\O})\,\mu_N(F_n^{\O_1}-F_n^{\O_2},F_n^{\O_1}-F_n^{\O_2})^{1/2}\\ 
&=& 2\epsilon + C(\tilde{\O})
\|k_N(F_n^{\O_1})-k_N(F_n^{\O_2})\|_{\H_N}\\
&\leq& 2\epsilon
+C(\tilde{\O})\l(\|k_N(F_n^{\O_1})-u\|_{\H_N}+\|u-k_N(F_n^{\O_2})\|_{\H_N}\r)
\\ 
&\leq& 2\epsilon (1+C(\tilde{\O})),
\end{eqnarray*}
hence $v^{\O_1}=v^{\O_2}$, and we denote this unique element of
$\tilde{\H}$ by $v$.\\
Since $v\in \bigcap_{\O\supset
S}\ol{\tilde{k}(\Gamma(\O))}^{\tilde{\H}}$ it follows from Lemma
\ref{Lemma33} that $v=0$ and therefore $\tilde{k}(F_n^\O)\to 0$ in
$\tilde{\H}$. Since the norms given by $k_N$ and $\tilde{k}$ are
equivalent on $\Gamma(\O)$ we also have $k_N(F_n^\O)\to 0$ in $\H_N$
and thus $u=\lim_{n\to \infty} k_N(F_n^\O)=0$, which proves the theorem.
\hfill$\blacksquare$\\
 
In the following theorem we show that the observable algebras ${\cal
R}_N(\O) :=
\pi_{\omega_N} (\A(\O))''$ generated by adiabatic vacuum states
(of order $N>3/2$) satisfy a certain maximality property, called Haag
duality. Due to the locality requirement it is clear that all
observables localized in spacelike separated regions of spacetime
commute. If $\O$
is some open, relatively compact subset of the Cauchy surface $\Sigma$ 
with smooth boundary, this means that
\beq
{\cal R}_N(\O^c) \subset {\cal R}_N(\O)', \label{13}
\eeq
where $\O^c:=\Sigma\setminus \ol{\O}$ and 
\beq
{\cal R}_N(\O^c) :=\l(\bigcup_{\ol{\O_1}\subset
\O^c}\pi_{\omega_N}(\A(\O_1))\r)'' \label{13a}
\eeq
is the von Neumann algebra generated by all
$\pi_{\omega_N}(\A(\O_1))$ with $\O_1$ bounded and 
$\ol{\O_1}\subset \O^c$. One says that 
Haag duality holds for the net of von Neumann algebras generated by a
pure state if \rf{13} is
even an equality. For mixed states (i.e.~reducible
GNS-representations) this can certainly not be true,
because in this case, by Schur's lemma \cite[Prop.~2.3.8]{BRI}, there
is a set ${\cal S}$ of non-trivial operators commuting with the
representation $\pi_{\omega_N}$, i.e.
\beq
{\cal S} \subset {\cal R}_N(\O)'\cap {\cal R}_N(\O^c)'. \label{13b}
\eeq
If equality held in \rf{13} then the right hand side of \rf{13b} would 
be equal to ${\cal R}_N(\O)'\cap {\cal R}_N(\O)$, i.e.~to the centre
of ${\cal R}_N(\O)$, which, however, is trivial due to the local
primarity (Theorem \ref{Theorem25}) of the representation $\pi_{\omega_N}$,
hence ${\cal S}\subset \C {\mathbf 1}$, a contradiction. Therefore, in 
the reducible case one has to take the intersection with ${\cal R}_N:= 
\pi_{\omega_N}(\A[\Gamma,\sigma])''$ 
on the right hand side of \rf{13} to get equality\footnote{We are
grateful to Fernando Lled{\'o} for pointing out to us this
generalization of Haag duality and discussion about this topic.} (in
the irreducible case, again by Schur's lemma, $\pi_{\omega_N}(\A)'=\C{\mathbf
1}\Rightarrow \pi_{\omega_N}(\A)''={\cal B}(\H_{\omega_N})$, hence the
intersection with ${\cal R}_N$ is redundant).
Haag duality is an
important assumption in the theory of superselection sectors
\cite{Haag96} and has therefore been checked in many models of
physical interest. For our situation at hand, Haag duality has been shown by
L{\"u}ders \& Roberts \cite{LR90} to hold for the GNS-representations of 
adiabatic vacua on
Robertson-Walker spacetimes and by 
Verch \cite{Verch93b, Verch97} for those of 
Hadamard Fock states. He also noticed that it extends to all 
Fock states that are locally quasiequivalent to Hadamard states,
hence, by our Theorem \ref{Theorem29}, to pure adiabatic states of
order $N>5/2$. Nevertheless, we present an independent proof of Haag
duality for adiabatic states that does not rely on
quasiequivalence but only on Theorem \ref{Theorem20} and also holds for
mixed states.
\begin{thm}\label{Theorem34}
Let $\omega_N$ be an adiabatic state of order $N>3/2$.\\
Then, for any open, relatively compact subset $\O\subset\Sigma$ with
smooth boundary, 
\[ {\cal R}_N(\O^c) = {\cal R}_N(\O)'\cap {\cal R}_N, \]
where $\O^c:=\Sigma\setminus \ol{\O}$ and ${\cal R}_N(\O^c)$ is
defined by \rf{13a}.
\end{thm}
\begin{beweis}
Denoting again by $(k_N,\H_N)$ the one-particle Hilbert space
structure of $\omega_N$, it follows from results of Araki
\cite{Araki63, LRT78} that the assertion is equivalent to the
statement
\[  \ol{k_N(\Gamma(\O^c))}= k_N(\Gamma(\O))^\vee\cap \ol{k_N(\Gamma)},\]
where the closure has to be taken w.r.t.~$\H_N$ and
$k_N(\Gamma(\O))^\vee$ was defined in Eq.~\rf{7a}. Since
\[ k_N(\Gamma(\O^c)) \subset k_N(\Gamma(\O))^\vee \cap k_N(\Gamma)\]
(due to the locality of $\sigma$, compare \rf{13} above), we only have 
to show that $k_N(\Gamma(\O^c))$ is dense in
$k_N(\Gamma(\O))^\vee \cap\ol{k_N(\Gamma)}$. This in turn is the case iff
\beq
k_N(\Gamma(\O)) +k_N(\Gamma(\O^c))\;\mbox{ is dense in}\;\ol{k_N(\Gamma)}
\label{14}
\eeq
(for the convenience of the reader, the argument will be given in
Lemma \ref{Lemma35} below). 
\rf{14} will follow if we show that every element $u=k_N (F)\in
k_N(\Gamma),\,F=(q,p)\in\Gamma$, can be approximated by a sequence in
$k_N(\Gamma(\O))+k_N(\Gamma(\O^c))$. \\
To this end we fix a bounded open set $\O_0\subset \Sigma$ with smooth 
boundary such that $\supp p$ and $\supp q \subset \O_0$. According to 
Lemma \ref{Lemma26} we find sequences $\{q_n\},\{p_n\}\subset \co{\O}, 
\{q_n^c\},\{p_n^c\}\subset \co{\O^c}$ such that
\beqa
q-(q_n+q_n^c) &\to& 0\;\mbox{in}\;H^{1/2}(\O_0) \label{19}\\
p-(p_n+p_n^c) &\to& 0\;\mbox{in}\;H^{-1/2}(\O_0). \label{20}
\eeqa
Note that it is no restriction to ask that the supports of all
functions are contained in $\O_0$. Let us denote by
$(\tilde{k},\tilde{\H})$ the one-particle Hilbert space structure
introduced in \rf{8a} with the real scalar product
$\tilde{\mu}$ given by \rf{8b}. The relations \rf{19} and \rf{20}
imply that
\[ \Gamma(\O_0)\ni F_n:=(q-(q_n+q_n^c),p-(p_n+p_n^c))\to 0 \]
with respect to the norm induced by $\tilde{\mu}$. According to
Theorem \ref{Theorem20} it also tends to zero with respect to the norm 
induced by $\mu_N$, in other words
\[ k_N(F_n)\to 0 \;\mbox{in}\;\H_N. \]
This completes the argument.
\end{beweis}

\begin{lemma}\label{Lemma35}
\[k_N(\Gamma(\O))+k_N(\Gamma(\O^c)) \;\mbox{is dense in}\;\ol{k_N(\Gamma)}
\Leftrightarrow
 k_N(\Gamma(\O^c))\;\mbox{is dense
in}\;k_N(\Gamma(\O))^\vee\cap\ol{k_N(\Gamma)}.\]
\end{lemma}
\begin{beweis}
$\Rightarrow$. Let $u\in k_N(\Gamma(\O))^\vee\cap \ol{k_N(\Gamma)}$, 
and choose $v_n\in
k_N(\Gamma(\O)),w_n\in k_N(\Gamma(\O^c))$ such that
\beq
v_n+w_n\to u\;\mbox{in}\;\H_N.\label{21}
\eeq
In view of the fact that $k_N(\Gamma(\O^c))\subset
k_N(\Gamma(\O))^\vee$ we have
\[ \ol{k_N(\Gamma(\O))}\cap \ol{k_N(\Gamma(\O^c))}\subset
\ol{k_N(\Gamma(\O))} \cap k_N(\Gamma(\O))^\vee = \{0\}. \]
Indeed, the last equality is a consequence of Theorem \ref{Theorem25},
cf.~\rf{8}. We can therefore define a continuous map
\begin{eqnarray*}
\pi: \ol{k_N(\Gamma(\O))}\oplus \ol{k_N(\Gamma(\O^c))} &\to& \H_N\\
 v\oplus w &\mapsto& v.
\end{eqnarray*}
Now \rf{21} implies that $\{v_n+w_n\}$ is a Cauchy sequence in $\H_N$, 
hence so are $\{v_n\}= \{\pi(v_n+w_n)\}$ and $\{w_n\}$. Let $v_0:=
\lim v_n\in \ol{k_N(\Gamma(\O))}, w_0:=\lim w_n\in
\ol{k_N(\Gamma(\O^c))}.$ By \rf{21},
\[ u-w_0=v_0\in k_N(\Gamma(\O))^\vee \cap \ol{k_N(\Gamma(\O))}=\{0\}.\]
Therefore $u=w_0\in \ol{k_N(\Gamma(\O^c))}.$\\
$\Leftarrow$. Denoting by $\bot$ the orthogonal complement in
$\ol{k_N(\Gamma)}$,
we clearly have from the definition \rf{7a} of $\vee$ that
$k_N(\Gamma(\O))^\bot \subset k_N(\Gamma(\O))^\vee\cap\ol{k_N(\Gamma)}$. Since
$\ol{k_N(\Gamma(\O))} +k_N(\Gamma(\O))^\bot =\ol{k_N(\Gamma)}$ it follows that
$k_N(\Gamma(\O))+\l(k_N(\Gamma(\O))^\vee\cap \ol{k_N(\Gamma)}\r)$ 
is dense in $\ol{k_N(\Gamma)}$. From the
assumption that $k_N(\Gamma(\O^c))$ is dense in
$k_N(\Gamma(\O))^\vee\cap \ol{k_N(\Gamma)}$ the assertion follows.
\end{beweis}

\section{Construction of adiabatic vacuum states} \label{Section5}
The existence of Hadamard states on arbitrary globally hyperbolic
spacetimes has been proven by Fulling, Narcowich \& Wald \cite{FNW81}
using an elegant deformation argument. Presumably, the existence of
adiabatic vacuum states could be shown in a similar way employing the
propagation of the Sobolev wavefront set, Proposition
\ref{PropositionA4}. Instead of a mere existence argument, however, we
prefer to explicitly construct a large class of adiabatic vacuum
states as it
is indispensable for the extraction of concrete
information in physically relevant situations to have available a
detailed construction of the solutions of the theory.\\
In Section \ref{Section5.1} we first present a parametrization of
quasifree states in terms of two operators $R$ and $J$ acting on the
$L^2$-Hilbert space w.r.t.\ a Cauchy surface $\Sigma$, Theorem
\ref{Theorem5.1}. The main technical result is Theorem
\ref{Theorem5.2} which gives a sufficient condition on $R$ and $J$
such that the associated quasifree states are adiabatic of a certain
order. In Section \ref{Section5.2} we construct operators $R$ and $J$
satisfying the above assumptions, Theorem \ref{Theorem5.7}.
\pagebreak
\subsection{Criteria for initial data of adiabatic vacuum
states}\label{Section5.1} 
We recall the following theorem from \cite[Thm.~3.11]{Junker96}:
\begin{thm} \label{Theorem5.1}
Let $(\M,g)$ be a globally hyperbolic spacetime with Cauchy surface
$\Sigma$. Let $J,R$ be operators on $\lq$ satisfying the following
conditions:\\
(i) $\co{\Sigma} \subset dom (J)$,\\
(ii) $J$ and $R$ map $\co{\Sigma,\R}$ to $L^2_{\R}(\Sigma,d^3\sigma)$,\\
(iii) $J$ is selfadjoint and positive with bounded inverse,\\
(iv) $R$ is bounded and selfadjoint.\\
Then 
\begin{eqnarray}
k:\Gamma&\to& \H:=\ol{k(\Gamma)}\subset \lq\nonumber\\
(q,p)&\mapsto& (2J)^{-1/2} \l[ (R-iJ)q-p\r]\label{5.000} 
\end{eqnarray}
is the one-particle Hilbert space structure of a pure quasifree state.
\end{thm}
Note that we can define the inverse square root by
\beq
(2J)^{-1/2}=\frac{1}{\pi}\int_0^\infty \lambda^{-1/2}(\lambda
+2J)^{-1}\,d\lambda .\label{5.00}
\eeq
The integral converges since $\lambda +2J\geq \lambda$ and hence
$(\lambda +2J)^{-1} \leq \lambda^{-1}$ for $\lambda \geq 0$. Therefore 
$(2J)^{-1/2}$ is a bounded operator on $\lq$. Moreover, $(2J)^{-1/2}$
maps $L^2_{\R}(\Sigma,d^3\sigma)$ to itself since $\lambda+2J$ and
therefore $(\lambda+2J)^{-1}$ commutes with complex conjugation
($\lambda \geq 0$). 

\begin{beweis}
A short computation shows that for $F_j=(q_j,p_j)\in \Gamma,\,j=1,2,$
we have
\begin{eqnarray*}
\sigma (F_1,F_2) &=& -\la q_1,p_2\ra +\la p_1,q_2\ra 
= 2 \mbox{Im}\,\la kF_1,kF_2\ra.
\end{eqnarray*}
Here, $\la\cdot,\cdot\ra$ is the scalar product of $\lq$. We then let
\[ \mu(F_1,F_2) := \mbox{Re}\,\la kF_1,kF_2\ra. \]
We note that
\begin{eqnarray*}
|\sigma(F_1,F_2)|^2 &\leq& 4 |\la kF_1,kF_2\ra|^2 \leq 4 \la
kF_1,kF_1\ra \la kF_2,kF_2\ra \\
&=&4 \mu (F_1,F_1)\mu(F_2,F_2);
\end{eqnarray*}
hence $k$ defines the one-particle Hilbert space structure of a
quasifree state with real scalar product $\mu$ (Definition
\ref{dfn1.2}) and one-particle
Hilbert space $\H=\ol{k\Gamma +ik\Gamma}$ (Proposition \ref{Proposition1.3}).\\
Let us next show that the state is pure, i.e.~$k\Gamma$ is dense in
$\H$ (see Proposition \ref{Proposition1.3}). 
We apply a criterion by Araki \& Yamagami \cite{AY82} and check
that the operator $S:\Gamma\to \lq \oplus \lq$ defined by $\la
kF_1,kF_2\ra =2\mu (F_1,S F_2)$ is a projection (cf.\ Eq.\ \rf{8c}). 
Indeed, this relation implies that 
\[ S=\frac{1}{2} \l(\begin{array}{cc} iJ^{-1}R+ {\mathbf 1} & -iJ^{-1}\\
iRJ^{-1}R +iJ & -iRJ^{-1}+ {\mathbf 1} \end{array}\r).\]
Therefore $S^2=S$, and the proof is complete.
\end{beweis}
\pagebreak

{\it The distinguished parametrices of the Klein-Gordon operator}\\
In the following we shall use the calculus of Fourier integral
operators of Duistermaat \& H\"ormander \cite{DH72} in order to analyze the
wavefront set of certain bilinear forms related to fundamental
solutions of the Klein-Gordon operator $P=\Box_g+m^2$. We recall from
\cite[Thm.~6.5.3]{DH72} that $P$
on a globally hyperbolic spacetime (which is known to be
pseudo-convex w.r.t.~$P$, see \cite{Radzikowski96a}) has $2^2=4$
orientations $C\setminus \mbox{diag}\, (C)=C_\nu^1 \dot{\cup} C_\nu^2$ of the
bicharacteristic relation $C$, Eq.~\rf{2.0a}. 
Here, $\nu$ is one of the four sets $\{\emptyset,N_+,N_-,N_+\cup
N_-\}$ of components of the lightcone $N:= \mbox{char}\,P$, with
$N_\pm := N\cap \{\xi_0 \gk 0\}$. $C_\nu^{1(2)}$ are the subsets of the
bicharacteristic relation which are denoted by $C_\nu^{+(-)}$ in
\cite{DH72}. Associated to these orientations there are four pairs
$E_\nu^1, E_\nu^2$ of distinguished parametrices with
\[ WF'(E_\nu^1) =\Delta^*\cup C_\nu^1,\quad WF'(E_\nu^2)=\Delta^*\cup
C_\nu^2 \]
where $\Delta^*$ is the diagonal in $(T^*X\setminus 0)\times
(T^*X\setminus 0)$. 
Moreover, Duistermaat \& H\"ormander show that every parametrix $E$
with $WF'(E)$ contained in $\Delta^*\cup C_\nu^1$ resp.~$\Delta^*\cup
C_\nu^2$ must be equal to $E_\nu^1$ resp.~$E_\nu^2$ modulo
$\ci$. In addition,
\[ E_\nu^1-E_\nu^2 \in I^{1/2-2} (\M\times\M,C') \]
and $E_\nu^1-E_\nu^2$ is noncharacteristic at every point of $C'$. 
Here, $I^\mu(X,\Lambda)$ denotes the space of Lagrangean distributions
of order $\mu$ over the manifold $X$ associated to the Lagrangean
submanifold $\Lambda\subset T^*X\setminus 0$, cf.\
\cite[Def.~25.1.1]{HormIV}. \\
We shall need three particular parametrices: For the forward light cone
$N_+$ we obtain $E_{N_+}^1=E^R$
(mod $\ci$),
the retarded Green's function, for the backward light cone
$N_-$ we have the advanced Green's
function $E_{N_-}^1=E^A$ (mod $\ci$) while $E_{N_+\cup N_-}^1$ is the
so-called Feynman parametrix $E^F$ (mod $\ci$).
We deduce that $E^1_{N_+}=E^2_{N_-}$ (mod $\ci$), in particular
\[ E=E^R-E^A \in I^{-3/2} (\M\times \M,C'). \]
We next write $E=E^+ +E^-$ with $E^+:=E^F-E^A, E^-:=E^R-E^F$. We
deduce from \cite[Thm.~6.5.7]{DH72} that
\beqa
E^-&=& E^R-E^F=E^1_{N_+}-E^1_{N_+\cup N_-}\in I^{-3/2}(\M\times \M,
(C^-)') \label{5.1}\\
E^+ &=& E^F-E^A = E^1_{N_+\cup N_-}-E^1_{N_-}\in I^{-3/2}(\M\times
\M,(C^+)') \label{5.2}
\eeqa
where $C^+=C\cap (N_+\times N_+),\, C^-=C\cap (N_-\times N_-)$ 
as in Eq.~\rf{2.0b}.\\ 
It follows from \cite[Thm.~5.1]{Radzikowski96a} that the two-point function
$\Lambda_H$ of every Hadamard state coincides with $iE^+$ (mod $\ci$).
(We define the physical
Feynman propagator by $F(x,y):=-i \la
T\Phi(x)\Phi(y)\ra$, i.e.\ $-i$ $\times$ the expectation value of the 
time ordered product of two field operators. From this choice it
follows that $iF=\Lambda_H+i E^A$ and hence $F=E^F$ (mod $\ci$) and
$\Lambda_H = iE^+$ (mod $\ci$).)
\begin{lemma}\label{Lemma5.1a}
For every Hadamard state $\Lambda_H$ we have
\[ WF'^s(\Lambda_H)=WF'^s(E^+)=\l\{\begin{array}{ll} \emptyset, &
s<-\frac{1}{2} \\ C^+,& s\geq -\frac{1}{2} \end{array}\r. .\]
\end{lemma}
\begin{beweis}
The statement for $s<-1/2$ follows from Eq.\ \rf{5.2} and Proposition
\ref{PropositionA10}. For $s\geq -1/2$ we rely on
\cite[Section 6]{DH72}. According to \cite[Eq.~(6.6.1)]{DH72}
\[E^1_{N_+\cup N_-}+E^1_\emptyset = E^1_{N_+} +E^1_{N_-}\;\mbox{mod}\,\ci,\]
so that, in the notation of \cite[Eq.~(6.6.3)]{DH72},
\[ E^+=E^1_{N_+}-E^1_\emptyset = S_{N_+}.\]
The symbol of $S_{N_+}$ is computed in \cite[Thm.~6.6.1]{DH72}. It
is non-zero on the diagonal $\Delta_N$ in $N\times N$.
Moreover, it satisfies a homogenous first order ODE along the
bicharacteristics of $P$ in each pair of variables, so that it is
non-zero everywhere on $C^+$. Hence $E^+$ is non-characteristic at
every point of $C^+$. Now Proposition \ref{PropositionA10} gives the
assertion. 
\end{beweis}

We fix a normal coordinate $t$ which allows us to
identify a neighborhood of $\Sigma$ in $\M$ with $(-T,T)\times
\Sigma =:\M_T$. We assume that $R_l=\{R_l(t);\;-T<t<T\}$ and
$J_l=\{J_l(t);\;-T<t<T\}, l=1,2,$ are smooth families of properly supported
pseudodifferential operators on $\Sigma$ with local symbols $r_l=r_l(t)\in 
\ci ((-T,T),S^{0}(\Sigma\times \R^3))$ and $j_l=j_l(t) \in \ci
((-T,T),S^{1}(\Sigma\times \R^3))$.
Moreover, let $H=\{H(t);\;-T<t<T\}$ be a smooth family of properly
supported pseudodifferential operators of order $-1$ on $\Sigma$. 
We can then also view $R_l, J_l$, and $H$ as
operators on, say, $\co{(-T,T)\times \Sigma}$. 

\begin{thm}\label{Theorem5.2}
Let $R_l, J_l$, and $H$ be as above, and let $Q_l$ be a properly supported
first order pseudodifferential operator on $(-T,T)\times \Sigma$ such that
\beq
Q_l(R_l-iJ_l-\partial_t)E^- = S_l^{(N)}E^-,\quad l=1,2, \label{5.0}
\eeq
with $S_l^{(N)}=S_l^{(N)}(t) 
\in \ci ((-T,T),L^{-N}(\Sigma))$ a smooth family of properly supported
pseudodifferential operators on $\Sigma$ of order $-N$. Moreover, we
assume that $Q_l$ has a real-valued principal symbol such that
\[ {\rm char}\,Q_l\cap N_- =\emptyset.\]
Then the distribution $D_N\in \Dp{\M\times \M}$, defined by
\[ D_N(f_1,f_2) =  \l\la \l[(R_1-iJ_1)\rho_0 -\rho_1\r] Ef_1,
H  \l[ (R_2-iJ_2)\rho_0 -\rho_1\r] Ef_2\r\ra \]
satisfies the relation
\beq
 WF'^s (D_N)\subset \l\{\begin{array}{ll}\emptyset,&s<-1/2\\
C^+, &-1/2\leq s<N+3/2. \end{array}\r. \label{5.0a}
\eeq
\end{thm}
Note that $D_N$ will in general not be a two-point function unless
$R_1=R_2$ and $J_1=J_2=H^{-1}$ are selfadjoint and $J$ is positive
(compare Theorem \ref{Theorem5.1}).

\begin{beweis}
Since $\rho_0$ commutes with $R_l,J_l$ and $H$ we have
\beq
D_N(f_1,f_2)=\l\la
\rho_0\l[R_1-iJ_1-\partial_t\r]Ef_1,\rho_0
H\l[R_2-iJ_2-\partial_t\r]Ef_2\r\ra.  
\label{5.0b}
\eeq
Denoting by $K_1$ and $K_2$ the distributional kernels of
$(R_1-iJ_1-\partial_t)E$ and $H(R_2-iJ_2-\partial_t)E$, respectively,
we see that 
\[ D_N=(\rho_0K_1)^*(\rho_0K_2). \]
We shall apply the calculus of Fourier integral operators
in order to analyze the composition $(\rho_0 K_1)^* (\rho_0 K_2)$. The
following lemma is similar in spirit to \cite[Thm.~25.2.4]{HormIV}.

\begin{lemma}\label{Lemma5.2a}
Let $X\subset \R^{n_1},Y\subset \R^{n_2}$ be open sets and $A\in
L^{k}(X)$ 
be a properly supported pseudodifferential operator with symbol
$a(x,\xi)$. Assume that $C$ is a homogeneous canonical relation from
$T^*Y\setminus 0$ to $T^*X\setminus 0$ and that $a(x,\xi)$ vanishes on a
conic neighborhood of the projection of $C$ in $T^*X\setminus 0$. If
$B\in I^{m}(X\times Y, C')$ then
\[ AB\in I^{-\infty}(X\times Y, C').\]
\end{lemma}
\begin{beweis}
The problem is microlocal, so we may assume that $B$ has the form
\[ Bu(x) = \int e^{i\phi (x,y,\xi)} b(x,y,\xi) u(y) \,dy\,d^N\xi, \]
where $\phi$ is a non-degenerate phase function on $X\times Y\times
(\R^N\setminus \{0\})$ and $b\in S^{m+(n_1+n_2-2N)/4}(X\times Y\times
\R^N)$ an amplitude. 
We know that $C=T_\phi (C_\phi)$, where
\[ C_\phi := \{(x,y,\xi)\in X\times Y\times(\R^N\setminus
\{0\});\;d_\xi \phi(x,y,\xi)=0\}, \]
and $T_\phi$ is the map
\begin{eqnarray*}
T_\phi :X\times Y\times (\R^N\setminus \{0\}) &\to& T^*(X\times
Y)\setminus 0\\
(x,y,\xi) &\mapsto& (x,d_x\phi;y,d_y\phi).
\end{eqnarray*}
We recall that 
ess supp$\,b$ is the smallest closed conic subset of $X\times Y\times
(\R^N\setminus \{0\})$ outside of which $b$ is of class $S^{-\infty}$
and that the wavefront set of the kernel of $B$ is contained in the
set
\[T_\phi (C_\phi\cap \mbox{ess supp}\,b), \] 
cf.~\cite[Thm.~2.2.2]{Duistermaat96}. Hence 
we may assume that $b$ vanishes outside a conic neighborhood $\cal N$
of $C_\phi$ in $X\times Y\times \R^N$. In fact we can choose
this neighborhood so small that $a(x,\xi')=0$ whenever $(x,\xi')$ lies 
in the projection of $T_\phi({\cal N}) \subset T^*X\times T^*Y$ onto
the first component (we call this projection $\pi_1$). Then
\[ ABu(x)=\int e^{i\phi(x,y,\xi)} c(x,y,\xi) u(y) \,dy\,d^N\xi \]
where \[ c(x,y,\xi)= e^{-i\phi(x,y,\xi)}
A(b(\cdot,y,\xi)e^{i\phi(\cdot,y,\xi)}). \]
According to \cite[Ch.~VIII, Eq.~(7.8)]{Taylor81}, $c$ has the asymptotic
expansion 
\beq
c(x,y,\xi)\sim \sum_{\alpha\geq 0 \atop \beta\leq \alpha} D^\alpha_\xi 
a(x,d_x\phi(x,y,\xi)) D^\beta_xb(x,y,\xi) \psi_{\alpha\beta}(x,y,\xi)
\label{5.2a} 
\eeq
where $\psi_{\alpha\beta}$ is a polynomial in $\xi$ of degree $\leq
|\alpha-\beta|/2$. Now from our assumptions on $a$ and $b$ it follows
that in Eq.~\rf{5.2a}
\begin{eqnarray*}
b(x,y,\xi)=0 &\mbox{if}& (x,y,\xi)\notin {\cal N}\\
a(x,\xi')=0 &\mbox{if}& (x,\xi')\in \pi_1 T_\phi({\cal N})\quad
\Rightarrow \quad
a(x,d_x\phi(x,y,\xi))=0 \quad\mbox{if}\;(x,y,\xi)\in {\cal N},
\end{eqnarray*}
and hence $c\sim 0$. This proves that $AB\in I^{-\infty}(X\times Y,C')$.
\end{beweis}

\begin{lemma}\label{Lemma5.2b}
Let $A\in \ci((-T,T),L^k(\Sigma))$ be properly supported and $B\in
I^m(\M\times\M,(C^\pm)')$. Then
\[ AB\in I^{m+k}(\M_T\times\M,(C^\pm)').\]
\end{lemma}
\begin{beweis}
Choosing local coordinates and a partition of unity we may assume that 
$\M=\R^4,\Sigma=\R^3\cong \R^3\times\{0\}\subset \R^4$ and that $A$ is
supported in a compact set. We let
$X=\mbox{op}\,\chi$ where $\chi=\chi(\tau,\xi)\in \ci(\R^4)$ vanishes
near $(\tau,\xi)=0$ and is
homogeneous of degree $0$ for $|(\tau,\xi)|\geq 1$ with $\chi(\tau,\xi)=1$
for $(\tau,\xi)$ in a conic neighborhood of $\{\xi=0\}$, and
$\chi(\tau,\xi)=0$ for $(\tau,\xi)$ outside a larger conic
neighborhood of $\{\xi=0\}$, such that, in particular,
$\chi(\tau,\xi)=0$ in a neighborhood of $\pi_1(C^\pm)$ (by $\pi_1$ we
denote the projection onto the first component in $T^*\M\times T^*\M$, 
i.e.~$\pi_1(x,\xi;y,\eta):=(x,\xi)$). We have 
\[ AB=AXB+A(1-X)B .\]
Denoting by $a(t,\vx,\xi)$ the local symbol of $A$, the operator
$A(1-X)$ has the symbol 
\[ a(t,\vx,\xi) (1-\chi(\tau,\xi))\in S^k(\R^4\times \R^4). \]
(Here we have used the fact that $(1-\chi(\tau,\xi))$ is non-zero only
in the area where $\la\tau\ra$ can be estimated by $\la\xi\ra$.) Hence
$A(1-X)$ is a properly supported pseudodifferential operator of order $k$
on $\M_T$. We may apply \cite[Thm.~25.2.3]{HormIV} with excess equal
to zero and obtain that $A(1-X)B\in I^{m+k}(\M_T \times \M,(C^\pm)')$.\\
On the other hand, $X$ is a pseudodifferential operator with symbol
vanishing in a neighborhood of $\pi_1(C^\pm)$. According to Lemma
\ref{Lemma5.2a}, $XB \in I^{-\infty}(\M\times\M, (C^\pm)')$. Hence
$XB$ is an integral operator with a smooth kernel on $\M\times\M$, and 
so is $AXB$, since $A$ is continuous on $\ci(\M)$.
\end{beweis}

\begin{lemma}\label{Lemma5.3}
(i) $(R_l-iJ_l-\partial_t)E^+ \in I^{-1/2}(\M_T\times\M, (C^+)'),\;l=1,2$;\\
(ii) $H(R_2-iJ_2-\partial_t)E^+\in I^{-3/2}(\M_T\times\M,(C^+)');$\\
(iii) $(R_l-iJ_l-\partial_t)E^- \in
I^{-N-5/2}(\M_T\times\M,(C^-)'),\;l=1,2; $\\
(iv) $H(R_2-iJ_2-\partial_t)E^-\in I^{-N-7/2}(\M_T\times\M,(C^-)').$
\end{lemma}
\begin{beweis}
(i) It follows from \rf{5.2} and Lemma \ref{Lemma5.2b} that
$(R_l-iJ_l)E^+\in I^{-1/2}(\M_T\times \M,(C^+)')$. Since
$\partial_t$ is a differential operator, the assumptions of the
composition theorem for Fourier integral operators
\cite[Thm.~25.2.3]{HormIV} are met with excess equal to zero, and we
conclude from \rf{5.2} that also $\partial_t E^+\in
I^{-1/2}(\M_T\times \M, (C^+)')$.\\
Since, by assumption, $H\in
\ci((-T,T),L^{-1}(\Sigma))$ is properly supported we also obtain (ii).\\
(iii) We know from \rf{5.0} that
\beq
 Q_l(R_l-iJ_l-\partial_t)E^-=S_l^{(N)}E^-. \label{5.3}
\eeq
Applying Lemma \ref{Lemma5.2b} and \rf{5.1}, 
the right hand side is an element of $I^{-3/2-N}(\M_T\times\M,(C^-)')$
(note that $S_l^{(N)}$ is properly supported).
We next observe that the question is local, so that we can
focus on a small neighborhood $U$ of a point $x_0\in \M$. Here, we
write $Q_l=Q_l^{(1)}+Q_l^{(2)}$ as a sum of two pseudodifferential
operators, where $Q_l^{(1)}$ is elliptic, and the essential support of 
$Q_l^{(2)}$ is
contained in the complement of $N_-$. To this end choose a real-valued
function $\chi\in
\ci(T^*U)$ with the following properties:\\
($\alpha$) $\chi(x,\xi)=0$ for small $|\xi|$,\\
($\beta$) $\chi$ is homogeneous of degree $1$ for $|\xi|\geq 1$,\\
($\gamma$) $\chi(x,\xi)\equiv 0$ on a conic neighborhood of $N_-$,\\
($\delta$) $\chi(x,\xi)\equiv |\xi|$ on a neighborhood of
 char$\,Q_l\cap \{|\xi|\geq 1\}$.\\
We denote the local symbol of $Q_l$ by $q_l$ and let
\[ Q_l^{(1)}:=op\,(q_l(x,\xi)+i\chi(x,\xi)), \quad Q_l^{(2)} :=
op\,(-i\chi(x,\xi)).\]
By the Lemmata \ref{Lemma5.2a} and \ref{Lemma5.2b} we have
\[ Q_l^{(2)}(R_l-iJ_l-\partial_t)E^-\in
I^{-\infty}(\M_T\times\M,(C^-)').\]
Moreover, $Q_l^{(1)}$ is elliptic of order $1$, since $q_l$ is real-valued
and $\chi(x,\xi)=|\xi|$ on char$\,Q_l$. 
We conclude from \rf{5.3} that 
\[ Q_l^{(1)}(R_l-iJ_l-\partial_t)E^- \in
I^{-3/2-N}(\M_T\times\M,(C^-)').\]
Multiplication by a parametrix to $Q_l^{(1)}$ shows that 
\[ (R_l-iJ_l-\partial_t)E^-\in I^{-5/2-N}(\M_T\times\M,(C^-)').\]
(iv) follows from (iii) and Lemma \ref{Lemma5.2b}.
\end{beweis}

We next analyze the effect of the restriction operator $\rho_0$. We
recall from \cite[p.~113]{Duistermaat96} that 
\beq
\rho_0\in I^{1/4}(\Sigma\times \M,R_0') \label{5.4}
\eeq
where \[ R_0:= \{(x_o,\xi_o;x,\xi)\in (T^*\Sigma\times
T^*\M)\setminus 0;\;x_o=x,\xi_o=\xi|_{T_{x_o}\Sigma}\}. \]
\begin{lemma}\label{Lemma5.4}
\begin{eqnarray*}
\rho_0 H(R_2-iJ_2-\partial_t)E^- &\in& I^{-N-13/4}(\Sigma\times \M,
(R_0\circ C^-)')\\
\rho_0(R_1-iJ_1-\partial_t)E^- &\in& I^{-N-9/4}(\Sigma\times
\M,(R_0\circ C^-)')\\
\rho_0 (R_1-iJ_1-\partial_t)E^+ &\in& I^{-1/4}(\Sigma\times
\M,(R_0\circ C^+)')\\
\rho_0 H(R_2-iJ_2-\partial_t)E^+ &\in& I^{-5/4}(\Sigma\times
\M,(R_0\circ C^+)').
\end{eqnarray*}
\end{lemma}
\begin{beweis}
All these statements follow from \rf{5.4}, Lemma \ref{Lemma5.3} and
the composition formula for Fourier integral operators
\cite[Thm.~25.2.3]{HormIV}, provided that the compositions $R_0\circ
C^-$ and $R_0\circ C^+$ of the canonical relations are clean, proper
and connected with excess zero (cf.~\cite[C.3]{HormIII} and
\cite[p.~18]{HormIV} for notation). We note that
\beqa
& &(R_0\times C^+)\cap (T^*\Sigma\times \mbox{diag}\,
(T^*\M)\times T^*\M)\label{5.5}\\
& &=\{(x_o,\xi_o; x,\xi;x,\xi;y,\eta);\;x=x_o,\xi_o=\xi|_{T_{x_o}\Sigma},
(x,\xi;y,\eta)\in C^+\}.\nonumber
\eeqa
Given $(x_o,\xi_o)\in T^*\Sigma\setminus 0$ there is precisely one
$(x,\xi)\in N_+$ such that $x=x_o$ and $\xi|_{T_{x_o}\Sigma}=\xi_o$;
given $(x,\xi)\in N_+$ there is a $1$-parameter family of $(y,\eta)$ such
that $(x,\xi;y,\eta)\in C^+$. We deduce that 
\begin{eqnarray*}
& &\mbox{codim}\,(R_0\times C^+) +\mbox{codim}\,(T^*\Sigma\times 
\mbox{diag}\,(T^*\M)\times T^*\M) \\
& &= 6\, \mbox{dim}\,\M-1 = \mbox{codim}\,(R_0\times C^+)\cap (T^*\Sigma\times
\mbox{diag}\,(T^*\M)\times T^*\M);
\end{eqnarray*}
here the codimension is taken in $T^*\Sigma\times (T^*\M)^3$. Hence
the excess of the intersection, i.e.~the difference of the left and
the right hand side, is zero. In particular, the intersection is
transversal, hence clean. Moreover, the fact that in \rf{5.5} the
$(x,\xi)$ is uniquely determined as soon as $(x_o,\xi_o)$ and
$(y,\eta)$ are given shows that the associated map 
\[ (x_o,\xi_o;x,\xi;x,\xi;y,\eta)\mapsto (x_o,\xi_o;y,\eta)\]
is proper (i.e.~the pre-image of a compact set is compact). Indeed,
the pre-image of a closed and bounded set is trivially closed; it is
bounded, because $|\xi|\leq C|\xi_o|$ for some constant $C$. Finally,
the pre-image of a single point $(x_o,\xi_o;y,\eta)$ is again a single 
point, in particular a connected set.\\
An analogous argument applies to $R_0\circ C^-$.
\end{beweis}

\begin{lemma}\label{Lemma5.5}
(i) $(\rho_0(R_1-iJ_1-\partial_t)E^-)^*(\rho_0
H(R_2-iJ_2-\partial_t)E^-) \in I^{-2N-11/2}(\M\times\M,(C^-)')$,\\
(ii) $(\rho_0(R_1-iJ_1-\partial_t)E^+)^*(\rho_0 H
(R_2-iJ_2-\partial_t)E^+\in I^{-3/2}(\M\times\M,(C^+)').$\\
Denoting by $D^{\pm}$ the relation $(R_0\circ
C^\mp)^{-1}\circ(R_0\circ C^\pm)$ we have\\
(iii) $(\rho_0(R_1-iJ_1-\partial_t)E^+)^*(\rho_0
H(R_2-iJ_2-\partial_t)E^-) \in I^{-N-7/2}(\M\times\M,(D^-)'),$\\
(iv) $(\rho_0(R_1-iJ_1-\partial_t)E^-)^*(\rho_0 H
(R_2-iJ_2-\partial_t)E^+) \in I^{-N-7/2}(\M\times\M, (D^+)').$
\end{lemma}
\begin{beweis}
(i) According to \cite[Thm.~25.2.2]{HormIV} and Lemma \ref{Lemma5.4}
\[ (\rho_0(R_1-iJ_1-\partial_t)E^-)^*\in I^{-N-9/4} (\M\times\Sigma,
((R_0\circ C^-)^{-1})').\]
We first note that the composition $(R_0\circ C^-)^{-1}\circ (R_0\circ 
C^-)$ equals $C^-$: In fact, $(R_0\circ C^-)^{-1}$ is the set of all
$(y,\eta;x_o,\xi_o)$, where $(x_o,\xi_o)\in T^*\Sigma$, $y$ is joined
to $x_o$ by a null geodesic $\gamma$, $\eta\in N_-$ is cotangent to
$\gamma$ at $y$ and the projection 
$P_\gamma(\eta)|_{T_{x_o}\Sigma}$ of the parallel transport of $\eta$
along $\gamma$ coincides with $\xi_o$.
The codimension of $(R_0\circ C^-)^{-1}\times (R_0\circ C^-)$ in
$T^*\M\times T^*\Sigma\times T^*\Sigma\times T^*\M$ therefore equals
$4\,$dim$\,\Sigma +2$, and we have
\begin{eqnarray*}
& &\mbox{codim}\,((R_0\circ C^-)^{-1}\times (R_0 \circ C^-)) + 
\mbox{codim} \,(T^*\M
\times \mbox{diag}\,(T^*\Sigma)\times T^*\M) \\ 
& &=6\,\mbox{dim}\,\Sigma +2 \\
& &=\mbox{codim}\,((R_0\circ 
C^-)^{-1} \times (R_0\circ C^-)) \cap (T^*\M \times
\mbox{diag}\,(T^*\Sigma)\times T^*\M).
\end{eqnarray*}
In particular, the intersection of $(R_0\circ C^-)^{-1}\times
(R_0\circ C^-)$ and $T^*\M\times \mbox{diag}\,(T^*\Sigma) \times T^*\M$ is
transversal in $T^*\M\times T^*\Sigma \times T^*\Sigma \times T^*\M$,
hence clean with excess $0$.\\
Given $(y,\eta;x_o,\xi_o;x_o,\xi_o;\tilde{y},\tilde{\eta})$ in the
intersection, the element $(x_o,\xi_o)$ is uniquely determined by
$(y,\eta)$ and $(\tilde{y},\tilde{\eta})$. The mapping
\[ (y,\eta;x_o,\xi_o;x_o,\xi_o;\tilde{y},\tilde{\eta})\mapsto
(y,\eta;\tilde{y},\tilde{\eta}) \]
therefore is proper. The pre-image of each element
is a single point, hence a connected set. We can apply the
composition theorem \cite[Thm.~25.2.3]{HormIV} and obtain the
assertion.\\
The proof of (ii), (iii) and (iv) is analogous.
\end{beweis}

We can now finish the proof of Theorem \ref{Theorem5.2}. According to
\rf{5.0b} and the following remarks we have to find the wavefront set
of
\[(\rho_0(R_1-iJ_1-\partial_t)(E^++E^-))^*(\rho_0
H(R_2-iJ_2-\partial_t)(E^+ +E^-)). \]
By Proposition \ref{PropositionA10} we have, for an arbitrary
canonical relation $\Lambda$,
\[ I^\mu (\M\times\M,\Lambda) \subset H^s(\M\times\M)\]
if $\mu+\frac{1}{2}\,\mbox{dim}\,\M +s<0$; moreover, the
wavefront set of elements of $I^\mu(\M\times\M,\Lambda)$ 
is a subset of $\Lambda$. Lemma \ref{Lemma5.5} therefore
immediately implies \rf{5.0a}.
\end{beweis}

\subsection{Construction on a compact Cauchy surface}\label{Section5.2}
Following the idea in \cite{Junker96} we shall now show that one can
construct adiabatic vacuum states on any globally hyperbolic spacetime 
$\M$ with compact Cauchy surface $\Sigma$.\\
In Gau{\ss}ian normal coordinates w.r.t.~$\Sigma$ the metric reads
\[ g_{\mu\nu}=\l(\begin{array}{cc}1 & \\ &
-h_{ij}(t,\vx)\end{array}\r) \]
and the Klein-Gordon operator reduces to
\[ \Box_g+m^2 =\frac{1}{\sqrt{\mathfrak h}}\partial_t (\sqrt{\mathfrak h}
\partial_t\cdot)
-\Delta_\Sigma +m^2,\]
where $h_{ij}$ is the induced Riemannian metric on $\Sigma$, $\mathfrak h$
its determinant and $\Delta_\Sigma$ the Laplace-Beltrami operator
w.r.t.~$h_{ij}$ acting on $\Sigma$. Following 
\cite[Eq.~(130)ff.]{Junker96} there exist
operators $P_1^{(N)}, P_2^{(N)},\,N=0,1,2,\ldots,$ of the form
\begin{eqnarray*}
P_1^{(N)} &=&-a^{(N)}(t,\vx,D_\vx)-\frac{1}{\sqrt{\mathfrak h}}\partial_t
\sqrt{\mathfrak h}
\\
P_2^{(N)} &=&a^{(N)}(t,\vx,D_\vx)-\partial_t
\end{eqnarray*}
with $a^{(N)}=a^{(N)}(t,\vx,D_\vx)\in \ci([-T,T],L^1(\Sigma))$ such
that 
\beq
P_1^{(N)}\circ P_2^{(N)} -(\Box_g+m^2) = s_N(t,\vx,D_\vx)
\label{5.10}
\eeq
with $s_N\in \ci([-T,T],L^{-N}(\Sigma))$. In fact one gets
\[ a^{(N)}(t,\vx,D_\vx)=-iA^{1/2}+\sum_{\nu=1}^{N+1} b^{(\nu)}(t,\vx,
D_\vx); \]
here $A$ is the self-adjoint extension of $-\Delta_\Sigma+m^2$ on
$L^2(\Sigma)$, so that $A^{1/2}$ is an elliptic pseudodifferential
operator of order $1$. The $b^{(\nu)}$ are elements of
$\ci([-T,T],L^{1-\nu}(\Sigma))$ defined recursively so that \rf{5.10}
holds. One then sets similarly as in \cite[Eq.~(134)]{Junker96}
\beqa
j^{(N)}(t,\vx,\mathbf{\xi}) &:=&
-\frac{1}{2i}\sum_{\nu=1}^{N+1}\l[b^{(\nu)}(t,\vx,\xi) -
\ol{b^{(\nu)}(t,\vx,-\xi)}\r]\in S^0 \label{5.11}\\
r^{(N)}(t,\vx,\xi) &:=& \frac{1}{2} \sum_{\nu=1}^{N+1}
\l[b^{(\nu)}(t,\vx,\xi) +\ol{b^{(\nu)}(t,\vx,-\xi)}\r] \in
S^0\label{5.12}\\
J(t) &:=&
A^{1/2} +\frac{1}{2}\l[j^{(N)}(t,\vx,D_\vx)+j^{(N)}(t,\vx,D_\vx)^*\r]\in 
L^1 \nonumber\\
R(t)&:=&
\frac{1}{2}\l[r^{(N)}(t,\vx,D_\vx)+r^{(N)}(t,\vx,D_\vx)^*\r]\in
L^0.\nonumber
\eeqa
\begin{lemma}\label{Lemma5.6}
We can change the operator $J$ defined above by a family of
regularizing operators  such that the assumptions of Theorem
\ref{Theorem5.1} are met.
\end{lemma}
\begin{beweis}
It is easily checked that a pseudodifferential operator on $\R^n$ with 
symbol $a(x,\xi)$ maps $\co{\R^n,\R}$ to $L^2_\R(\R^n)$ (i.e.~commutes
with complex conjugation) iff $a(x,\xi)=\ol{a(x,-\xi)}$. The symbols
$j^{(N)}$ and $r^{(N)}$ have this property by construction.\\ 
The operator family $R(t)
=\frac{1}{2}\l(r^{(N)}(t,\vx,D_\vx)+r^{(N)}(t,\vx,D_\vx)^*\r) \in L^0$ is
bounded and symmetric, hence
selfadjoint. Moreover, it commutes with complex conjugation: If $v\in
L^2_\R(\Sigma,d^3\sigma)$, then $R(t)v$ is real-valued, since for every
$u\in L^2_\R(\Sigma,d^3\sigma)$
\[2\la u, R(t) v\ra =
\la u,(r^{(N)}+r^{(N)*})v\ra = \la u, r^{(N)} v\ra+\la
r^{(N)}u,v\ra\in \R. \]
The operator $A^{1/2}$ maps $\co{\Sigma,\R}$ to
$L^2_\R(\Sigma,d^3\sigma)$ by \rf{5.00}; it is selfadjoint on ${\cal
D}(A^{1/2}) = H^1(\Sigma)$. Hence $J$ defines a selfadjoint family of
pseudodifferential operators of order $1$; it is invariant under
complex conjugation. Moreover, its principal symbol is
$\sqrt{h^{ij}\xi_i\xi_j}>0$. According to \cite[Ch.~II, Lemma 6.2]{Taylor81}
there exists a family of regularizing operators $J_\infty =
J_\infty(t)$ such that $J+J_\infty$ is strictly positive. Replacing
$J_\infty$ by $\frac{1}{2}(J_\infty + CJ_\infty C)$, where $C$ here
denotes the operator of complex conjugation, we obtain an operator
which is both strictly positive and invariant under complex
conjugation. It differs from $J$ by a regularizing family.

\end{beweis}

\begin{thm}\label{Theorem5.7}
For $N=0,1,2,\ldots$ we let
\[\Lambda_N(f_1,f_2)=\frac{1}{2}\l\la\l[(R-iJ)\rho_0-\rho_1\r]Ef_1,
J^{-1} \l[ (R-iJ)\rho_0-\rho_1\r]Ef_2\r\ra \]
with $J$ modified as in Lemma \ref{Lemma5.6}. Then $\Lambda_N$ 
is the two-point function of a (pure) adiabatic vacuum state of order $N$.
\end{thm}
\begin{beweis}
By Theorem \ref{Theorem5.1} and Lemma \ref{Lemma5.6}, $\Lambda_N$
defines the two-point
function of a (pure) quasifree state. We write
\begin{eqnarray*}
R(t) &=& \frac{1}{2}r^{(N)}(t,\vx,D_\vx)+\frac{1}{2}
r^{(N)}(t,\vx,D_\vx)^*\\
J(t) &=& \frac{1}{2}\l(A^{1/2} +j^{(N)}(t,\vx,D_\vx)\r)+\frac{1}{2}\l(A^{1/2}+
j^{(N)}(t,\vx,D_\vx)^*\r)+j_\infty (t,\vx,D_\vx)
\end{eqnarray*}
with $r^{(N)}, j^{(N)}$ as in \rf{5.11}, \rf{5.12} and
$j_\infty$ the regularizing modification of Lemma
\ref{Lemma5.6}. We shall use Theorem \ref{Theorem5.2} to analyze the
wavefront set of $\Lambda_N$. We decompose
\beqa
& &\Lambda_N(f_1,f_2) =\nonumber\\
& &\frac{1}{8}
\l\la\l[\l(r^{(N)}(t,\vx,D_\vx)-i(A^{1/2}
+j^{(N)}(t,\vx,D_\vx)+2j_\infty (t,\vx,D_\vx))\r)\rho_0-\rho_1\r]Ef_1\r.
\nonumber\\
& &+\l[\l(r^{(N)}(t,\vx,D_\vx)^*-i(A^{1/2}
+j^{(N)}(t,\vx,D_\vx)^*)\r)\rho_0-\rho_1\r]Ef_1,\nonumber\\
& &J(t)^{-1}\l[\l(r^{(N)}(t,\vx,D_\vx)-i(A^{1/2}
+j^{(N)}(t,\vx,D_\vx)+2j_\infty (t,\vx,D_\vx))\r)\rho_0-\rho_1\r]Ef_2
\nonumber\\
& &\l.+ J(t)^{-1}\l[\l(r^{(N)}(t,\vx,D_\vx)^*-i(A^{1/2}
+j^{(N)}(t,\vx,D_\vx)^*)\r)\rho_0-\rho_1\r]Ef_2\r\ra.\label{5.12a}
\eeqa
Now we let
\beqa
\tilde{Q}_1(t) &:=& A^{1/2} +i 
\l(r^{(N)}(t,\vx,D_\vx)-ij^{(N)}(t,\vx,D_\vx)\r)+\frac{i}{\sqrt{\mathfrak
h}}\partial_t 
\sqrt{\mathfrak h} \nonumber\\
&=&i\l(a^{(N)}(t,\vx,D_\vx)+\frac{1}{\sqrt{\mathfrak h}}\partial_t
\sqrt{\mathfrak h}\r) = -i P_1^{(N)} 
\label{5.12c}
\eeqa
and
\beq
\tilde{Q}_2(t) := A^{1/2} +i 
\l(r^{(N)}(t,\vx,D_\vx)^* -ij^{(N)}(t,\vx,D_\vx)^*\r)+
\frac{i}{\sqrt{\mathfrak h}}\partial_t \sqrt{\mathfrak h}. \label{5.12d}
\eeq
Equation \rf{5.10} implies that
\begin{eqnarray}
& &i\tilde{Q}_1(t)
\l(r^{(N)}(t,\vx,D_\vx)-i(A^{1/2}+j^{(N)}(t,\vx,D_\vx)) 
-2ij_\infty(t,\vx,D_\vx)-\partial_t\r)\nonumber\\
& &=P_1^{(N)}\l(P_2^{(N)}-2 ij_\infty(t,\vx,D_\vx)\r)\nonumber\\
& &=\Box_g+m^2+\tilde{s}_N(t,\vx,D_\vx) \label{5.12b}
\end{eqnarray}
where $\tilde{s}_N$ differs from $s_N$ in \rf{5.10} by an element in
$\ci([-T,T], L^{-\infty}(\Sigma))$. 
Next we note that \rf{5.10} is equivalent to the identity
\[ \l(-r^{(N)}+i(A^{1/2}+j^{(N)})-\frac{1}{\sqrt{\mathfrak
h}}\partial_t\sqrt{\mathfrak h}\r)
\l(r^{(N)}-i(A^{1/2}+j^{(N)})-\partial_t\r) =\Box_g+m^2 +s_N \] 
which in turn is equivalent to
\begin{eqnarray*}
& &\l(-r^{(N)}+i(A^{1/2}+j^{(N)})\r)\l(r^{(N)}-i(A^{1/2}+j^{(N)})\r)
-\frac{1}{\sqrt{\mathfrak h}} \partial_t\l(\sqrt{ \mathfrak h}
\l(r^{(N)}-i(A^{1/2}+j^{(N)})\r)\r) \\
& &= -\Delta_\Sigma+m^2+s_N 
\end{eqnarray*}
or - taking adjoints and conjugating with the operator $C$ of complex
conjugation - 
\begin{eqnarray}
& & -C\l(r^{(N)*}+i(A^{1/2}+j^{(N)*})\r) C
C\l(r^{(N)*}+i(A^{1/2}+j^{(N)*})\r) C\nonumber \\
& &\quad\quad - \frac{1}{\sqrt{\mathfrak h}}\partial_t\l(\sqrt{\mathfrak h}
C\l(r^{(N)*}+i(A^{1/2}+j^{(N)*})\r) C\r) \nonumber\\
& &=C(-\Delta_\Sigma +m^2 +s_N^*)C =-\Delta_\Sigma +m^2 +Cs_N^*C.\label{5.13}
\end{eqnarray}
Here we have used the fact that 
\[\l[\partial_t\l(\sqrt{\mathfrak h}\l(r^{(N)}-i(A^{1/2}+j^{(N)})\r)\r)\r]^* =
\partial_t\l[\sqrt{\mathfrak h} \l(r^{(N)}-i(A^{1/2}+j^{(N)})\r)\r]^*. \]
Using that $r^{(N)*},\,j^{(N)*}$ and $A^{1/2}$ commute with $C$,
\rf{5.13} reads
\begin{eqnarray*}
& &
-\l(r^{(N)*}-i(A^{1/2}+j^{(N)*})\r)\l(r^{(N)*}-i(A^{1/2}+j^{(N)*})\r)
-\frac{1}{\sqrt{\mathfrak h}}\partial_t
\l(\sqrt{\mathfrak h}\l(r^{(N)*}-i(A^{1/2}+j^{(N)*})\r)\r) \\
& &= -\Delta_\Sigma +m^2 +Cs_N^*C.
\end{eqnarray*}
Adding the time derivatives, it follows that
\beqa
& & i \tilde{Q}_2(t) \l(r^{(N)*}-i(A^{1/2}+j^{(N)*})-\partial_t\r)
\nonumber \\
& &=\l(-r^{(N)*}+i(A^{1/2}+j^{(N)*})-\frac{1}{\sqrt{\mathfrak
h}}\partial_t\sqrt{\mathfrak h}\r)
\l(r^{(N)*}-i(A^{1/2}+j^{(N)*})-\partial_t\r)\nonumber\\
& &= \Box_g+m^2 +Cs_N^*C.\label{5.14}
\eeqa 
Note that the operators $\tilde{Q}_1$ and $\tilde{Q}_2$, defined by
Eq.s \rf{5.12c} and \rf{5.12d}, are not yet pseudodifferential
operators since their symbols will not decay in the covariable of $t$, 
say $\tau$, if we take derivatives w.r.t.~the covariables of $\vx$,
say $\xi$. To make them pseudodifferential operators we choose a
finite number of coordinate neighborhoods $\{U_j\}$ for $\Sigma$,
which yields finitely many coordinate neighborhoods for $(-T,T)\times
\Sigma$. As $(t,\vx)$ varies over $(-T,T)\times U_j$, the negative
light cone will not intersect a fixed conic neighborhood $\cal N$ of
$\{\xi=0\}$ in $T^*((-T,T)\times U_j)$. We choose a real-valued 
function $\chi$
which is smooth on $T^*((-T,T)\times U_j)$, zero for $|(\tau,\xi)|\leq 
1/2$, homogeneous of degree zero for $|(\tau,\xi)|\geq 1$ such that
\beqa
\chi(t,\vx,\tau,\xi) &=& 0\quad \mbox{on a conic neighborhood
of}\;\{\xi=0\} \label{5.15}\\
\chi(t,\vx,\tau,\xi) &=& 1\quad\mbox{outside}\;{\cal N}.\nonumber
\eeqa
We now let $X:=$ op$\,\chi$. Then
\[ Q_1:=X\tilde{Q}_1\quad\mbox{and}\quad Q_2:=X\tilde{Q}_2\]
are pseudodifferential operators due to \rf{5.15}. Their principal
symbols are $((h^{ij}\xi_i \xi_j)^{1/2}-\tau)\chi(t,\vx,\tau,\xi)$, 
so that their characteristic set does not intersect 
$N_-$. Eq.s \rf{5.12b}, \rf{5.14} and the fact that
$(\Box_g+m^2)E^-=0$ show that the assumption of 
Theorem \ref{Theorem5.2} is satisfied for each of the four terms
arising from the decomposition of $\Lambda_N$ in \rf{5.12a}. This
yields the assertion.
\end{beweis}

Lemma \ref{Lemma5.5} explicitly shows that the non-Hadamard like
singularities of the two-point function $\Lambda_N$ in Theorem
\ref{Theorem5.7} (i.e.\ those not contained in the canonical relation
$C^+$) are either pairs of purely negative frequency singularities
lying on a common bicharacteristic ($C^-$) or pairs of mixed
positive/negative frequency singularities ($D^\pm$) which lie on
bicharacteristics that are ``reflected'' by the Cauchy surface. They
may have spacelike separation.\\
For the states constructed in Theorem \ref{Theorem5.7} one can
explicitly find the Bogoljubov $B$-operator, which was introduced in
the proof of Theorem \ref{Theorem29}(ii), in terms of the operators
$R$ and $J$.
Applying the criterion of Lemma \ref{Lemma31}(iii) one can check
the unitary equivalence of the GNS-representations generated by these states.
A straightforward (but tedious) calculation shows that
unitary equivalence already holds for $N\geq 0$, thus improving the
statement of Theorem \ref{Theorem29}(ii) for these particular examples.

\section{Adiabatic vacua on Robertson-Walker spaces}
\label{Section6}
By introducing adiabatic vacua on Robertson-Walker spaces Parker
\cite{Parker69} was among the first to construct a quantum field
theory in a non-trivial background spacetime. A mathematically precise
version of his construction and an analysis of the corresponding
Hilbert space representations along the same lines as in our Section
\ref{Section3} were given by L\"uders \& Roberts \cite{LR90}. Relying
on their work we want to show in this section that these adiabatic
vacua on Robertson-Walker spaces are indeed adiabatic vacua in the
sense of our Definition \ref{Definition1}. This justifies our naming
and gives a mathematically intrinsic meaning to the ``order'' of an
adiabatic vacuum. In \cite{Junker96} one of us had claimed to have
shown that all adiabatic vacua on Robertson-Walker spaces are Hadamard
states. This turned out to be wrong in general, when the same question
was investigated for Dirac fields \cite{Hollands01}\footnote{We want
to thank S.~Hollands for discussions about this point.}. So the present
section also serves to correct this mistake. Our presentation follows
\cite{Junker96}. \\
In order to be able to apply our Theorem \ref{Theorem5.2} without
technical complications we restrict
our attention to Robertson-Walker spaces with compact spatial
sections. These are the 4-dim.~Lorentz manifolds $\M= \R\times \Sigma$
where $\Sigma$ is regarded as being embedded in $\R^4$ as
\[ \Sigma =\{x\in \R^4;\;(x^0)^2 + \sum_{i=1}^3 (x^i)^2 =1\} \cong S^3,
\]
and $\M$ is endowed with the homogeneous and isotropic metric
\beq
ds^2=dt^2-a(t)^2 \l[\frac{dr^2}{1-r^2} +r^2 (d\theta^2
+\sin^2\theta\,d\varphi^2)\r]; \label{6.1}
\eeq
here $\varphi\in[0,2\pi],\theta \in [0,\pi],r\in[0,1)$
are polar coordinates for the unit ball in $\R^3$, and $a$ is a
strictly positive smooth function.
In \cite{Junker96} it was shown that an adiabatic vacuum state of
order $n$ (as defined in \cite{LR90}) is a pure quasifree state on the
Weyl algebra of the Klein-Gordon field on the spacetime \rf{6.1}
given by a one-particle Hilbert space structure w.r.t.~a fixed Cauchy
surface $\Sigma_t = \{t= \mbox{const.}\} = \Sigma \times \{t\}$ 
(equipped with the induced metric from \rf{6.1})
\begin{eqnarray*}
k_n: \Gamma &\to& \H_n:= \ol{k_n(\Gamma)} \subset L^2(\Sigma_t)\\
(q,p) &\mapsto& (2J_n)^{-1/2} \l[(R_n-iJ_n)q-p\r]
\end{eqnarray*}
of the form \rf{5.000} of Theorem \ref{Theorem5.1}, where the
operator families $R_n(t),J_n(t)$ 
acting on $L^2(\Sigma,d^3\sigma)$ are defined in
the following way:
\beqa
(R_nf)(t,\vx) &:=& -\frac{1}{2} \dm\l(3\frac{\dot{a}(t)}{a(t)}
+ \frac{\dot{\Omega}^{(n)}_k(t)}{\Omega_k^{(n)}(t)}\r) \tilde{f}(t,\vk)
\ph(\vx) \nonumber\\
(J_nf)(t,\vx) &:=& \dm \Omega_k^{(n)}(t) \tilde{f}(t,\vk) \ph
(\vx),\label{6.2} 
\eeqa
with $t$ in some fixed finite interval $I\subset \R$, say.
Here, $\{\ph,\; \vec{k}:=(k,l,m),\; k=0,1,2,\ldots;\; l=0,1,\ldots,k;\;
m=-l,\ldots,l\}$ 
are the $t$-independent eigenfunctions of the
Laplace-Beltrami operator $\Delta_\Sigma$ w.r.t.~the Riemannian metric
\beq
s_{ij}=\l(\begin{array}{ccc} \frac{1}{1-r^2} & & \\
        & r^2 & \\
        & & r^2\sin^2\theta \end{array}\r) \label{6.2a}
\eeq
on the hypersurface $\Sigma$: 
\[ \Delta_\Sigma \ph \equiv \l\{ (1-r^2)\ab{^2}{r^2} +\frac{2-3r^2}{r}
\ab{}{r} +\frac{1}{r^2} \Delta(\theta,\varphi)\r\} \ph = - k(k+2)
\ph, \]
where $\Delta(\theta,\varphi):= \ab{^2}{\theta^2}
+\cot\theta \ab{}{\theta} +\frac{1}{\sin^2\theta} \ab{^2}{\varphi^2}$
is the Laplace operator on $S^2$. They form an orthonormal basis of
$L^2(\Sigma,d^3\sigma)$ with $d^3\sigma := r^2(1-r^2)^{-1/2}
dr\,\sin\theta d\theta\,d\varphi$. $\tilde{}$ denotes the generalized Fourier
transform 
\beqa
\tilde{}: L^2(\Sigma,d^3\sigma) &\to& L^2(\tilde{\Sigma},d\mu(\vk)) 
\nonumber\\
f&\mapsto& \tilde{f}(\vk) := \int_\Sigma \!d^3\sigma\,\ol{\ph(\vx)}
f(\vx), \label{6.4}
\eeqa
which is a unitary map from $L^2(\Sigma,d^3\sigma)$ to
$L^2(\tilde{\Sigma},d\mu(\vk))$ where $\tilde{\Sigma}$ is the space of
values of $\vk=(k,l,m)$ equipped with the measure $\dm
:=\sum_{k=0}^\infty \sum_{l=0}^k \sum_{m=-l}^l$ \cite{LR90}. 
(Note that \rf{6.4} is defined w.r.t.~$\Sigma$ with the
metric $s_{ij}$, Eq.~\rf{6.2a}, and not w.r.t.~$\Sigma_t$ with the
metric $a^2(t) s_{ij}$.)
The inverse is given by
\[f(\vx)=\dm \tilde{f}(\vk) \ph(\vx) .\]
Using duality and interpolation of the Sobolev spaces one deduces
\begin{lemma}\label{Lemma6.4}
\[ H^s(\Sigma) =\{f=\dm \tilde{f}(\vk)\ph;\; \dm
(1+k^2)^s |\tilde{f}(\vk)|^2<\infty\}.\]
\end{lemma}
The Klein-Gordon operator associated to the metric \rf{6.1} is given
by 
\beq
\Box_g+m^2= \dd{t} +3\frac{\dot{a}}{a}\d{t}-\frac{1}{a^2}\Delta_\Sigma
+m^2. \label{6.4a}
\eeq
The functions $\Omega_k^{(n)}(t),\,n\in \N_0$, in \rf{6.2} were
introduced by L\"uders \& Roberts. 
$\Omega_k^{(n)}(t)$ is strictly positive and plays the role of a
generalized frequency which is determined by a WKB approximation to
Eq.~\rf{6.4a}. It is iteratively defined by the following
recursion relations 
\beqa
(\Omega_k^{(0)})^2 &:=& \omega_k^2 \equiv \frac{k(k+2)}{a^2} +m^2
\nonumber\\
(\Omega_k^{(n+1)})^2 &=& \omega_k^2 -\frac{3}{4}
\l(\frac{\dot{a}}{a}\r)^2 -\frac{3}{2}\frac{\ddot{a}}{a}
+\frac{3}{4}\l(\frac{\dot{\Omega}_k^{(n)}}{\Omega_k^{(n)}}\r)^2
-\frac{1}{2} \frac{\ddot{\Omega}_k^{(n)}}{\Omega_k^{(n)}}. \label{6.5}
\eeqa
In the following we shall study the analytic properties of these
functions. We shall see that, using \rf{6.5}, we may express $\on$ as
a function of $t$ and $\omega_k$. As a function of these two variables,
it turns out to behave like a classical pseudodifferential symbol. In
Lemma \ref{Lemma6.1} below we shall derive the corresponding estimates
and expansions for $(t,\omega_k)\in I\times \R_+$. It is a little
unusual to consider a `covariable' in $\R_+$; in later applications,
however, we will have $\omega_k=\sqrt{k(k+2)/a^2 +m^2}$ bounded away
from zero, so that the behaviour of $\omega_k$ near zero is irrelevant.\\
We first observe that $(\Omega_k^{(n+1)})^2$ can be determined by
an iteration involving only $(\Omega_k^{(n)})^2$
and its time derivatives: Since, for an arbitrary $F$, we have
$\partial_t (F^2)/ F^2 = 2 \dot{F}/F$, we obtain
\beqa
(\Omega_k^{(n+1)})^2 &=& 
\omega_k^2 -\frac{3}{4} \l(\frac{\dot{a}}{a}\r)^2 -\frac{3}{2}
\frac{\ddot{a}}{a} +\frac{1}{16} \l(\frac{\frac{d}{dt}
\l((\on)^2\r)}{\on^2} \r)^2 -\frac{1}{4} \frac{d}{dt}
\l(\frac{\frac{d}{dt} \l((\on)^2\r)}{(\on)^2}\r). \label{6.6}
\eeqa
An induction argument shows that $(\on)^2-\omega_k^2$ is a
rational function in $\omega_k$ of degree $\leq 0$ with coefficients
in $\ci(I)$. Indeed this is trivially true for $n=0$. Suppose it is
proven for some fixed $n$. We write 
\beq
(\on)^2-\omega_k^2 =
r(t,\omega_k) = \frac{p(t,\omega_k)}{q(t,\omega_k)} \label{6.7}
\eeq
with polynomials $p$ and $q$ in $\omega_k$ 
such that deg\,$(p)\leq \mbox{deg}\,(q)$ and the leading coefficient
of $q$ is $1$. Then
\beq
\frac{\frac{d}{dt}\l((\on)^2\r)}{(\on)^2} = \frac{2\omega_k\dot{\omega}_k
+\dot{r}} {\omega_k^2+r}. \label{6.8}
\eeq
In view of the fact that 
\[\dot{\omega}_k =
-\frac{\dot{a}}{a}\l(\omega_k- \frac{m^2}{\omega_k}\r) 
\quad\mbox{and}\quad\dot{r}
= \frac{q \partial_{\omega_k}p -p\partial_{\omega_k}q}{q^2} \dot{\omega}_k+
\frac{q\partial_t p -p\partial_t q}{q^2},\] 
\rf{6.8} is again rational
of degree $0$ and the leading coefficient of the polynomial in the
denominator again equals $1$. The same is true for the time derivatives of
\rf{6.8}. The recursion formula \rf{6.6} then shows the assertion for
$n+1$. \\
Next we observe that also $\frac{d^l}{dt^l}\l((\on)^2 -\omega_k^2\r)$ is a
rational function of $\omega_k$ with coefficients in $\ci(I)$ of
degree $\leq 0$. Moreover, this shows that, for sufficiently large
$\omega_k$ (equivalently for sufficiently large $k$), $(\on)^2$ is
a strictly positive function (uniformly in $t\in I$). We may redefine
$(\on)^2$ for small values of $\omega_k$ so that it is strictly
positive and bounded away from zero on $I\times\R_+$. It makes sense 
to take its square root, and in the
following we shall denote this function by $\on$. We note:
\begin{lemma}\label{Lemma6.1}
$\on$, considered as a function of $(t,\omega)\in I\times \R_+$, is an
element of $S_{{cl}}^1(I\times\R_+)$, i.e.
\beq
\partial_t^l\partial_\omega^m \on(t,\omega) = \O(\la\omega\ra^{1-m})
\label{6.9} 
\eeq
and, in addition, $\on$ has an asymptotic expansion $\on \sim
\sum_{j=0}^\infty (\on)_j$ into symbols $(\on)_j \in S^{1-j}$
which are positively homogeneous for large $\omega$. Its principal
symbol is $\omega$. With the same
understanding 
\beqa
\frac{\dot{\Omega}_k^{(n)}}{\on} &\in& S^0_{{cl}} (I\times \R_+),
\label{6.10}\\
(\Omega_k^{(n+1)})^2-(\on)^2 &\in& S_{{cl}}^{-2n} (I\times
\R_+). \label{6.11}
\eeqa
\end{lemma}

\begin{beweis}
By induction, \rf{6.9} is immediate from \rf{6.7} together with the
formulae 
\[ \partial_t \sqrt{\omega^2+r(t,\omega)} = \frac{1}{2}
\frac{\partial_t r(t,\omega)}{\sqrt{\omega^2+r(t,\omega)}}
\quad\mbox{and}\quad \partial_\omega\sqrt{\omega^2 +r(t,\omega)}
=\frac{1}{2} \frac{2\omega +\partial_\omega
r(t,\omega)}{\sqrt{\omega^2 +r(t,\omega)}}. \]
Relation \rf{6.10} is immediate from \rf{6.8}, noting that
$2 \dot{\Omega}_k^{(n)}/ \on = \frac{d}{dt}\l((\on)^2\r)/(\on)^2$. In both
cases the existence of the asymptotic expansion follows from 
\cite[Ch.~II, Thm.~3.2]{Taylor81} and the expansion
\begin{eqnarray*}
\sqrt{\omega^2+r(t,\omega)} = \omega\sqrt{1+\frac{r(t,\omega)}{\omega^2}}
= \omega \sum_{j=0}^\infty {1/2 \choose j}
\l(\frac{r(t,\omega)}{\omega^2}\r)^j 
\end{eqnarray*}
valid for large $\omega$. The principal symbol of $\on$
is $\omega$ since $(\on)^2=\omega^2$ modulo rational functions of
degree $\leq 0$, as shown above (cf.\ Eq.~\rf{6.7}).
In order to show \rf{6.11} we write,
following L\"uders \& Roberts,
\[ (\Omega_k^{(n+1)})^2 =(\on)^2 (1+\epsilon_{n+1});\]
this yields \cite[Eq.~(3.9)]{LR90}
\begin{eqnarray*}
\epsilon_{n+1} &=&
\frac{1}{\omega^2}\frac{1}{(1+\epsilon_1)\cdots(1+\epsilon_n)}
\l(\frac{1}{4}\frac{\dot{\omega}}{\omega}\frac{\dot{\epsilon}_n}
{1+\epsilon_n}+\frac{1}{8}\frac{\dot{\epsilon}_1}{1+\epsilon_1}
\frac{\dot{\epsilon}_n}{1+\epsilon_n} +\ldots \r. \\
& & \l. +\frac{1}{8}\frac{\dot{\epsilon}_{n-1}}{1+\epsilon_{n-1}}
\frac{\dot{\epsilon}_n}{1+\epsilon_n}
+\frac{5}{16}\frac{\dot{\epsilon}_n^2} {1+\epsilon_n}-\frac{1}{4}
\frac{\ddot{\epsilon}_n}{1+\epsilon_n} \r). 
\end{eqnarray*}
We know already that $(\Omega_k^{(1)})^2-(\Omega_k^{(0)})^2
=(\Omega_k^{(1)})^2- \omega^2$ is rational in $\omega$ of degree $\leq
0$, hence $\epsilon_1$ is rational of degree $-2$.
Noting that $\dot{\epsilon}_n
=(\partial_{\omega}\epsilon_n)\dot{\omega} + \partial_t \epsilon_n$,
we deduce from the above recursion that $\epsilon_n$ is rational of
degree $-2n$. This completes the argument.
\end{beweis}

The operators $R_n$ and $J_n$, Eq.~\rf{6.2}, are unitarily equivalent
to multiplication operators on $L^2(\tilde{\Sigma},d\mu(\vk))$. 
From the fact that $\dot{\Omega}_k^{(n)}/\on$ is bounded and 
$\on$ is strictly positive with principal symbol $\omega_k$
(Lemma \ref{Lemma6.1}) we can immediately deduce that the
assumptions of Theorem \ref{Theorem5.1} are satisfied if we let
dom\,$J(t) =H^1(\Sigma)$ for $t\in I$.\\
We are now ready to state the theorem that connects the adiabatic
vacua of L\"uders \& Roberts \cite{LR90} to our more general Definition
\ref{Definition1}:
\begin{thm}\label{Theorem6.2}
For fixed $t$ let
\[ \Lambda_n(f,g) := \la (R_n-iJ_n-\partial_t) Ef, J_n^{-1}
(R_n-iJ_n-\partial_t)Eg\ra_{L^2(\Sigma_t)} \]
be the two-point function of a pure quasifree state of the
Klein-Gordon field on the Robertson-Walker spacetime \rf{6.1} with
$R_n,J_n$ given by Eq.s \rf{6.2} and \rf{6.5}. Then
\[ WF'^s(\Lambda_n)\subset \l\{\begin{array}{ll}\emptyset,& s<-\frac{1}{2} \\
                            C^+,& -\frac{1}{2}\leq s<2n+\frac{3}{2}
\end{array},\r.   \]
i.e.~$\Lambda_n$ describes an adiabatic vacuum state of order
$2n$ in the sense of our Definition \ref{Definition1}.
\end{thm}
To prove the theorem we shall need the following observations:
\begin{lemma}\label{Lemma6.3}
Let $m\in\R$. Let $M$ be a compact manifold and $A: \D{M}\to
\Dp{M}$ a linear operator. Suppose that, for each $k\in \N$, we can
write
\beq
A=P_k+R_k \label{6.12}
\eeq
where $P_k$ is a pseudodifferential operator of order $m$ and $R_k$ is
an integral operator with a kernel function in ${\cal
C}^k(M\times M)$. 
Then $A$ is a pseudodifferential operator of order $m$.
\end{lemma}
\begin{beweis}
Generalizing a result by R.~Beals \cite{Beals77}, Coifman \& Meyer showed the
following: A linear operator $T:\D{M}\to \Dp{M}$ is a
pseudodifferential operator of order $0$ if and only if $T$ as well as
its iterated commutators with smooth vector fields are bounded on
$L^2(M)$ \cite[Thm.~III.15]{CM78}. As a corollary, $T$ is a
pseudodifferential operator of order $m$ if and only if $T$ and its
iterated commutators induce bounded maps $L^2(M)\to
H^{-m}(M)$. Given the iterated commutator of $A$ with, say, $l$
vector fields $V_1,\ldots,V_l$, we write $A=P_k+R_k$ with $k\geq
l+|m|$. The iterated commutator $[V_1,[\ldots[V_l,P_k]\ldots]]$ is a
pseudodifferential operator of order $m$ and hence induces a bounded
map $L^2(M)\to H^{-m}(M)$. The analogous commutator with $R_k$ has
an integral kernel in ${\cal C}^{k-l}(M\times M)$. As $k-l\geq|m|$,
it furnishes even a bounded operator $L^2(M)\to H^{|m|}(M)$.
\end{beweis}

\begin{lemma}\label{Lemma6.5}
Let $\mu\in\Z$ and $b=b(t,\tau)\in S^\mu_{cl}(I\times \R)$ with
principal symbol $b_{-\mu}(t) \tau^\mu$. Replacing $\tau$ by $\omega_k(t)
= \l(\frac{k(k+2)}{a^2(t)}+m^2\r)^{1/2}$, $b$ defines a family
$\{B(t);\;t\in I\}$ of operators $B(t):\D{\Sigma}\to \Dp{\Sigma}$ by
\[ (\widetilde{B(t)f})(t,\vk) := b(t,\omega_k(t)) \tilde{f}(\vk).\]
We claim that this is a smooth family of pseudodifferential operators
of order $\mu$ with principal symbol
\beq
\sigma^{(\mu)}(B(t))=b_{-\mu}(t)
(|\xi|_\Sigma/a(t))^{\mu},\label{6.13}
\eeq
where the length $|\xi|_\Sigma$ of a covector $\xi$ is taken w.r.t.\ the
(inverse of the) metric \rf{6.2a}.
\end{lemma}
Note that for the definition of $B(t)$ we only need to know
$b(t,\tau)$ for $\tau\geq m$. We may therefore also apply this result
to the symbols that appear in Lemma \ref{Lemma6.1}.

\begin{beweis}
The fact that $b$ is a classical symbol allows us to write, for each
$N$,
\beq
b(t,\tau)= \sum_{j=-\mu}^N b_j(t)\tau^{-j}
+b^{(N)}(t,\tau),\label{6.14}
\eeq
where $b_j\in \ci(I)$ and
$|\partial_t^j\partial_\tau^lb^{(N)}(t,\tau)|\leq
C_{jl}(1+|\tau|)^{-N}$ for all $t\in I$ and $\tau\geq\epsilon$,
$\epsilon>0$ fixed. (Note that we will not obtain the estimates for
all $\tau$, since we have a fully homogeneous expansion in \rf{6.14},
but as we shall substitute $\tau$ by $\omega_k$ and $\omega_k$ is
bounded away from $0$, this will not be important.) Equation \rf{6.14}
induces an analogous decomposition of $B$:
\[ B(t) =\sum_{j=-\mu}^N B_j(t) +B^{(N)}(t),\]
where $B_j(t)$ is given by 
\[ (\widetilde{B_j(t)f})(t,\vk) = b_j(t) \omega_k(t)^{-j} \tilde{f}(\vk) \]
and $B^{(N)}(t)$ by 
\[ (\widetilde{B^{(N)}(t)f})(t,\vk) = b^{(N)}(t,\omega_k(t)) \tilde{f}(\vk).\]
In view of the fact that $\Delta_\Sigma \ph =-k(k+2) \ph$, we have 
\[ B_j(t) = b_j(t) \l(m^2-\Delta_\Sigma/a^2(t)\r)^{-j/2}. \]
According to Seeley \cite{Seeley68}, $B_j$ is a smooth family of
pseudodifferential operators of order $-j$. Next, we observe that, by
\rf{6.9}, $\partial_t^l\omega_k =\O(\omega_k)$ and hence, for each
$l\in \N$,
\[|\partial_t^l\l( b^{(N)}(t,\omega_k(t))\r)|\leq C(1+\omega_k(t))^{-N}\leq
C'(1+k)^{-N} \]
for all $t\in I$. Lemma \ref{Lemma6.4} therefore shows that, for each
$s\in \R$
\beq
\partial_t^l B^{(N)}(t): H^s(\Sigma)\to H^{s+N}(\Sigma) \label{6.15}
\eeq
is bounded, uniformly in $t\in I$. On the other hand, it is well known
that a linear operator $T$ which maps $H^{-s-k}(\Sigma)$ to
$H^{s+k}(\Sigma)$ for some $s>3/2$ (dim $\Sigma =3$) has an integral
kernel of class ${\cal C}^k$ on $\Sigma\times\Sigma$. It is given by
$K(x,y) =\la T\delta_y,\delta_x\ra$. Choosing $N>3+2k$, the family
$B^{(N)}$ will therefore have integral kernels of class ${\cal
C}^k$. Now we apply Lemma \ref{Lemma6.3} to conclude that $B$ is a
smooth family of pseudodifferential operators of order $\mu$. Since
$B_j$ is of order $-j$ the principal symbol is that of
$B_{-\mu}=b_{-\mu}(t) \l(m^2-\Delta_\Sigma/a^2(t)\r)^{-\mu/2}$. This yields
\rf{6.13}.
\end{beweis}

Now let us define the family of operators $A_n(t)$ acting on
$L^2(\Sigma,d^3\sigma)$ by
\[ (A_nf)(t,\vx) := \dm a^{(n)}(t,k)\tilde{f}(t,\vk)\ph(\vx), \]
with $a^{(n)}$ given by the function
\[ a^{(n)}(t,k):=\frac{3}{2}\frac{\dot{a}}{a} -\frac{1}{2}
\frac{\dot{\Omega}_k^{(n)}} {\on}-i\on .\]
Moreover let 
\[Q_n := iX (\partial_t +A_n(t)),\]
where $X:=\mbox{op}\,\chi$ and $\chi=\chi(t,\vx,\tau,\xi)$ is as in
\rf{5.15}. 
\begin{lemma}\label{Lemma6.6}
$A_n\in\ci(I,L^1_{cl}(\Sigma))$ with principal symbol
$\sigma^{(1)}(A_n(t))=i|\xi|_\Sigma/a(t)$. $Q_n$ is a
pseudodifferential operator of order $1$ on $I\times\Sigma$ with
real-valued principal symbol whose characteristic does not intersect
$N_-$. 
\end{lemma}
\begin{beweis}
We apply Lemma \ref{Lemma6.5} in connection with Lemma \ref{Lemma6.1}
to see that $A_n\in \ci(I,L^1_{cl}(\Sigma))$. The operator $Q_n$
clearly is an element of $L^1_{cl}(I\times\Sigma)$. 
Outside a small neighborhood of $\{\xi=0\}$ its characteristic set is
\[\{(t,\vx,\tau,\xi)\in
T^*(I\times\Sigma);\;-\tau+|\xi|_\Sigma/a(t) =0\} . \]
Since $N_-=\{\tau=-|\xi|_\Sigma/a(t)\}$, the intersection is empty.
\end{beweis}

{\it Proof of Theorem \ref{Theorem6.2}:} In view of Theorem
\ref{Theorem5.2} we only have to check that 
\beq
Q_n(R_n-iJ_n-\partial_t)E^-=S^{(2n)} E^- \label{6.16}
\eeq
for a pseudodifferential operator $S^{(2n)}$ of order $-2n$. A
straightforward computation shows that
$(\partial_t+A_n)(R_n-iJ_n-\partial_t)$ is the operator defined by
\[ -\l(\partial_t^2+3\frac{\dot{a}}{a}\partial_t +\omega_k^2 +(\on)^2
-(\Omega_k^{(n+1)})^2\r). \]
Now $\partial_t^2+3\frac{\dot{a}}{a}\partial_t +\omega_k^2$ induces
$\Box_g+m^2$, Eq.\ \rf{6.4a}, 
while, by Lemma \ref{Lemma6.1} combined with Lemma
\ref{Lemma6.5}, $(\on)^2-(\Omega_k^{(n+1)})^2$ induces an element of
$\ci(I,L^{-2n}(\Sigma))$. Composing with the operator $X$ from the
left and noting that $(\Box_g+m^2)E^-=0$, we obtain
\rf{6.16}.\hfill$\blacksquare$ 

\section{Physical interpretation}\label{Section7}
Using the notion of the Sobolev wavefront set (Definition
\ref{DefinitionA1}) we have generalized in this paper the previously
known positive frequency conditions to define a large new class of
quantum states for the Klein-Gordon field on arbitrary globally
hyperbolic spacetime manifolds (Definition
\ref{Definition1}). Employing the techniques of pseudodifferential and
Fourier integral operators we have explicitly constructed examples of
them on spacetimes with a compact Cauchy surface (Theorem
\ref{Theorem5.7}). We call these states adiabatic vacua because on
Robertson-Walker spacetimes they include a class of quantum states
which is already well-known under this name (Theorem
\ref{Theorem6.2}). We order the adiabatic vacua by a real number $N$
which describes the Sobolev order beyond which the positive frequency
condition may be perturbed by singularities of a weaker nature. Our
examples show that these additional singularites may be of negative
frequency or even non-local type (Lemma \ref{Lemma5.5}). Hadamard
states are naturally included in our definition as the adiabatic
states of infinite order.\\
To decide which orders of adiabatic vacua are physically admissible we
have investigated their corresponding GNS-representations: Adiabatic
vacua of order $N>5/2$ generate a quasiequivalence class of local
factor representations (in other words, a unique local primary
folium). For pure states on a spacetime with compact Cauchy surface -
a case which often occurs in applications - this holds true already
for $N>3/2$ (Theorems \ref{Theorem25} and
\ref{Theorem29}). Physically, locally quasiequivalent states can be
thought of as having a finite energy density relative to each
other. Primarity means that there are no classical observables
contained in the local algebras. Hence there are no local
superselection rules, i.e. the local states can be coherently
superimposed without restriction. For $N>3/2$ the local von Neumann
algebras generated by these representations contain no observables
which are localized at a single point (Theorem \ref{Theorem32}).
Together with quasiequivalence this implies that all the states become
indistinguishable upon measurements in smaller and smaller spacetime
regions (Corollary \ref{Corollary32a}). This complies well with the
fact that the correlation functions have the same leading
short-distance singularities, whence the states should have the same
high energy behaviour. Finally, the algebras are
maximal in the sense of Haag duality (Theorem \ref{Theorem34}) and
additive (Lemma \ref{Lemma1.5}). For a more thorough discussion of all
these properties in the framework of algebraic quantum field theory we
refer to \cite{Haag96}. Taken together, all these results suggest that
adiabatic vacua of order $N>5/2$ are physically meaningful states.\\
Furthermore we expect that the energy momentum tensor of the
Klein-Gordon field can be defined in these states by an appropriate
regularisation generalizing the corresponding results for Hadamard
states \cite{BFK96,Wald94} and adiabatic vacua on Robertson-Walker
spaces \cite{PF74}.\\

However, all the mentioned physical properties of the
GNS-representations are of a rather universal nature and therefore
cannot serve to distinguish between different types of states. How can
we physically discern an adiabatic state of order $N$ from one of
order $N'$, say, or from an Hadamard state? To answer this question we
investigate the response of a quantum mechanical model detector (a
so-called Unruh detector \cite{BD82,Unruh76}) to the coupling with the
Klein-Gordon field in an $N$-th order adiabatic vacuum state. So let
us assume we are given an adiabatic state $\omega_N$ of order $N$ of
the Klein-Gordon quantum field $\hat{\Phi}$ on the spacetime $\M$ and
its associated GNS-triple
$(\H_{\omega_N},\pi_{\omega_N},\Omega_{\omega_N})$ as in
Proposition \ref{Proposition1.3}(b). We consider a detector that moves
on a wordline $\gamma:\R\to\M,\,\tau\mapsto x(\tau)$, in $\M$ and is
described as a quantum mechanical system by a Hilbert space $\H_D$ and
a free time evolution w.r.t.\ proper time $\tau$. It shall be determined
by a free Hamiltonian $H_0$ with a discrete energy spectrum
$E_0<E_1<E_2<\ldots,\, E_0$ being the groundstate energy of $H_0$
(e.g.\ a harmonic oscillator). We assume that the detector has
negligible extension and is coupled to the quantum field $\hat{\Phi}$
via the interaction Hamiltonian
\beq
H_I:=\lambda M(\tau) \hat{\Phi}(x(\tau))\chi(\tau) \label{7.1}
\eeq
acting on $\H_D\otimes \H_{\omega_N}$, where $\lambda \in \R$ is a
small coupling constant, $M(\tau)=e^{iH_0\tau}M(0)e^{-iH_0\tau}$ the
monopole moment operator characterizing the detector and
$\chi\in\co{\R}$ a cutoff function that describes the adiabatic
switching on and off of the interaction. To calculate transition
amplitudes between states of $\H_D\otimes \H_{\omega_N}$ under the
interaction \rf{7.1} one uses most conveniently the interaction
picture, in which the field $\hat{\Phi}$ and the operator $M$ evolve
with the free time evolution (but the full coupling to the
gravitational background) whereas the time evolution of the states 
is determined by the interaction $H_I$. In this formulation the
perturbative $S$-matrix is given by \cite{BD82,BF00}
\beqa
S&=&\mathbf{1}+\sum_{j=1}^\infty \frac{(-i)^j}{j!} \int\!\!d\tau_1\ldots
\int\!\!d\tau_j \,T[H_I(\tau_1)\ldots H_I(\tau_j)]\nonumber\\
&=&\mathbf{1}+\sum_{j=1}^\infty
\frac{(-i\lambda)^j}{j!}\int\!\!d\tau_1\chi(\tau_1) \ldots
\int\!\!d\tau_j\chi(\tau_j)\,T[M(\tau_1)\ldots M(\tau_j)]\, T[\hat{\Phi}
(x(\tau_1))\ldots \hat{\Phi}(x(\tau_j))],\nonumber\\
&&\hfill\label{7.2}
\eeqa
where $T$ denotes the operation of time ordering. Let us assume that
the detector is prepared in its ground state $|E_0\ra$ prior to
switching on the interaction, and calculate in first order perturbation
theory ($j=1$ in \rf{7.2}) the transition amplitude between the
incoming state $\psi_{in}:=|E_0\ra \otimes \Omega_{\omega_N}\in
\H_D\otimes \H_{\omega_N}$ and some outgoing state
$\psi_{out}:=|E_n\ra \otimes \psi$, $n\not= 0$, where $|E_n\ra\in\H_D$
is the eigenstate of $H_0$ corresponding to the energy $E_n$ and
$\psi$ some one-particle state in the Fock space 
$\H_{\omega_N}$ (the scalar products of $\hat{\Phi}(x)\Omega_{\omega_N}$ 
with other states vanish in a quasifree representation): 
\[\la\psi_{out}, S\psi_{in}\ra = -i\lambda\la
E_n|M(0)|E_0\ra\int\!\!d\tau\,\chi(\tau)e^{i(E_n-E_0)\tau}
\la\psi|\hat{\Phi}(x(\tau))\Omega_{\omega_N}\ra .\]
From this we obtain the probability $P(E_n)$ that a transition 
to the state $|E_n\ra$ occurs in the detector by summing over a complete set of
(unobserved) one-particle states in $\H_{\omega_N}$:
\begin{eqnarray*}
P(E_n)&=& \lambda^2 |\la E_n|M(0)|E_0\ra|^2 \int\!\!d\tau_1\int\!\!d\tau_2\,
e^{-i(E_n-E_0)(\tau_1-\tau_2)}\chi(\tau_1)\chi(\tau_2)
\Lambda_N(x(\tau_1),x(\tau_2))\\
&=& \lambda^2 |\la E_n|M(0)|E_0\ra|^2 {\cal F}(E_n-E_0).
\end{eqnarray*}
Here, $|\la E_n|M(0)|E_0\ra|^2$ describes the model dependent
sensitivity of the detector, whereas
\[{\cal F}(E):=
\int_{-\infty}^\infty\!\!d\tau_1\,\int_{-\infty}^\infty\!\!d\tau_2\,
e^{-iE(\tau_1-\tau_2)}
\chi(\tau_1)\chi(\tau_2)\Lambda_N(x(\tau_1),x(\tau_2)) \]
is the well-known expression for the detector response function
depending on the two-point function $\Lambda_N$ of the adiabatic state
$\omega_N$. Inspection of the formula shows that it is in fact 
obtained from $\Lambda_N\in\Dp{\M\times\M}$
by restricting $\Lambda_N$ to $\gamma\times\gamma\subset\M\times\M$,
multiplying this restricted distribution pointwise 
by $\chi\otimes\chi$ and taking the Fourier transform at $(-E,E)$:
\[
{\cal F}(E) = 2\pi \l((\Lambda_N|_{\gamma\times\gamma})\cdot
(\chi\otimes\chi)\r)^\wedge(-E,E).
\]
It follows from the very definition of $\Lambda_N$ (Definition
\ref{Definition1}) and Proposition \ref{PropositionA6} that
$\Lambda_N|_{\gamma\times\gamma}$ is a well-defined distribution on
$\R\times\R$ if $N>3/2$, since $N^*(\gamma)$ consists only of
space-like covectors. It holds
\[WF^{\prime s}(\Lambda_N|_{\gamma\times\gamma})\subset \varphi_* (C^+)
\quad\mbox{for}\,s<N-3/2,\]
where $\varphi_*$ is the pullback of the embedding
$\varphi:\gamma\times\gamma\to\M\times\M$. We now observe that
\beq
\{(\tau_1,-E;\tau_2,E)\in\R^4;\;E\geq 0\}\cap
\varphi_*(C^+)'=\emptyset \label{7.4}
\eeq
(this observation was already made by Fewster \cite{Fewster00} in the
investigation of energy mean values of Hadamard states). Hence there is
an open cone $\Gamma$ in $\R^2\setminus \{0\}$ containing
$(-E,E),E>0$, such that $WF^s(\Lambda_N|_{\gamma\times\gamma})\cap
\Gamma =\emptyset$. By \rf{A1a} we can write
$(\Lambda_N|_{\gamma\times\gamma})\cdot(\chi\otimes\chi) =u_1+u_2$
with $u_1\in H^s_{loc}(\R^2)$ for $s<N-3/2$ and $WF(u_2)\cap\Gamma
=\emptyset$. Since
$(\Lambda_N|_{\gamma\times\gamma})\cdot(\chi\otimes\chi)$ has compact
support we can assume without loss of generality that also $u_1$ and
$u_2$ have compact supports. From $WF(u_2)\cap \Gamma=\emptyset$ it
follows then that 
\[\hat{u}_2(\xi) =\O(\la\xi\ra^{-k})\quad\forall
k\in\N\;\forall\xi\in\Gamma, \]
whereas $u_1\in H^s_{comp}(\R^2)$ implies that
\begin{eqnarray*}
& &D^\alpha u_1 \in L^2_{comp}(\R^2)\quad \mbox{for}\;|\alpha|\leq
s<N-3/2,\,\mbox{cf.\ Prop.\ \ref{PropositionA3}}\\
&\Rightarrow& (D^\alpha u_1)^\wedge (\xi) = \xi^\alpha
\hat{u}_1(\xi)\;\mbox{is bounded}\\
&\Rightarrow& \hat{u}_1(\xi) =\O(\la\xi\ra^{-|\alpha|}).
\end{eqnarray*}
Taken together, we find that
$((\Lambda_N|_{\gamma\times\gamma})\cdot(\chi\otimes\chi))^\wedge(\xi)
=\O(\la\xi\ra^{-[N-3/2]})$ for $\xi\in\Gamma$, where
$[N-3/2]:=\max\{n\in\N_0;\;n<N-3/2\}$. Since $(-E,E)\in\Gamma,
E>0$, we can now conclude that
\[{\cal F}(E)=\O(\la E\ra^{-[N-3/2]})\]
for an adiabatic vacuum state of order $N>3/2$. (Note that this
estimate could be improved for the states constructed in Section
\ref{Section5} by taking into account that for them the singularities
of lower order are explicitly known, cf.\ Lemma \ref{Lemma5.5}, and
the sub-leading singularities also satisfy relation \rf{7.4}.)
This means that the
probability of a detector, moving in an adiabatic vacuum of order $N$,
to get excited to the energy $E$ decreases like $E^{-[N-3/2]}$ for
large $E$, in an Hadamard state it decreases faster than any inverse
power of $E$. We can therefore interpret adiabatic states of lower
order as higher excited states of the quantum field. One should
however keep in mind that all the states usually considered in
elementary particle physics (on a static spacetime, say) are of the
Hadamard type: ground states and thermodynamic equilibrium states are
Hadamard states \cite{Junker96}, particle states satisfy the
microlocal spectrum condition  (the generalization of the Hadamard
condition to higher $n$-point functions) \cite{BFK96}. We do not know
by which physical operation an adiabatic state of finite order 
could be prepared.\\

Although all results in this paper are concerned with the free
Klein-Gordon field, it is clear that our Definition \ref{Definition1}
is capable of a generalization to higher $n$-point functions of an
interacting quantum field theory in analogy to the microlocal spectrum
condition of Brunetti et al.\ \cite{BFK96}. 
In order to treat the pointwise product $(\Lambda_N)^2$ we write
\[
(\Lambda_N)^2 =(\Lambda_H)^2 +(\Lambda_N-\Lambda_H)^2 +2
(\Lambda_N-\Lambda_H) \Lambda_H,
\] 
where $\Lambda_H$ is the two-point function of any Hadamard state. It
follows from Proposition \ref{PropositionA5}, Lemma \ref{Lemma5}, and
Lemma \ref{Lemma5.1a} that these pointwise products are well-defined
if $N>-1$. It is well-known that
\[ WF'((\Lambda_H)^2) \subset C^+\oplus C^+,
\]
where $C^+\oplus C^+ :=\{(x_1,\xi_1+\eta_1;x_2,\xi_2+\eta_2)\in
T^*(\M\times\M)\setminus 0;\;(x_1,\xi_1;x_2,\xi_2),
(x_1,\eta_1;x_2,\eta_2)\in C^+\}$ \cite[Thm.~8.2.10]{HormI}. The
regularity of the other two terms can be estimated by Theorems 8.3.1
and 10.2.10 in \cite{Horm97} such that finally
\[
WF'^s((\Lambda_N)^2) \subset C^+\oplus C^+\quad\mbox{if}\;s<N-3.
\]
Higher powers of $\Lambda_N$ can be treated similarly. Thus, for
sufficiently large adiabatic order $N$, finite Wick powers (with
finitely many derivatives) can be defined. This should be sufficient
for the perturbative construction of a quantum field theory with an
interaction Lagrangian involving a fixed number of derivatives and
powers of the fields. Obviously one has to require more and
more regularity of the states if one wants to define higher and higher
Wick powers. This complies with a recent result of Hollands \&
Ruan \cite{HR01}.\\ It is also
clear that the notion of adiabatic vacua can be extended to other
field theory models than merely the scalar field. A first step in this
direction has been taken by Hollands \cite{Hollands01} for Dirac fields.\\
Finally we want to point out that, although the whole analysis in this
paper has been based on a given $\ci$-manifold $\M$ with smooth
Lorentz metric $g$, the notion of adiabatic vacua should be
particularly relevant for manifolds
with ${\cal C}^k$-metric. Typical examples that occur in general
relativity are star models: here the metric outside the star satisfies
Einstein's vacuum field equations and is matched on the boundary
${\cal C}^1$ to the metric inside the star where it satisfies Einstein's
equations with an energy momentum tensor of a suitable matter model as
a source term. In such a situation Hadamard states cannot even be
defined on a part of the spacetime that contains the boundary of the
star, whereas adiabatic states up to a certain order should still be
meaningful. This remark could e.g.\ be relevant for the
derivation of the Hawking radiation from a realistic stellar collapse
to a black hole.

\appendix
\section{Sobolev spaces}\label{AppendixA}
$H^s(\R^n),\,s\in \R$, is the set of all tempered distributions $u$ on
$\R^n$ whose Fourier transforms $\hat{u}$ are regular distributions
satisfying
\[ \| u\|^2_{H^s(\R^n)} := \int\!\la\xi\ra^{2s} |\hat{u}(\xi)|^2\,d^n\xi <
\infty.\]
For a domain ${\cal U}\subset \R^n$ we let
\[H^s({\cal U}) :=\{r_{\cal U}u;\;u\in H^s(\R^n)\}\]
be the space of all restrictions to ${\cal U}$ of $H^s$-distributions
on $\R^n$, equipped with the quotient topology
\[ \|u\|_{H^s({\cal U})} :=\inf \{\|U\|_{H^s(\R^n)};\;U\in H^s(\R^n),
r_{\cal U}U=u\}.\]
Moreover, we denote by $H_0^s(\ol{\cal U})$ the space of all elements
in $H^s(\R^n)$ whose support is contained in $\ol{\cal U}$. If ${\cal
U}$ is bounded with smooth boundary, then it follows from
\cite[Thm.~B.2.1]{HormIV} that $\co{{\cal U}}$ is dense in
$H_0^s(\ol{\cal U})$ for every $s$ and that $H_0^s(\ol{\cal U})$ is the
dual space of $H^s({\cal U})$ with respect to the extension of the
sesquilinear form
\[ \int \bar{u}v\,d^nx,\quad u\in \co{{\cal U}},v\in \ci({\ol{\cal
U}}).\]
If $\Sigma$ is a compact manifold without boundary we choose a
covering by coordinate neighborhoods with associated coordinate maps,
say $\{U_j,\kappa_j\}_{j=1,\ldots,J}$ with a subordinate partition of
unity $\{\varphi_j\}_{j=1,\ldots,J}$. Given a distribution $u$ on
$\Sigma$, we shall say that $u\in H^s(\Sigma)$ if, for each $j$, the
push-forward of $\varphi_j u$ under $\kappa_j$ is an element of
$H^s(\R^n)$. It is easy to see that this definition is independent of
the choices made for $U_j,\kappa_j$ and $\varphi_j$. The space
$H^s(\Sigma)$ is a Hilbert space with the norm
\[ \|u\|_{H^s(\Sigma)}:= \l(\sum_{j=1}^J
\|(\kappa_j)_*(\varphi_ju)\|^2_{H^s(\R^n)}\r)^{1/2}.\]
We denote by $\Delta$ the Laplace-Beltrami operator with respect to an
arbitrary metric on $\Sigma$. Then we have
\[H^{2k}(\Sigma)= \{u\in L^2(\Sigma);\;(1-\Delta)^ku\in L^2(\Sigma)\}
\]
for $k=0,1,2,\ldots$: Clearly, the left hand side is a subset of the
right hand side. Conversely, we may assume that $u$ has support in a
single coordinate neighborhood, so that we can look at the
push-forward $u_*$ under the coordinate map. The fact that both $u_*$
and $((1-\Delta)^ku)_*$ belong to $L^2(\R^n)$ implies that $u_*\in
H^{2k}(\R^n)$, hence $u\in H^{2k}(\Sigma)$. Moreover, this
consideration shows that the two topologies are equivalent (and in
particular independent of the choice of metric on $\Sigma$).\\
We may identify $H^{-s}(\Sigma)$ with the dual of $H^s(\Sigma)$ with
respect to the $L^2$-inner product in $\Sigma$.\\

Now let $\Sigma$ be a (possibly) non-compact Riemannian manifold which
is geodesically complete. The Laplace-Beltrami operator
$\Delta:\co{\Sigma}\to \co{\Sigma}$ is essentially selfadjoint by
Chernoff's theorem \cite{Chernoff73}. We can therefore define the
powers $(1-\Delta)^{s/2}$ for all $s\in\R$. By $H^s(\Sigma)$ we denote
the completion of $\co{\Sigma}$ with respect to the norm 
\[ \|u\|_{H^s(\Sigma)} := \|(1-\Delta)^{s/2}u\|_{L^2(\Sigma)}.\]
For $s\in 2\N_0$, this shows that $H^{2k}(\Sigma)$ is the set of all
$u\in L^2(\Sigma)$ for which $(1-\Delta)^ku\in L^2(\Sigma)$. We deduce
that this definition coincides with the previous one if $\Sigma$ is
compact and $s=2k$; using complex interpolation, cf.\ \cite[Ch.~I,
Thm.~4.2]{Taylor81}, equality holds for all $s\geq 0$. Moreover, we can
define a sesquilinear form
\[\la.,.\ra: H^{-s}(\Sigma)\times H^s(\Sigma)\to \C\]
by letting
\[ \la u,v\ra
:=\l((1-\Delta)^{-s/2}u,(1-\Delta)^{s/2}v\r)_{L^2(\Sigma)}.\]
This allows us to identify $H^{-s}(\Sigma)$ with the dual of
$H^s(\Sigma)$, as in the compact case. In particular, the definition
of the Sobolev spaces on compact manifolds coincides also for negative $s$.\\

Now suppose that $\O$ is a relatively compact subset of $\Sigma$. 
We let $H^s(\O):=r_\O H^s(\Sigma)$, the restriction to $\O$ of
$H^s$-distributions on $\Sigma$, endowed with the quotient topology
\[ \|u\|_{H^s(\O)}:=\inf \{\|U\|_{H^s(\Sigma)};\;U\in H^s(\Sigma),
r_\O U=u\}.\]
This definition is local:
If $\O'$ is another relatively compact subset with smooth boundary
containing $\ol{\O}$, then we can find a function $f\in \co{\O'}$
with $f\equiv 1$ on $\O$. Hence, whenever there exists a $U\in
H^s(\Sigma)$ with $r_\O U=u$, then there is a $U_1\in H^s(\Sigma)$
with supp\,$U_1\subset \O'$ and $r_\O U_1 =u$, namely $U_1=fU$. 
We therefore obtain the same space and the same topology, if we
replace the right hand side by
\[ \inf\{\|U\|_{H^s(\Sigma)};\;U\in H^s(\Sigma), \supp U\subset
{\O'}, r_\O U=u\}.\]
Indeed, both definitions yield the same space, which also is a Banach
space with respect to both norms. As the first norm can be estimated
by the second, the open mapping theorem shows that both are
equivalent. Note that $H^s(\O)$ is independent of the particular
choice of $\O'$.\\
On $\co{\O}$ the topology of $H^s(\Sigma)$ is independent of the
choice of the Riemannian metric; moreover it coincides with that
induced from $H^s(\R^n)$ via the coordinate maps: This follows from
the fact that, for $s=0,2,4,\ldots,$ the spaces $H^s(\Sigma)$ are the
domains of powers of the Laplacian, together with interpolation and
duality. As a consequence, $H^s(\O)$ does not depend on the choice of
the metric, and its topology is that induced by the Euclidean
$H^s$-topology. \\

Finally we define the local Sobolev spaces
\begin{eqnarray*}
H^s_{loc} (\Sigma) &:=& \{u\in \Dp{\Sigma};\;\int\!\la\xi\ra^{2s}
|\widehat{\kappa_* (\varphi u) }(\xi)|^2\,d^n\xi <\infty \;
\mbox{for all coordinate maps}\\
& &\kappa:{\cal U}\to\R^n,\,{\cal U}\subset\Sigma,\;
\mbox{and all}\;\varphi\in\co{{\cal U}}\} \\
H^s_{comp}(\Sigma) &:=& \{u\in H^s_{loc}(\Sigma);\;\supp u\;
\mbox{compact}\}. 
\end{eqnarray*}
We have the following inclusions of sets
\[ H^s_{comp}(\O)\subset 
H^s_{comp}(\Sigma)\subset H^s_{loc}(\Sigma)\subset
H^s(\O)\subset H^s_{loc}(\O)\]
for any relatively compact subset $\O$ of $\Sigma$.

\section{Microlocal analysis with finite Sobolev regularity}\label{AppendixB}
The $\ci$-wavefront set $WF$ of a distribution $u$ characterizes the
directions in Fourier space which cause the appearance of
singularities of $u$. It does however not specify the strength with
which the different directions contribute to the singularities. To
give a precise quantitative measure of the strength of singular
directions of $u$ the notion of the $H^s$-wavefront set 
$WF^s$ was introduced by Duistermaat \& H\"ormander \cite{DH72}. 
It is the mathematical tool which we use in the main part of
the paper to characterize the adiabatic vacua of a quantum field on a
curved spacetime manifold. To make the paper reasonably self-contained 
we present the definition of $WF^s$ and collect some results of the
calculus related to it which are otherwise spread over the
literature. They are mainly taken from \cite{DH72,Garding87, 
HormIII,Horm97,Taylor81}. All other notions from microlocal analysis which we 
use can also be found there or, in a short synopsis, in
\cite{Junker96}.\\
In the following let $X$ denote an open subset of $\R^n$.
\begin{dfn}\label{DefinitionA1}
Let $u\in \Dp{X},\;x_0\in X,\;\xi_0\in\R^n\setminus\{0\},\;s\in\R$. We 
say that $u$ is  $H^s$ (microlocally) in $(x_0,\xi_0)$ or that
$(x_0,\xi_0)$ is not in the {\bf $H^s$-wavefront set} of $u$
($(x_0,\xi_0)\notin WF^s(u)$) if there is a test function
$\varphi\in\co{X}$ with $\varphi(x_0)\not= 0$ and an open conic
neighborhood $\Gamma$ of $\xi_0$ in $\R^n\setminus\{0\}$ such that
\beq
\int_\Gamma\!\la\xi\ra^{2s} |\widehat{\varphi u}(\xi)|^2\,d^n\xi <\infty,
\label{A1}
\eeq
where $\la\xi\ra:=(1+|\xi|^2)^{1/2}$.
\end{dfn}
Note that, since $\varphi u\in {\cal E}'(X)$, there is for all
$(x,\xi)\in X\times \R^n\setminus 0$ a sufficiently small $s\in
\R$ such that $(x,\xi)\notin WF^s(u)$.
From the definition the following {\bf properties of $WF^s$} are
immediate:
\begin{itemize}
\item[(i)] $WF^s(u)$ is a local property of $u$, depending only on an
infinitesimal neighborhood of a point $x_0$, in the following sense:\\
If $u\in \Dp{X},\;\varphi\in\co{X}$ with $\varphi(x_0)\not= 0$ then
\[ (x_0,\xi_0)\in WF^s(u) \Leftrightarrow (x_0,\xi_0)\in WF^s(\varphi u)
\]
\item[(ii)] $WF^s(u)$ is a closed cone in
$X\times(\R^n\setminus\{0\})$, i.e.~in particular
\[ (x,\xi)\in WF^s(u) \Rightarrow (x,\lambda \xi)\in
WF^s(u)\quad\forall \lambda>0.\]
\item[(iii)] \[ WF^s(u)=\emptyset \Leftrightarrow u\in H^s_{loc}(X) \]
\item[(iv)] 
\beq
(x,\xi)\in WF^s(u)\Leftrightarrow \forall v\in
H^s_{loc}(X):\;(x,\xi) \in WF(u-v) \label{A1a}
\eeq
\item[(v)] \[ WF^{s_1}(u)\subset WF^{s_2}(u)\subset
WF(u)\quad\forall\,s_1\leq s_2 
\]
\item[(vi)] \[ WF^s(u_1+u_2)\subset WF^s(u_1)\cup WF^s(u_2)
\]
\item[(vii)] \[ WF(u)= \ol{\bigcup_{s\in\R}WF^s(u)} \]
\end{itemize}
As an example consider the $\delta$-distribution in $\Dp{\R^n}$. One
easily calculates from the criterion of the definition
\beq
WF^s(\delta) = \l\{ \begin{array}{ll} \emptyset,&s<-n/2\\
                    \{(0,\xi);\;\xi\in\R^n\setminus\{0\}\},&s\geq
-n/2. \end{array}\r.\label{A1b}
\eeq
The following proposition gives an important characterization of the
$H^s$-wavefront set in terms of pseudodifferential operators. Remember 
that $S^m_{\rho,\delta}(X\times\R^n)$ is the space of symbols of order 
$m$ and type $\rho,\delta$ ($m\in\R,\;0\leq\delta,\rho \leq 1$), and
$L^m_{\rho,\delta}(X)$ the corresponding space of pseudodifferential
operators on $X$.
\begin{prop}\label{PropositionA2}
Let $u\in\Dp{X}$. Then
\beq
WF^s(u) = \bigcap_{\begin{array}{c}A\in L^0_{1,0}\\Au\in H^s_{loc}(X)
\end{array}} char\, A =\bigcap_{\begin{array}{c}A\in L^s_{1,0}\\Au\in
L^2_{loc}(X) \end{array}} char\, A, \label{A2} 
\eeq
where the intersection is taken over all properly supported classical
pseudodifferential operators $A$ (having principal symbol $a(x,\xi)$)
and $char A:= a^{-1}(0)\setminus 0$ is the characteristic
set of $A$. 
\end{prop}
Also the {\bf pseudolocal property} of pseudodifferential operators can be
stated in a refined way taking into account the finite Sobolev
regularity:
\begin{prop}\label{PropositionA3}
If $A\in L^m_{\rho,\delta}(X)$ is properly supported, with
$0\leq\delta<\rho\leq 1$, and $u\in\Dp{X}$, then 
\[ WF^{s-m}(Au)\subset WF^s(u) \]
for all $s\in \R$, in particular \[A:H^s_{loc}(X)\to H^{s-m}_{loc}(X).\]
\end{prop}
From Propositions \ref{PropositionA2} and \ref{PropositionA3} we can
draw the following important {\bf conclusions}:
\begin{itemize}
\item[(i)] Since the principal symbol of a pseudodifferential operator 
is an invariant function on the cotangent bundle $T^*X$ we see from
\rf{A2} that $WF^s(u)$ is well-defined as a subset of
$T^*X\setminus 0$, i.e.~does not depend on a particular choice of
coordinates. By a partition of unity one can therefore define
$WF^s(u)$ for any paracompact smooth manifold $\M$ and $u\in\Dp{\M}$ as a
subset of $T^*\M\setminus 0$ and all results in this appendix
remain valid when replacing $X$ by $\M$.
\item[(ii)] If $Au\in H^s_{loc}(X)$ for some properly supported $A\in
L^m_{1,0}(X)$ then
\beq
 WF^{s+m}(u)\subset char A.\label{A3}
\eeq
This follows from Proposition \ref{PropositionA2} because, choosing
some elliptic $B\in L^{-m}_{1,0}(X)$, we have $BA\in L^0_{1,0}(X)$ and, 
by Proposition \ref{PropositionA3}, $BAu\in H^{s+m}_{loc}(X)$, and
therefore, by \rf{A2}, $WF^{s+m}(u)\subset char(BA)=char (A)$.
\item[(iii)] If $A\in L^{-\infty}(X)$, then $WF(Au)=\emptyset$ and
hence $WF^s(Au)=\emptyset$ for all $s\in\R$.
\item[(iv)] If $A\in L^m_{\rho,\delta}(X),\,0\leq\delta<\rho\leq 1$,
is a properly supported elliptic pseudodifferential operator, $u\in
\Dp{X}$, then \[ WF^{s-m}(Au)=WF^s(u)\]
for all $s\in\R$.\\
This is a consequence of the fact that an elliptic pseudodifferential
operator has a parametrix, i.e.~there is a properly supported $Q\in
L^{-m}_{\rho,\delta}(X)$ with $QAu=u+Ru$ and $AQu=u+R'u$ for some
$R,R'\in L^{-\infty}(X)$. Therefore, by Proposition \ref{PropositionA3},
\[ WF^s(u)=WF^s(QAu)\subset WF^{s-m}(Au)\subset WF^s(u).\]
\end{itemize}
The behaviour of $WF^s(u)$ for hyperbolic operators (like the
Klein-Gordon operator, which plays an important role in this work) 
is determined by the theorem of {\bf propagation of singularities} due 
to Duistermaat \& H{\"o}rmander \cite[Thm.~6.1.1']{DH72}. 
It states in particular
that, if $u$ satisfies $Au\in H^s_{loc}(X)$ for $A\in L^m_{1,0}(X)$
with real principal symbol $a(x,\xi)$ which is homogeneous of degree $m$,
then $WF^{s+m-1}(u)$ consists of complete bicharacteristics of $A$,
i.e.~complete integral curves in $a^{-1}(0)\subset 
T^*X$ of the Hamiltonian vectorfield 
\[ H_a(x,\xi) := \sum_{i=1}^n \l[\frac{\partial a(x,\xi)}{\partial
x^i} \frac{\partial}{\partial \xi_i}-\frac{\partial a(x,\xi)}{\partial 
\xi_i} \frac{\partial}{\partial x^i}\r]. \]
The precise statement is as follows:

\begin{prop}\label{PropositionA4}
Let $A\in L^m_{1,0}(X)$ be a properly supported pseudodifferential
operator with real principal symbol $a(x,\xi)$ which is homogeneous of 
degree $m$. If $u\in \Dp{X}$ and $Au=f$ it follows for any $s\in \R$
that
\[ WF^{s+m-1}(u)\setminus WF^s(f)\subset a^{-1}(0)\setminus 0\]
and $WF^{s+m-1}(u)\setminus WF^s(f)$ is invariant under the
Hamiltonian vectorfield $H_a$.
\end{prop}
It is well-known that the wavefront set gives sufficient criteria when 
two distributions can be pointwise multiplied, composed or restricted
to submanifolds. We reconsider these operations from the point of view 
of finite Sobolev regularity and obtain weaker conditions in terms of
$WF^s$.
We start with the regularity of the tensor product of two
distributions:
\begin{prop}\label{PropositionA7}
Let $X\subset \R^n,Y\subset \R^m$ be open sets and $u\in \Dp{X},
v\in\Dp{Y}$. \\
Then the tensor product $w:=u\otimes v\in \Dp{X\times Y}$ satisfies
\begin{eqnarray*}
 WF^r(w) &\subset& WF^s(u)\times WF(v)\cup WF(u)\times WF^t(v)\\
& &\cup\l\{
\begin{array}{ll} (\supp u\times \{0\})\times WF(v)\cup
WF(u)\times(\supp v \times \{0\}), & r=s+t \\
(\supp u\times \{0\})\times WF^t(v)\cup
WF^s(u)\times(\supp v \times \{0\}), & r=\min \{s,t,s+t\}.
\end{array} \r.
\end{eqnarray*}
\end{prop} 
The proof of this proposition can be adapted from the proof of Lemma
11.6.3 in \cite{Horm97}.\\
The pointwise product of two distributions $u_1,u_2\in \Dp{X}$ -- if
it exists -- is defined by convolution of Fourier transforms as the
distribution $v\in \Dp{X}$ such that $\forall x\in X\,\exists f\in
\D{X}$ with $f=1$ near $x$ such that for all $\xi\in \R^n$
\[
\widehat{f^2v}(\xi) =\frac{1}{(2\pi)^{n/2}}\int_{\R^n}
\widehat{fu_1}(\eta)\widehat{fu_2} (\xi-\eta)\,d^n\eta
\]
with absolutely convergent integral.
It is clear that for the integral to be absolutely convergent it is
sufficient that $\widehat{fu_1}(\eta)$ and $\widehat{fu_2}(\xi-\eta)$
decay sufficiently fast in the opposite directions $\eta$
resp.~$-\eta$, i.e.~that $u_1$ and $u_2$ are in Sobolev spaces of
sufficiently high order at $(x,\eta)$ resp.~$(x,-\eta)$. The precise
condition is the following:
\begin{prop}\label{PropositionA5}
Let $u_1,u_2\in \Dp{X}$. Suppose that $\forall (x,\xi)\in T^*X\setminus 
0\;\exists s_1,s_2\in\R$ with $s_1+s_2\geq 0$ such that
$(x,\xi)\notin WF^{s_1}(u_1)$ and $(x,-\xi)\notin WF^{s_2}(u_2)$.\\
Then the pointwise product $u_1 u_2$ exists.
\end{prop}
For a proof see \cite{Oberguggenberger86}.\\
Next we consider the restriction of distributions to submanifolds. Let 
$\Sigma$ be an $(n-1)$-dimensional hypersurface of $X$ (i.e.~there
exists a $\ci$-embedding $\varphi:\Sigma \to X$) with conormal bundle
\[
N^*\Sigma := \{(\varphi(y),\xi)\in
T^*X;\;y\in\Sigma,\varphi_*(\xi)=0\}.
\]
We can define the restriction
$u_\Sigma \in \Dp{\Sigma}$ of $u\in \Dp{X}$ to $\Sigma$ -- if it
exists -- as the mapping $f\mapsto \la u\cdot(f\delta_\Sigma),1\ra$, where
$f\delta_\Sigma: \ci(X)\to\C$ is the distribution given by
$(f\delta_\Sigma)(g):=\int_\Sigma fg,\,f\in \D{\Sigma}$. If $\Sigma$ is
locally given by $x^0=0$ then $f\delta_\Sigma$ is locally given by
$f(\vx)\delta(x^0)$, where $\delta(x^0)$ is the delta-function in the
$x^0$-variable. By a consideration analogous to \rf{A1b} we see that
\beq
WF^s(f\delta_\Sigma) \subset \l\{\begin{array}{ll}\emptyset, &s<-1/2\\
N^*\Sigma,&s\geq -1/2 \end{array}\r. .\label{A4}
\eeq
We obtain
\begin{prop}\label{PropositionA6}
Let $u\in\Dp{X}$ with $WF^s(u)\cap N^*\Sigma=\emptyset$ for some
$s>1/2$.\\
Then the restriction $u_\Sigma$ of $u$ is a well-defined distribution
in $\Dp{\Sigma}$, and
\[ WF^{r-1/2}(u_\Sigma)\subset \varphi_*WF^r(u) :=\{(y,\varphi_*
(\xi))\in T^*\Sigma;\; (\varphi(y),\xi)\in WF^r(u)\} \]
for all $r>1/2$.
\end{prop}
\begin{beweis}
Let $s>1/2$ and $WF^s(u)\cap N^*\Sigma =\emptyset$. 
It follows from \rf{A4} and Proposition \ref{PropositionA5} that the
product $u\cdot f\delta_\Sigma$ is defined. Suppose that $(y,\eta)\in
WF^{r-1/2}(u_\Sigma)$ for some $r>1/2$. By \rf{A1a} we have
$(y,\eta)\in WF(u_\Sigma -w)$ for each $w\in
H^{r-1/2}_{loc}(\Sigma)$. Since the restriction operator $H^r_{loc}(X)\to
H^{r-1/2}_{loc}(\Sigma)$ is onto \cite[Ch.~I, Thm.~3.5]{Taylor81}, there
exists a $v\in H^r_{loc}(X)$ for each $w$ such that
$w=v_\Sigma$. Hence we have for every $v\in H^r_{loc}(X)$
\[ (y,\eta)\in WF(u_\Sigma -v_\Sigma)=WF((u-v)_\Sigma)\subset
\varphi_* WF(u-v)\]
where we have used the standard result on the wavefront set of a
restricted distribution \cite[Thm.~2.5.11']{Horm71}. Applying \rf{A1a}
again we obtain the assertion.
\end{beweis}

The proposition can easily be generalized to submanifolds of higher
codimension by repeated projection.
From Proposition \ref{PropositionA7} and \ref{PropositionA6} one can
get an estimate for the $H^s$-wavefront set of the pointwise product
in Proposition \ref{PropositionA5} when noticing that $u_1u_2$ is the 
pull-back of $u_1\otimes u_2$ under the map $\varphi:X\to X\times
X,\,x\mapsto (x,x)$ and that
$\varphi_*(\xi_1,\xi_2)=\xi_1+\xi_2$. This estimate, however, is
rather poor and we will not present it here, 
better information on the regularity of products can be
gained e.g.~from \cite[Thm.~8.3.1 and Thm.~10.2.10]{Horm97}.\\
\begin{prop}\label{PropositionA8}
Let $X\subset \R^n, Y\subset \R^m$ be open sets, $u\in \co{Y}$ and let
${\cal K}\in \Dp{X\times Y}$ be the kernel of the continuous map
$K:\co{Y}\to \Dp{X}$.\\
Then we have for all $s\in \R$
\[
WF^s(Ku)\subset WF_X^s({\cal K}):=\{(x,\xi)\in T^*X\setminus
0;\;(x,\xi;y,0)\in WF^s({\cal K})\;\mbox{for some}\;y\in Y\}.
\]
\end{prop}
\begin{beweis}
Assume that $(x,\xi;y,0)\notin WF^s({\cal K})$ for some $(x,\xi)\in
T^*X\setminus 0, y\in Y$. By \rf{A1a} we can write ${\cal
K}={\cal K}_1+{\cal K}_2$ with ${\cal K}_1\in H^s_{loc}(X\times Y)$ and
$(x,\xi;y,0)\notin WF({\cal K}_2)$. Since $Ku=K_1u+K_2u$ and
$WF(K_2u)\subset WF_X({\cal K}_2)$ it follows that $(x,\xi)\notin
WF(K_2u)$. It remains to be shown that $K_1u \in H^s_{loc}(X)$,
because then it follows from \rf{A1a} that $(x,\xi)\notin WF^s(Ku)$,
i.e.~$WF^s(Ku)\subset WF^s_X({\cal K})$.\\
 To this end we localize ${\cal K}_1$ with test functions $\varphi\in
\co{X}$ and $\psi\in \co{Y}$ such that $\psi=1$ on $\supp u$ and
estimate for $\varphi (K_1u)= \varphi(K_1\psi u) = \int{\cal
K}_1'(x,y)u(y)\,d^my \in {\cal E}'(X)$, where ${\cal K}_1'(x,y) :=
\varphi(x){\cal K}_1(x,y)\psi(y)$:
\begin{eqnarray*}
\int d^n\xi\,(1+|\xi|^2)^s |\widehat{\varphi(K_1u)}(\xi)|^2 &=& \int
d^n\xi\, (1+|\xi|^2)^s \l|\int d^m\eta\, \hat{\cal K}_1'(\xi,-\eta)
\hat{u}(\eta)\r|^2 \\
&\leq& \int d^n\xi\, (1+|\xi|^2)^s \int d^m\eta\, (1+|\eta|^2)^t
|\hat{\cal K}_1'(\xi,-\eta)|^2 \\
& &\int d^m\theta\, (1+|\theta|^2)^{-t}
|\hat{u}(\theta)|^2 \\
&=& C\int d^n\xi \int d^m\eta\, (1+|\xi|^2)^s (1+|\eta|^2)^t |\hat{\cal
K}_1'(\xi,-\eta)|^2 \\
&\leq& C\int d^n\xi \int d^m\eta\, (1+|\xi|^2+|\eta|^2)^s |\hat{\cal
K}_1'(\xi,-\eta)|^2 
\end{eqnarray*}
which is finite since ${\cal K}_1'\in H_{comp}^s(X\times Y)$. The last
estimate was obtained by putting $t:=0$ if $s\geq 0$, and $t:=s$ if
$s<0$.
\end{beweis}

In the next proposition we generalize this result to the case where $u$ 
is a distribution in ${\cal E}'(Y)$. Then $Ku$ -- if it exists -- is
defined as the distribution in $\Dp{X}$ such that, for $\varphi\in
\co{X}$, 
\[
\la Ku,\varphi\ra = \la {\cal K}(1\otimes u),\varphi\otimes 1\ra. 
\]
\begin{prop}\label{PropositionA9}
Let $X\subset \R^n, Y\subset \R^m$ be open sets, ${\cal K}\in
\Dp{X\times Y}$ be the kernel of the continuous map $K:\co{Y}\to
\Dp{X}, u\in {\cal E}'(Y)$ and denote
\[
WF_Y^{\prime s}({\cal K}):=\{(y,\eta)\in T^*Y\setminus 0;\;(x,0;y,-\eta)\in 
WF^s({\cal K})\;\mbox{for some}\;x\in X\}. 
\]
If $\forall (y,\eta)\in T^*Y\setminus 0\,\exists s_1,s_2\in \R$
with $s_1+s_2\geq 0$ such that 
\beq
(y,\eta)\notin WF_Y^{\prime s_1}({\cal K})\cap WF^{s_2}(u), \label{A.20}
\eeq
then $Ku$ exists. If, in addition, $WF_Y({\cal K})=\emptyset$ and
$K(H^s_{\rm comp} (Y))\subset H^{s-\mu}_{loc}(X)$, then
\[
WF^{s-\mu}(Ku)\subset WF'({\cal K})\circ WF^s(u)\cup WF_X({\cal K}),
\]
where $WF'({\cal K}):= \{(x,\xi;y,-\eta)\in T^*X\times
T^*Y;\;(x,\xi;y,\eta)\in WF({\cal K})\}$ is to be regarded as a relation 
mapping elements of $T^*Y$ to elements in $T^*X$.
\end{prop}
\begin{beweis}
For the first part of the statement we only have to check that the
product ${\cal K}(1\otimes u)$ exists. Indeed, by Proposition
\ref{PropositionA7} we have $WF^{s_2}(1\otimes u)\subset (X\times
\{0\})\times WF^{s_2}(u)$ and, because of \rf{A.20}, for no point
$(y,\eta)\in T^*Y\setminus 0$ is $(x,0;y,-\eta)$ in
$WF^{s_1}({\cal K})$ and at the same time $(x,0;y,\eta)$ in
$WF^{s_2}(1\otimes u)$. Therefore, according to Proposition
\ref{PropositionA5}, the pointwise product ${\cal K}(1\otimes u)$
exists.\\
Given an open conic neighborhood $\Gamma$ of $WF^s(u)$ in $T^*Y$, we
can write $u=u_1+u_2$ with $u_1\in H^s_{loc}(Y)$ and
$WF(u_2)\subset\Gamma$. This is immediate from \rf{A1a} with the help
of a microlocal partition of unity. By assumption we have $Ku_1\in
H^{s-\mu}_{loc}(Y)$, and hence, by \cite[Thm.~8.2.13]{HormI}, 
\begin{eqnarray*}
WF^{s-\mu}(Ku)&\subset& WF(Ku_2)\subset WF^\prime ({\cal K}) \circ
WF(u_2)\cup WF_X({\cal K}) \\
&\subset& WF^\prime ({\cal K})\circ \Gamma \cup WF_X({\cal K}).
\end{eqnarray*}
Since $\Gamma$ was arbitrary, we obtain
\[ WF^{s-\mu}(Ku)\subset WF^\prime ({\cal K})\circ WF^s(u) \cup
WF_X({\cal K}). \]
\end{beweis}

The assumptions in the last proposition are tailored for application
to the case that $\cal K$ is the kernel of a Fourier integral
operator. Indeed, if $K\in I^\mu_\rho (X\times Y,C'),\,1/2<\rho \leq
1,$ where $C$ is locally the graph of a canonical transformation
from $T^*Y\setminus
0$ to $T^*X\setminus 0$, then $WF({\cal K})\subset C'$
\cite[Thm.~3.2.6]{Horm71} and
$K(H^s_{comp}(Y))\subset H^{s-\mu}_{loc}(X)$
\cite[Cor.~25.3.2]{HormIV} and the proposition
applies. For pseudodifferential operators we have $C=id$ and hence we get 
back the result of
Proposition \ref{PropositionA3}. In the next proposition we give
information about the smoothness of the kernel $\cal K$ itself:
\begin{prop}\label{PropositionA10}
Let $K\in I^\mu_\rho (X\times Y,\Lambda),\,1/2<\rho\leq 1,$ where
$\Lambda$ is a closed Lagrangean submanifold of $T^*(X\times
Y)\setminus 0$, and ${\cal K}\in \Dp{X\times Y}$ its kernel. Then
$WF^s({\cal K}) \subset WF({\cal K})\subset \Lambda$, more precisely
\begin{eqnarray*}
WF^s({\cal K}) = \emptyset &\mbox{if}& s<-\mu-\frac{n+m}{4},\\
\lambda \in WF^s({\cal K}) &\mbox{if}& s\geq
-\mu-\frac{n+m}{4} \;\mbox{and}\;\lambda \in \Lambda
\;\mbox{is a non-characteristic point of}\;K.
\end{eqnarray*}
\end{prop}
$K\in I^\mu_\rho(X\times Y,\Lambda)$ is said to be non-characteristic
at a point $\lambda\in \Lambda$ if the principal symbol has an inverse 
(as a symbol) in a conic neighborhood of $\lambda$. A proof of the
proposition can be found in \cite[Thm.~5.4.1]{DH72}.
\vspace{2cm}
\begin{center}\bf Acknowledgements
\end{center}
We want to thank Stefan Hollands, Fernando Lled\'o, J{\"o}rg Seiler
and Ingo Witt for helpful discussions. W.J.\ is grateful to the DFG for
financial support, to Bernd Schmidt for moral support,
and to Prof.\ B.-W.\ Schulze for the hospitable
reception in his ``Arbeitsgruppe Partielle Differentialgleichungen und
komplexe Analysis'' at Potsdam University.
%

%
\end{document}